\documentclass[twocolumn]{aastex62}

\let\pwiflocal=\iffalse \let\pwifjournal=\iffalse
\usepackage[utf8]{inputenc}
\usepackage{textcomp}
\usepackage{amsmath,amssymb,gensymb,times,graphicx,morefloats,color}
\usepackage{afterpage, bm}
\usepackage{natbib,hyperref}
\usepackage{mathrsfs}
\usepackage{mathtools}
\usepackage{url}
\usepackage{rotating}
\newcommand{\teff}{\ensuremath{T_{\text{eff}}}}

\newcommand{\fbol}{$F_{\mathrm{bol}}$}
\newcommand{\rsun}{$R_\odot$}

\newcommand{\msun}{$\rm{M}_\odot$}
\newcommand{\lsun}{$L_\odot$}
\newcommand{\rhosun}{$\rho_\odot$}
\newcommand{\degs}{$^\circ$}

\newcommand{\vsini}{$v$sin$i$}

\newcommand{\lii}{Li {\scshape i}}

\newcommand{\kms}{km s$^{-1}$}
\newcommand{\gaia}{{\it Gaia}}
\newcommand{\kepler}{{\it Kepler}}
\newcommand{\ktwo}{{\it K}2}
\newcommand{\tess}{{\it TESS}}

\newcommand{\spitzer}{{\it Spitzer}}

\newcommand{\targ}{HD~110082}

\shortauthors{Tofflemire et al.}
\shorttitle{THYME V: A Sub-Neptune Transiting a Young Star}

\received{Nov 24, 2020}
\revised{Jan 19, 2021}
\accepted{Jan 23, 2021}
\submitjournal{The Astronomical Journal}

\begin{document}

\title{\tess\ Hunt for Young and Maturing Exoplanets (THYME) V: \\ A Sub-Neptune Transiting a Young Star in a Newly Discovered 250 Myr Association}

\correspondingauthor{Benjamin M.\ Tofflemire}
\email{tofflemire@utexas.edu}

\author[0000-0003-2053-0749]{Benjamin M. Tofflemire}
\altaffiliation{51 Pegasi b Fellow}
\affiliation{Department of Astronomy, The University of Texas at Austin, Austin, TX 78712, USA}

\author[0000-0001-9982-1332]{Aaron C.\ Rizzuto}
\altaffiliation{51 Pegasi b Fellow}
\affiliation{Department of Astronomy, The University of Texas at Austin, Austin, TX 78712, USA}

\author[0000-0003-4150-841X]{Elisabeth R.\ Newton}
\affiliation{Department of Physics and Astronomy, Dartmouth College, Hanover, NH 03755, USA}
\affiliation{Department of Physics and Kavli Institute for Astrophysics and Space Research, Massachusetts Institute of Technology, Cambridge, MA 02139, USA}

\author[0000-0001-9811-568X]{Adam L.\ Kraus}
\affiliation{Department of Astronomy, The University of Texas at Austin, Austin, TX 78712, USA}

\author[0000-0003-3654-1602]{Andrew W.\ Mann}
\affiliation{Department of Physics and Astronomy, The University of North Carolina at Chapel Hill, Chapel Hill, NC 27599, USA} 

\author[0000-0001-7246-5438]{Andrew Vanderburg}
\altaffiliation{NASA Sagan Fellow}
\affiliation{Department of Astronomy, The University of Texas at Austin, Austin, TX 78712, USA}
\affiliation{Department of Astronomy, The University of Wisconsin-Madison, Madison, WI 53706, USA}

\author[0000-0003-3707-5746]{Tyler Nelson}
\affiliation{Department of Astronomy, The University of Texas at Austin, Austin, TX 78712, USA}

\author[0000-0002-1423-2174]{Keith Hawkins}
\affiliation{Department of Astronomy, The University of Texas at Austin, Austin, TX 78712, USA}

\author[0000-0001-7336-7725]{Mackenna L. Wood}
\affiliation{Department of Physics and Astronomy, The University of North Carolina at Chapel Hill, Chapel Hill, NC 27599, USA}

\author[0000-0002-4891-3517]{George Zhou}
\affiliation{Center for Astrophysics | Harvard \& Smithsonian, 60 Garden Street, Cambridge, MA 02138, USA}

\author[0000-0002-8964-8377]{Samuel N.\ Quinn}
\affiliation{Center for Astrophysics | Harvard \& Smithsonian, 60 Garden Street, Cambridge, MA 02138, USA}

\author[0000-0002-2532-2853]{Steve B.\ Howell}
\affiliation{NASA Ames Research Center, Moffett Field, CA 94035, USA}

\author[0000-0001-6588-9574]{Karen A.\ Collins}
\affiliation{Center for Astrophysics | Harvard \& Smithsonian, 60 Garden Street, Cambridge, MA 02138, USA}

\author[0000-0001-8227-1020]{Richard P.\ Schwarz}
\affiliation{Patashnick Voorheesville Observatory, Voorheesville, NY 12186, USA}

\author[0000-0002-3481-9052]{Keivan G.\ Stassun}
\affiliation{Department of Physics and Astronomy, Vanderbilt University, VU Station 1807, Nashville, TN 37235, USA}

\author[0000-0002-0514-5538]{Luke G.\ Bouma}
\affiliation{Department of Astrophysical Sciences, Princeton University, 4 Ivy Lane, Princeton, NJ 08544, USA}

\author[0000-0002-2482-0180]{Zahra Essack}
\affiliation{Department of Earth, Atmospheric and Planetary Sciences, Massachusetts Institute of Technology, Cambridge, MA 02139, USA}
\affiliation{Kavli Institute for Astrophysics and Space Research, Massachusetts Institute of Technology, Cambridge, MA 02139, USA}

\author[0000-0002-4047-4724]{Hugh Osborn}
\affiliation{NCCR/PlanetS, Centre for Space \& Habitability, University of Bern, Bern, Switzerland}
\affiliation{Department of Physics and Kavli Institute for Astrophysics and Space Research, Massachusetts Institute of Technology, Cambridge, MA 02139, USA}

\author[0000-0003-0442-4284]{Patricia~T.~Boyd}
\affiliation{Astrophysics Science Division, NASA Goddard Space Flight Center, Greenbelt, MD 20771, USA}

\author{G{\'a}bor~F{\H u}r{\'e}sz}
\affiliation{Department of Physics and Kavli Institute for Astrophysics and Space Research, Massachusetts Institute of Technology, Cambridge, MA 02139, USA}

\author[0000-0002-5322-2315 ]{Ana~Glidden}
\affiliation{Department of Earth, Atmospheric and Planetary Sciences, Massachusetts Institute of Technology, Cambridge, MA 02139, USA}
\affiliation{Department of Physics and Kavli Institute for Astrophysics and Space Research, Massachusetts Institute of Technology, Cambridge, MA 02139, USA}

\author[0000-0002-6778-7552]{Joseph D.\ Twicken}
\affiliation{SETI Institute, Mountain View, CA 94043, USA}
\affiliation{NASA Ames Research Center, Moffett Field, CA 94035, USA}

\author[0000-0002-5402-9613]{Bill Wohler}
\affiliation{SETI Institute, Mountain View, CA 94043, USA}
\affiliation{NASA Ames Research Center, Moffett Field, CA 94035, USA}

\author[0000-0002-8058-643X]{Brian McLean}
\affiliation{Mikulski Archive for Space Telescopes}

\author[0000-0003-2058-6662]{George~R.~Ricker}
\affiliation{Department of Physics and Kavli Institute for Astrophysics and Space Research, Massachusetts Institute of Technology, Cambridge, MA 02139, USA}

\author[0000-0001-6763-6562]{Roland~Vanderspek}
\affiliation{Department of Physics and Kavli Institute for Astrophysics and Space Research, Massachusetts Institute of Technology, Cambridge, MA 02139, USA}

\author[0000-0001-9911-7388]{David~W.~Latham}
\affiliation{Center for Astrophysics | Harvard \& Smithsonian, 60 Garden Street, Cambridge, MA 02138, USA}

\author[0000-0002-6892-6948]{S.~Seager}
\affiliation{Department of Physics and Kavli Institute for Astrophysics and Space Research, Massachusetts Institute of Technology, Cambridge, MA 02139, USA}
\affiliation{Department of Earth, Atmospheric and Planetary Sciences, Massachusetts Institute of Technology, Cambridge, MA 02139, USA}
\affiliation{Department of Aeronautics and Astronautics, MIT, 77 Massachusetts Avenue, Cambridge, MA 02139, USA}

\author[0000-0002-4265-047X]{Joshua~N.~Winn}
\affiliation{Department of Astrophysical Sciences, Princeton University, 4 Ivy Lane, Princeton, NJ 08544, USA}

\author[0000-0002-4715-9460]{Jon~M.~Jenkins}
\affiliation{NASA Ames Research Center, Moffett Field, CA 94035, USA}


\begin{abstract}

\noindent The detection and characterization of young planetary systems offers a direct path to study the processes that shape planet evolution. We report on the discovery of a sub-Neptune-size planet orbiting the young star HD 110082 (TOI-1098). Transit events we initially detected during \tess\ Cycle 1 are validated with time-series photometry from \spitzer. High-contrast imaging and high-resolution, optical spectra are also obtained to characterize the stellar host and confirm the planetary nature of the transits. The host star is a late F dwarf ($M_\star = 1.2 M_\odot$) with a low-mass, M dwarf binary companion ($M_\star = 0.26 M_\odot$) separated by nearly one arcminute ($\sim$6200 AU). Based on its rapid rotation and Lithium absorption, HD 110082 is young, but is not a member of any known group of young stars (despite proximity to the Octans association). To measure the age of the system, we search for coeval, phase-space neighbors and compile a sample of candidate siblings to compare with the empirical sequences of young clusters and to apply quantitative age-dating techniques. In doing so, we find that HD 110082 resides in a new young stellar association we designate MELANGE-1, with an age of 250$^{+50}_{-70}$ Myr. Jointly modeling the \tess\ and \spitzer\ light curves, we measure a planetary orbital period of 10.1827 days and radius of $R_p = 3.2\pm0.1 R_\earth$. HD 110082 b's radius falls in the largest 12\% of field-age systems with similar host star mass and orbital period. This finding supports previous studies indicating that young planets have larger radii than their field-age counterparts. 

\end{abstract}


\section{Introduction}
\label{intro}

The demographics of short-period transiting planets have been well-studied using data from the \kepler\ mission \citep{Howardetal2012,Dressing&Charbonneau2015,Fultonetal2017}. Overwhelmingly dominated by field stars ($\tau>1$ Gyr), they reflect the outcome of the evolutionary processes that define the planetary characteristics and architectures in which these systems will spend the majority of their lives. The demographics of these planets provide hints as to which processes may have dominated \citep[e.g.,][]{Lopez&Fortney2013,Owen&Wu2017,Fulton&Petigura2018}, but are ultimately limited in answering how these planets arrived in their current state. 

From their formation, planets have the potential to migrate, either by disk interactions  \citep{Lubow&Ida2010,Kley&Nelson2012} or planet-planet interactions \citep{Fabryckyetal2007,Chatterjeeetal2008}, thermally contract \citep[e.g.,][]{Fortneyetal2011}, and/or experience atmospheric losses \citep{Lammeretal2003,Lopez&Fortney2013,Ginzburgetal2018}, all on timescales where there are few observational constraints. With many of these processes theorized to operate on similar timescales, the most direct way to assess their relative impact is to find and characterize young planets as they pass through these evolutionary phases. 

The discovery of young ($\tau \lesssim 700$ Myr), transiting exoplanets was thrust forward with the re-purposed \kepler\ \ktwo\ mission. Pointed campaigns monitoring young open clusters and associations in the ecliptic plane found planets in the Hyades \citep{Mannetal2016a,Mannetal2018,Livingstonetal2018}, Praesepe \citep{Obermeieretal2016,Mannetal2017,Pepperetal2017,Rizzutoetal2018,Livingstonetal2019}, Upper Sco \citep{Mannetal2016b,Davidetal2016}, and Taurus \citep{Davidetal2019}, as well as planet-hosts associated with less-well-studied groups \citep{Davidetal2018a,Davidetal2018b}. In aggregate, these young planets have larger radii than their field-age counterparts \citep[e.g.,][and other works cited above]{Mannetal2018,Rizzutoetal2018,Bergeretal2018}, suggesting ongoing radial evolution (e.g., contraction, atmospheric loss) throughout the first several hundred million years. 

The \textit{Transiting Exoplanet Survey Satellite} (\tess; \citealt{Rickeretal2015}) offers the opportunity to expand the K2 population on two fronts. First, the \ktwo\ sample is comprised primarily of two ages, $\sim$10 and $\sim$700 Myr. As \tess\ surveys the northern and southern ecliptic hemispheres, it monitors the members of many young moving groups that bridge the current age gap \citep[e.g.,][]{Gagneetal2018}. Second, many of these groups are closer and therefore brighter than the \ktwo\ clusters/association, which will enable extensive follow-up observations. Measurements of planetary mass and atmospheric properties in particular will provide the best leverage for constraining evolutionary processes. 

This potential motivates the \tess\ Hunt for Young and Maturing Exoplanets (THYME) collaboration, which has reported on planets in Upper Sco \citep[$\sim$20 Myr;][]{Rizzutoetal2020}, the Tuc-Hor association ($\sim$40 Myr; \citealt{Newtonetal2019}; reported independently by \citealt{Benattietal2019}), the Ursa-Major moving group \citep[$\sim$400 Myr;][]{Mannetal2020}, and the Pisces–Eridanus stream ($\sim$120 Myr; \citealt{Newtonetal2021}). Our work complements those of others, such as discovery of the AU Mic planetary system \citep[$\sim$20 Myr;][]{Plavchanetal2020}, and the discoveries from  CDIPS \citep{Boumaetal2019,Boumaetal2020} and PATHOS \citep{Nardielloetal2019,Nardielloetal2020}. 

In this paper we report on a sub-Neptune-size planet transiting the young field star HD 110082 (TOI-1098). The transit discovery is made with \tess\ light curves (identified by the SPOC pipeline; \citealt{Jenkins2015,Jenkinsetal2016}) and confirmed with follow-up transit observations from the \textit{Spitzer} Space Telescope. Although previously indicated as a likely member of the $\sim$40 Myr Octans moving group \citep{Murphyetal2015}, our characterization of the stellar host (and its wide binary companion) reveals it to be more evolved, though still generally young. Without membership to a well-studied association that would supply precise stellar parameters (e.g., age, metallicity), we develop a scheme to find and characterize coeval, phase-space neighbors (``siblings'') that enables a more precise and robust age measurement than would be possible for a lone field star. Given that many stars in the solar neighborhood exhibit signatures of youth with no apparent association membership \citep[e.g.,][]{Bowleretal2019}, this approach provides useful age priors for young, unassociated planet hosts, allowing the systems to be used to benchmark planet evolution theory. 

The paper is outlined as follows. In Section \ref{obs} we describe the \tess\ discovery light curve, follow-up observations, and their reduction. Section \ref{meas} presents the characterization of the host star and its wide binary companion. In Section \ref{Age} we describe our scheme for finding coeval neighbors to HD 110082 and how they are used to constrain the age of the system. In Section \ref{analysis} we jointly model the \tess\ and \spitzer\ light curves to measure the planet parameters, address false-positive scenarios, and place limits on additional planets in the system. Finally, in Section \ref{discussion} we summarize our results and place HD 110082 b in context with other young exoplanets. 

\begin{deluxetable}{l c c}
\tablecaption{Stellar Properties of HD 110082}
\label{tab:stellar_props}
\tablewidth{0pt}
\tabletypesize{\footnotesize}
\tablecolumns{3}
\phd
\tablehead{
  \colhead{Parameter} &
  \colhead{Value} &
  \colhead{Source}
}
\startdata
\multicolumn{3}{c}{\textbf{Identifiers}} \\
HD        & 110082               & \citet{CannonPickering1920-12-14} \\
2MASS     & J12502212-8807158    & 2MASS \\
\gaia\ DR2& 5765748511163751936  & \gaia\ DR2 \\
TIC       & 383390264            & \citet{Stassunetal2018} \\
TOI       & 1098                 &  \\
\hline
\multicolumn{3}{c}{\textbf{Astrometry}} \\
$\alpha$ RA (J2000)          & 12:50:22.020 & \gaia\ DR2\\
$\delta$ Dec (J2000)         & $-$88:07:15.72 & \gaia\ DR2\\
$\mu_\alpha$ (mas yr$^{-1}$) & $-$18.7758 $\pm$ 0.0394 & \gaia\ DR2\\
$\mu_\delta$ (mas yr$^{-1}$) & $-$18.0863 $\pm$ 0.0394 & \gaia\ DR2\\
$\pi$ (mas)                  & 9.48625 $\pm$ 0.02391 & \gaia\ DR2\\
\hline
\multicolumn{3}{c}{\textbf{Photometry}} \\
{\it B} (mag)   & 9.755 $\pm$ 0.036 & Tycho-2\\
{\it V} (mag)   & 9.23 $\pm$ 0.03 & Tycho-2\\
{\it G$_{\rm BP}$} (mag)  & 9.39966 $\pm$ 0.002275 & \gaia\ DR2 \\
{\it G} (mag)   & 9.11523 $\pm$ 0.00034 & \gaia\ DR2\\
{\it G$_{\rm RP}$} (mag)  & 8.72138 $\pm$ 0.002098 & \gaia\ DR2\\
\tess\ (mag)    & 8.758 $\pm$ 0.006 & TIC\\
{\it J} (mag)   & 8.272 $\pm$ 0.039 & 2MASS\\
{\it H} (mag)   & 8.014 $\pm$ 0.049 & 2MASS\\
{\it K$_s$} (mag)   & 8.002 $\pm$ 0.031 & 2MASS\\
IRAC 4.5 $\mu m$ (mag) & 8.07 $\pm$ 0.01 & This Work \\
{\it W}1 (mag)  & 7.965 $\pm$ 0.025 & {\it WISE}\\
{\it W}2 (mag)  & 7.986 $\pm$ 0.021 & {\it WISE}\\
{\it W}3 (mag)  & 7.983 $\pm$ 0.019 & {\it WISE}\\
{\it W}4 (mag)  & 7.851 $\pm$ 0.135 & {\it WISE}\\
\hline
\multicolumn{3}{c}{\textbf{Kinematics \& Positions}} \\
RV (\kms)  & 3.63 $\pm$ 0.06 & This Work\\
U (\kms)               & $-$8.13 $\pm$ 0.06 & This Work\\
V (\kms)               & $-$4.58 $\pm$ 0.08 & This Work\\
W (\kms)               & $-$9.74 $\pm$ 0.05 & This Work\\
X (pc)                 & 51.66 $\pm$ 0.13 & This Work \\
Y (pc)                 & $-$79.79 $\pm$ 0.20 & This Work \\
Z (pc)                 & $-$44.83 $\pm$ 0.11 & This Work \\
Distance (pc)          & 105.10 $\pm$ 0.27 & \citet{Bailer-Jonesetal2018} \\
\hline
\multicolumn{3}{c}{\textbf{Physical Parameters}} \\
$P_{\rm rot}$ (d)      & 2.34 $\pm$ 0.07 & This Work\\
\vsini\ (\kms)         & 25.3 $\pm$ 0.3  & This Work\\
$F_{\rm Bol}$ (ergs\,s$^{-1}$\,cm$^{-2}$)& $5.56\pm0.28\times10^{-9}$  & This Work \\
$T_{\rm eff}$ (K)      & 6200 $\pm$ 100 & This Work\\
$M_\star$ (\msun)      & 1.21 $\pm$ 0.06 & This Work\\
$R_\star$ (\rsun) & 1.19 $\pm$ 0.06 & This Work\\
$L_\star$ (\lsun)      & 1.91 $\pm$ 0.04 & This Work\\
$\rho_\star$ (\rhosun) & 0.7 $\pm$ 0.1& This Work\\
Spectral Type          & F8V $\pm$ 1 & This Work \\
\lii\ EW (\AA)         & 0.09 $\pm$ 0.02 & This Work \\
log $g$                & 4.2 $\pm$ 0.3 & This Work \\
$[$Fe/H$]$             & 0.08 $\pm$ 0.05 & This Work\\
$[$Mg/Fe$]$             & -0.17 $\pm$ 0.02 & This Work \\
A(Li) (dex)            & 3.08 $\pm$ 0.044 & This Work \\
Age (Myr)              & 250$^{+50}_{-70}$ & This Work \\
\enddata
\end{deluxetable}

\section{Observations \& Data Reduction}
\label{obs}

\subsection{Photometry}
\label{phot}

\subsubsection{\tess}

The \tess\ primary mission surveyed the northern and southern ecliptic hemispheres in 26 sectors, each covering $24^\circ \times 96^\circ$ of sky with near-continuous photometric coverage over 27 days. Near the ecliptic poles, the footprints of individual sectors overlap providing extended temporal coverage. HD 110082 was observed by \tess\ with 2-m cadence as part of the Candidate Target List -- a pre-selected target list prioritized for the detection and confirmation of small planets \citep{Stassunetal2018}. Observations took place in Sectors 12 and 13 (May 24 through July 17, 2019). During the \tess\ extended mission, HD 110082 was observed again in Cycle 3, Sector 27 (July 5 through July 29, 2020).

\begin{figure*}[!t]
    \centering
    \includegraphics[width=1.0\textwidth]{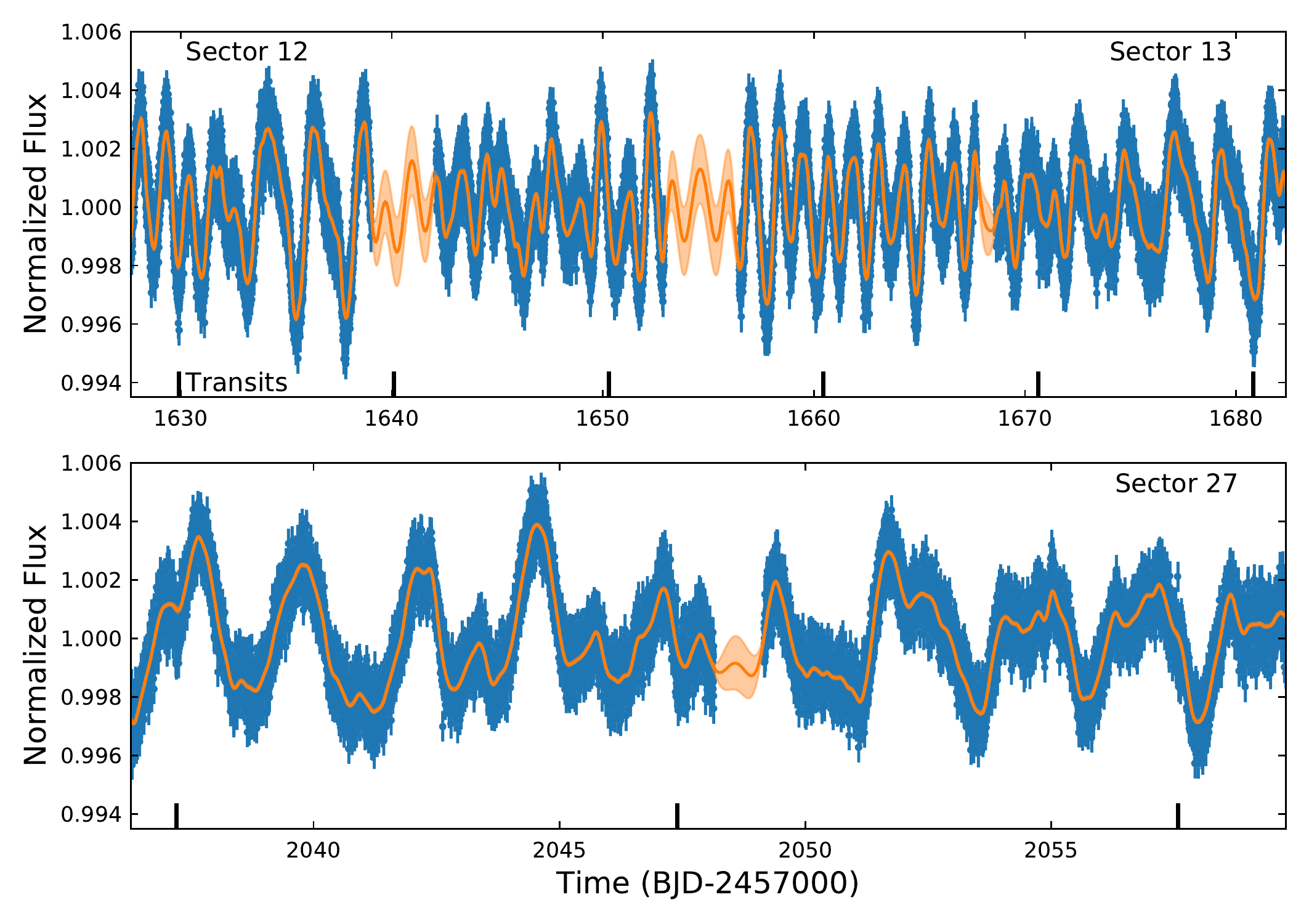}
    \caption{\tess\ light curve of HD 110082. Sectors 12 and 13 are presented in the top panel, with Sector 27 in the bottom panel. Flux-normalized, 2m cadence flux measurements are presented as blue points. A Gaussian-process model for the stellar variability is presented in orange (see Section \ref{rotp}). Transits are labeled with vertical ticks along the x-axes.}
    \label{fig:tess}
\end{figure*}

Raw \tess\ data are processed by the Science Processing Operations Center (SPOC) pipeline \citep{Jenkins2015,Jenkinsetal2016}, which performs pixel calibration, light curve extraction, deblending from near-by stars, and removal of common-mode systematic errors. We use the presearch data conditioning simple aperture photometry (PDCSAP) light curve \citep{StumpeKepler2012,SmithKepler2012, StumpeMultiscale2014}. The flux-normalized light curves are presented in Figure \ref{fig:tess}. Three large gaps are present in the top panel: two data down-links near the middle of sectors 12 and 13, and during the transition between sectors 12 and 13. The bottom panel presents Sector 27 with its own data down-link gap.

The SPOC Transiting Planet Search (TPS; \citealt{Jenkins2002,Jenkinsetal2010}) pipeline searches for ``threshold crossing events'' (TCEs) in the PDCSAP light curve after applying a noise-compensating matched filter to account for stellar variability and residual observation noise. TCEs with a $\sim$10.18 day period were detected independently in the SPOC transit search of the Sector 13 light curve and the combined light curves from Sectors 12-13. HD 110082 was alerted as a \tess\ Object of Interest (TOI), TOI-1098, in September, 2019 \citep{Guerrero2020}. 

\subsubsection{\spitzer}
\label{spitzer}

Due to possible membership to the Octans moving group, its rapid rotation period, and high-confidence transit detection from \tess, we obtained follow-up observations with the \spitzer\ {\it Space Telescope} to monitor two transits of HD 110082\,b. These observations occurred on November 25 and December 5, 2019 as part of our time-of-opportunity program dedicated to young planet-host follow-up (Program ID 14011, PI: Newton). Transit detections at infrared wavelengths are less affected by stellar variability and limb-darkening, while also providing refined ephemerides and the opportunity to search for transit-timing variations (TTVs). 

Our observation strategy followed \citet{Ingallsetal2012, Ingallsetal2016}, placing HD 110082 on the Infrared Array Camera (IRAC; \citealt{Fazioetal2004}) Channel 2 (4.5 $\mu$m) ``sweet spot'', using the ``peak-up'' pointing mode to limit pointing drift. Based on the source's brightness, we used the $32\times32$ pixel sub-array with 0.4 second exposure times. The observing sequence consisted of a 30-minute settling dither, an 8.5 hour stare covering the transit with $\sim$5.5 hours of out-of-transit baseline coverage, followed by a 10 minute post-stare dither. 

We process the \spitzer\ images of \targ~ to produce light curves using the Photometry for Orbits, Eccentricities, and Transits pipeline (POET\footnote{\url{https://github.com/kevin218/POET}}, \citealt{StevensonTransit2012,Campo_2011}). We first flag and mask bad pixels and calculate Barycentric Julian Dates for each frame. The centroid position of the star in each image is then estimated by fitting a 2-dimensional, elliptical Gaussian to the PSF in a 15 pixel square centered on the target position \citep{stevenson2010}. Raw photometry is produced using simple aperture photometry with apertures of 3.0, 3.25, 3.5, 3.75 and 4.0 pixel diameters, each with a sky annulus of 9 to 15 pixels. Upon inspection of the resulting raw light curves, we see no significant difference based on the choice of aperture size, and so the 3.5\,pixel aperture is used for the rest of the analysis. 

\begin{figure*}
    \centering
    \includegraphics[width=0.49\textwidth]{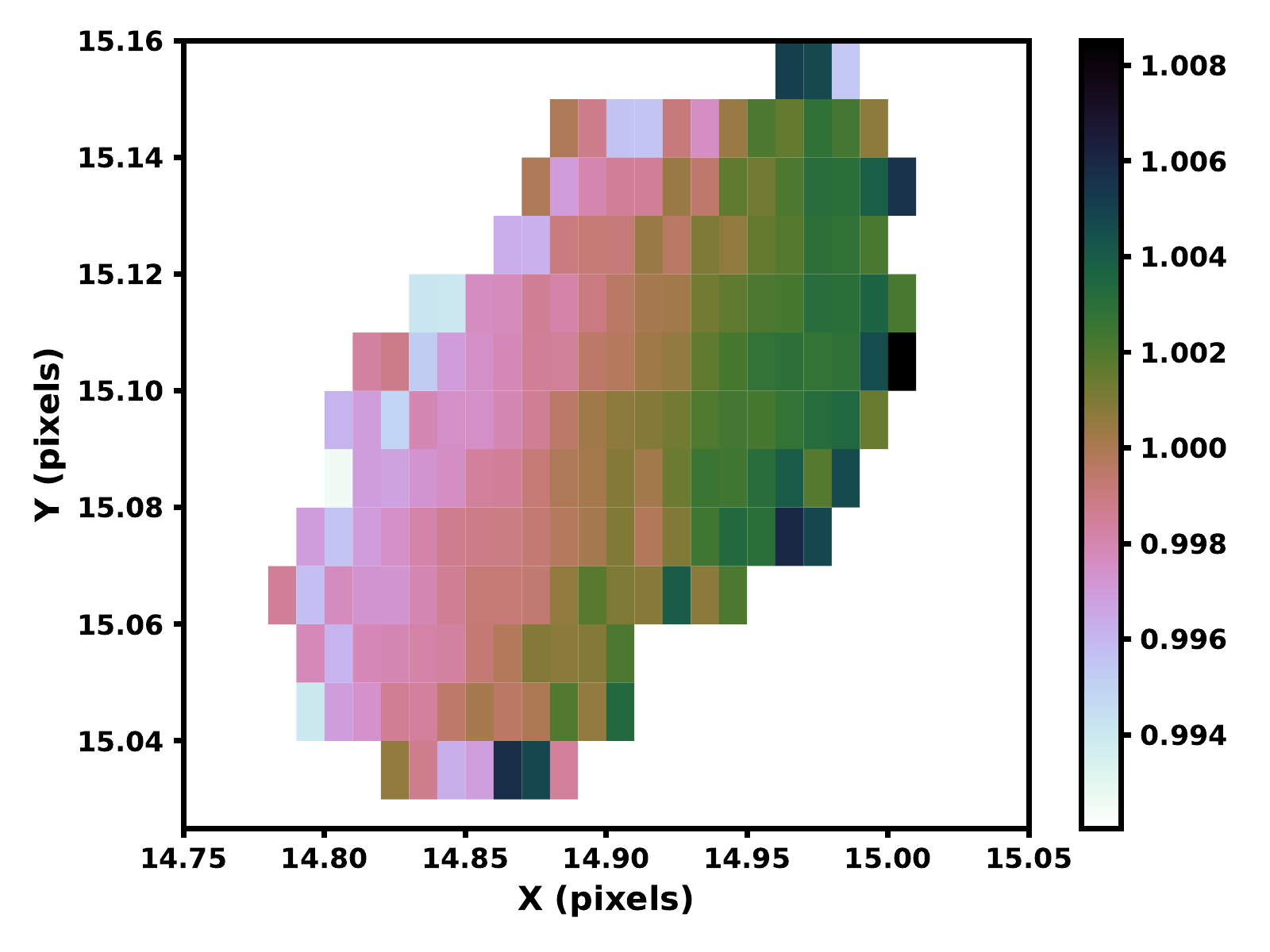}
    \includegraphics[width=0.49\textwidth]{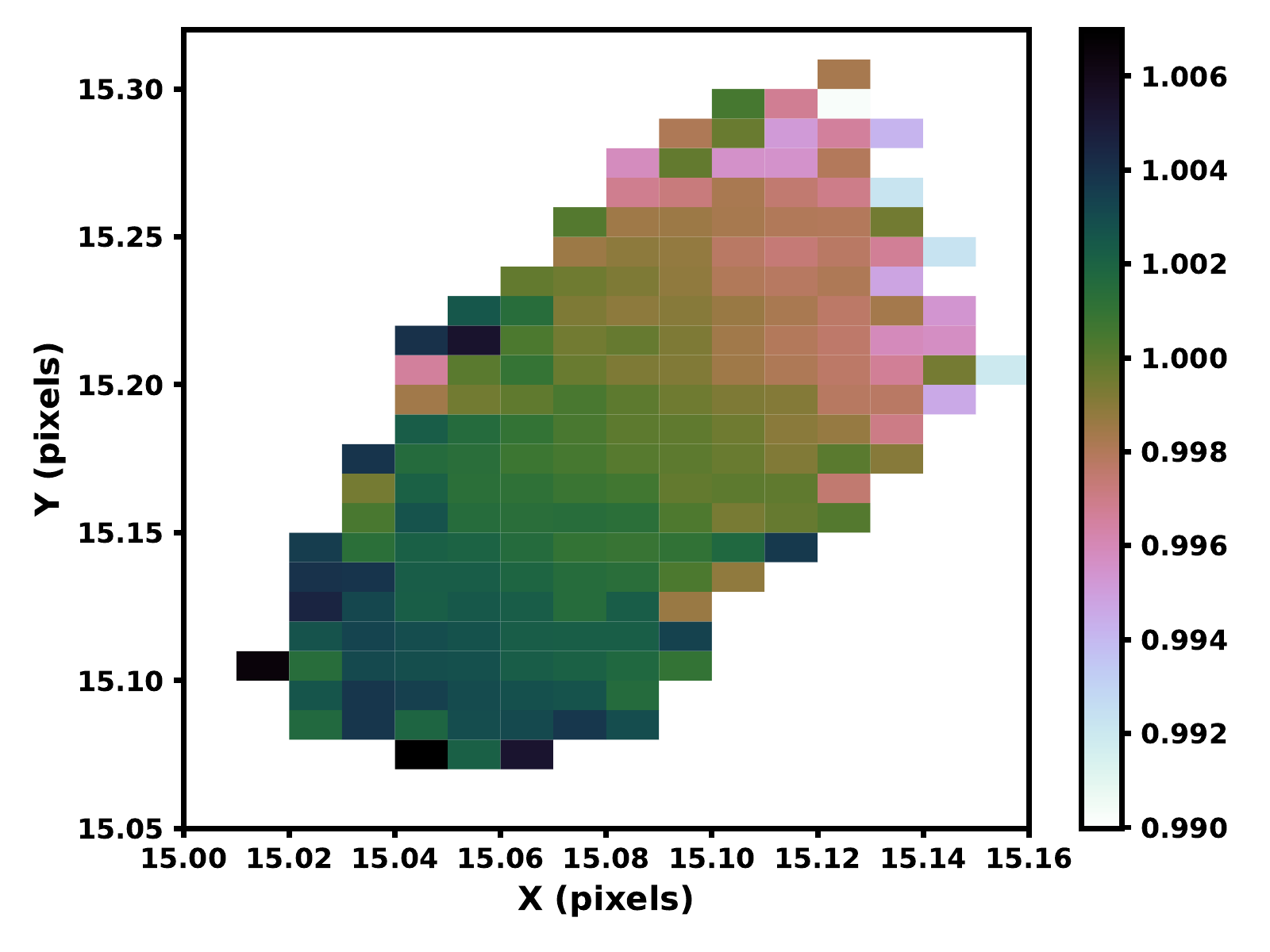}\\
    \includegraphics[width=0.48\textwidth]{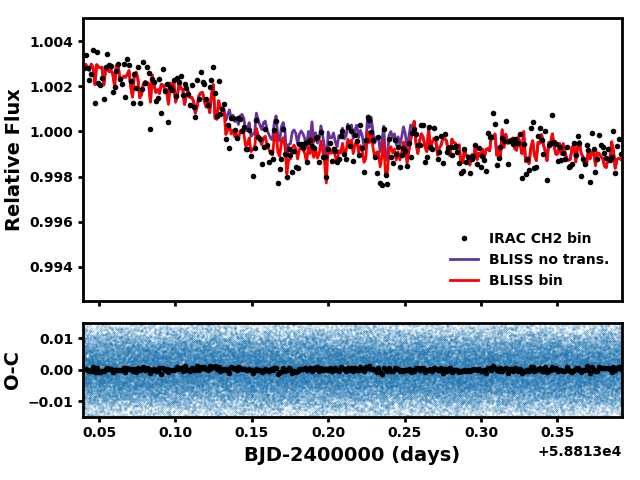}
    \includegraphics[width=0.48\textwidth]{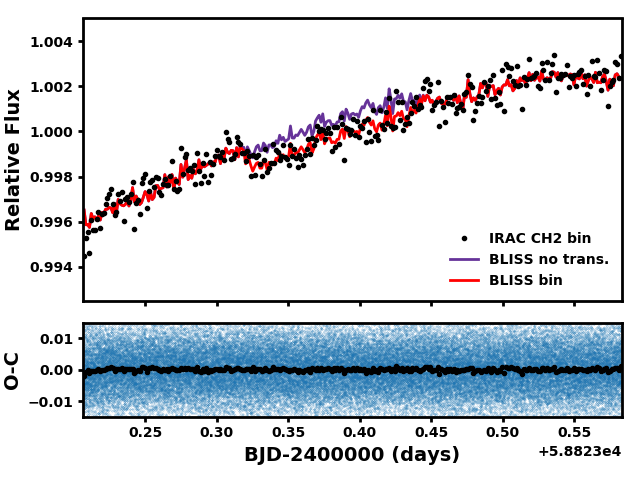}
\caption{\textbf{Top:} BLISS maps of the relative sub-pixel sensitivity for each transit of \targ\,b observed with \spitzer. \textbf{Bottom:} \spitzer{} IRAC Channel 2 data of two transit of \targ\,b. Black circles are the raw \spitzer\ data binned to 300 s, and the red line is the best fit BLISS model consisting of a quadratic ramp, transit, and subpixel sensitivity map binned to the same timesteps. The purple section shows the model without the transit included. We also show the residuals after subtraction of the best fit model, where blue points are the unbinned \spitzer\ residuals.}
\label{fig:spitzerbliss}
\end{figure*}

\spitzer{} detectors have significant intra-pixel sensitivity variations that can produce time dependent variability in the photometry of a target star as the centroid position of the star moves on the detector \citep{IngallsIntrapixel2012}. To correct for this source of potential systematics, we apply the BiLinearly Interpolated Subpixel Sensitivity (BLISS) Mapping technique of \citet{StevensonTransit2012}, which is provided as an optional part of the POET pipeline. BLISS fits a model to the transit of an exoplanet that consists of a series of time-dependent functions, including a ramp and a transit model, and a spatially dependent model that maps sensitivity to centroid position on the detector. There are several choices of ramp models that can be used to model the out-of-transit variability, and usually a linear or quadratic is used for IRAC Channel 2. We find that the quadratic ramp model fit both transits of \targ\,b well, with red noise within the expected bounds. This model does not include stellar variability, but given the low amplitude expected in the IR and the short observation time, any stellar variability present appears well-fit with the quadratic ramp. We model the planet transit as a symmetric eclipse without limb-darkening (expected to be minimal at this wavelength), as the purpose of this model is to remove time-dependent systematics and leave a spatially dependent sensitivity map. The time-dependent component of the model consists of the mid-transit time ($T_0$), the transit duration ($t_{14}$) and ingress and egress times ($t_{12/34}$) in units of the orbital phase, planet to star radius ratio ($R_P/R_\star$), system flux, a  quadratic ($r_2$) and linear ($r_1$) ramp coefficient, a constant ramp term ($r_0$; fixed to unity), and a ramp x offset in orbital phase ($\phi$). These parameters are explored with an Markov Chain Monte Carlo (MCMC) process, using 4 walkers with 100,000 steps and a burn in region of 50,000 steps. At each step, the BLISS map is computed after subtraction of the time dependent model components. 

\begin{deluxetable}{ccc}
\tabletypesize{\scriptsize}
\tablewidth{0pt}
\tablecaption{\spitzer\ IRAC Channel 2 BLISS model fit parameters for \targ\,b. \label{tab:blissfit}}
\tablehead{\colhead{Parameter} & \colhead{\targ\,b Tr. 1}& \colhead{\targ\,b Tr. 2}}
\startdata
$T_0$ (BJD)          & 2458813.19396$^{+0.00168}_{-0.00100}$ & 2458823.3808$^{+0.0018}_{-0.0018}$  \\ 
$t_{14}$ (phase)       & 0.01263$^{+0.00066}_{-0.00041}$       & 0.01288$^{+0.00068}_{-0.00052}$     \\ 
$R_P/R_\star$           & 0.0257$^{+0.0016}_{-0.0018}$          & 0.0278$^{+0.0013}_{-0.0014}$        \\ 
$t_{12/34}$ (phase)            & 0.00054$^{+0.00082}_{-0.00031}$       & 0.00168$^{+0.00082}_{-0.00046}$     \\ 
System Flux ($\mu$Jy)  & 105495$^{+114}_{-444}$                & 105660$^{+109}_{-262}$              \\ 
$r_2$                   &-0.49$^{+0.50}_{-0.57}$                & -3.59$^{+0.43}_{-0.42}$             \\ 
$r_1$                   & 0.042$^{+0.081}_{-0.040}$             & 0.138$^{+0.095}_{-0.071}$           \\ 
$r_0$                   & 1.0                                   & 1.0                                 \\ 
$\phi$                  & 0.940$^{+0.041}_{-0.052}$             & 0.985$^{+0.010}_{-0.013}$           \\ 
\tess\ T$_0^a$ (BJD)           & 2458813.202$^{+0.017}_{-0.013}$ & 2458823.385$^{+0.018}_{-0.013}$     \\ 
\enddata
\tablenotetext{a}{From the transit fit to only the \tess{} mission light curve}
\end{deluxetable}

Table \ref{tab:blissfit} lists the best fit parameters for each of the two transits of \targ\, observed with \spitzer{}. Figure \ref{fig:spitzerbliss} shows the intra-pixel sensitivity BLISS map, the \spitzer\ light curve of HD 110082 b with the best fit BLISS model, and light curve residuals. We find that the center of the transit in the \spitzer\ data agrees with the expected position based on our model of the \tess\ light curve (see Section \ref{analysis}). We then subtract the spatial component of the BLISS model, namely the sub-pixel sensitivity map, yielding a light curve corrected for positional systematics, which we use in a combined transit fit with the \tess\ light curve (Section \ref{analysis}).

\subsubsection{Las Cumbres Observatory -- SAAO 1.0 m}
\label{lco}

On 2020 May 16, HD 110082 was observed for $\sim$7 hr from the Las Cumbres Observatory Global Telescope (LCOGT) \citep{Brownetal2013} 1.0\,m network node at the Southern African Astronomical Observatory (SAAO) in Sutherland, South Africa. The observation centered temporally on a HD 110082 b transit, obtaining Sloan $i'$-band imaging using the 4096$\times$4096 pixel Sinistro CCD imager (0.39\arcsec\  pixel$^{-1}$). In total, 176 images were obtained, each with 110 s exposure times and 32 s inter-exposure times. The images were calibrated with the standard LCO BANZAI pipeline \citep{McCullyetal2018}, and photometric data were extracted with {\tt AstroImageJ} \citep{Collinsetal2017}. The long exposures saturated HD 110082 on the Sinistro detector, but allowed for photometry of the fainter neighboring stars to search for nearby eclipsing binaries (NEBS) that could be the source of the TESS detection. The TESS photometric apertures generally extend to $\sim1\arcmin$ from the target star. To account for possible contamination from the wings of neighboring star PSFs, we searched for NEBs from the inner edge of the saturated profile ($\sim2\farcs5$) out to $2\farcm5$ from the target star. If fully blended in the TESS aperture, a neighboring star that is fainter than the target star by 7.9 magnitudes in TESS-band could produce the SPOC-reported flux deficit at mid-transit (assuming a 100\% eclipse). To account for possible delta-magnitude differences between TESS-band and Sloan $i'$-band, we included an extra 0.5 magnitudes fainter (down to \textit{TESS}-band magnitude 17.2). Our search ruled out NEBs in all 15 neighboring stars that meet our search criteria by verifying that the 10-minute binned RMS of each star's light curve is at least a factor of five times smaller than the eclipse depth required in the star to produce the TESS detection. We also visually inspect each light curve to ensure that there is no obvious eclipse event. By process of elimination, we conclude that the transit is indeed occurring on HD 110082, or a star so close to HD 110082 that it was not detected by Gaia DR2. 

\subsection{High Contrast Imaging}
\label{zorro}
HD 110082 was observed with the Gemini South speckle imager, Zorro \citep{Matsonetal2019}, on 2020 January 14. Zorro provides simultaneous two-color, diffraction-limited imaging, reaching angular resolutions down to $\sim$0.02\arcsec. Observations of HD 110082 were obtained with the standard speckle imaging mode in the narrow-band 5620 \AA\ and 8320 \AA\ filters. These data were observed by the `Alopeke-Zorro visiting instrument team and reduced with their standard pipeline \citep[e.g.,][]{Howelletal2011}. 

\begin{figure}[!t]
    \centering
    \includegraphics[width=0.5\textwidth]{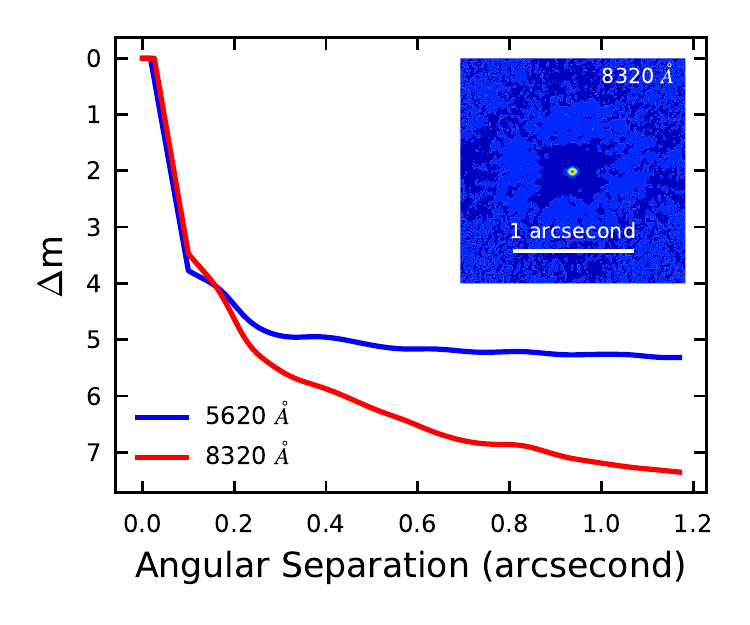}
    \caption{Detection limits for companions to HD 110082 from the Zorro speckle imager. Results for the narrow-band 5620 \AA\ and 8320 \AA\ filters are shown in blue and red, respectively. The inset shows the narrow-band 8320 \AA, reconstructed image.}
    \label{fig:zorro}
\end{figure}

Figure \ref{fig:zorro} presents the contrast curve for each filter where no additional sources are detected within the sensitivity limits. For an assumed age of $\tau = 300$ Myr at $D = 105$ pc, the evolutionary models of \citet{Baraffeetal2015} indicate corresponding physical limits: equal-mass companions at $\rho \sim 2.6$ AU; $M \sim 0.62 M_{\odot}$ at $\rho \sim 10.5$ AU; $M \sim 0.51 M_{\odot}$ at $\rho \sim 21$ AU; $M \sim 0.30 M_{\odot}$ at $\rho \sim 42$ AU; and $M \sim 0.20 M_{\odot}$ at $\rho > 105$ AU.

\subsection{Spectroscopy}
\label{spec}

\subsubsection{SMARTS -- CHIRON}

Five reconnaissance spectra of HD 110082 were obtained with the CHIRON spectrograph \citep{Tokovininetal2013} on the Small and Moderate Aperture Research Telescope System (SMARTS) 1.5m telescope located at the Cerro Tololo Inter-American Observatory (CTIO), Chile. Observations were obtained with 5 minute integrations between 2019 December 5 and 2020 February 1. The CHIRON spectrograph, operated by the SMARTS Consortium, is a fiber-fed, cross-dispersed echelle spectrometer that achieves a spectral resolution of $R\sim$ 80,000, in the image-slicer mode, with a wavelength range covering 4100--8700 \AA. Spectra are reduced with the standard data reduction pipeline, with wavelength calibration from Th-Ar calibration frames before and after each observation. The BJD of our observations are provided in Table \ref{tab:rv}.

Due to the short exposure times and the high airmass of our observations (owning to HD 110082's very low declination), each epoch provides a signal-to-noise ratio (SNR) of $\lesssim$ 15 (assessed near $\sim$6600 \AA). This SNR is sufficient for RV and \vsini\ determinations (Section \ref{rv}), but is too low to derive the stellar metallicity from a single epoch. As such, we combine the CHIRON spectra after correcting for heliocentric and systemic velocities, weighted by their pre-blaze-corrected flux. The combined spectrum has a SNR $\sim$ 25, which is sufficient to determine the metallicity as described in Section \ref{metal}.

\subsubsection{SOAR -- Goodman}
\label{soar}
On the nights of 2020 October 25 and 26 (UT) we observed seven stars in the neighborhood of HD 110082 (see Section \ref{neighbors} and Appendix \ref{appendix}) with the Goodman High-Throughput Spectrograph \citep{Goodman} on the Southern Astrophysical Research (SOAR) 4.1\,m telescope atop Cerro Pachón, Chile. The objective of these observations was to measure the \lii\ equivalent width (EW) of stars kinematically associated with HD 110082 (see Section \ref{liew_compare} and Appendix \ref{app:li}).
For this, we used the red camera, the 1200 l/mm grating, the M5 setup, and a 0.46\arcsec\ slit rotated to the parallactic angle, which provides a resolution of $R\simeq$5000 spanning 6350--7500\AA\ (covering both Li and H$\alpha$). For each target, we took five spectra with exposure times varying from 10\,s to 300\,s each. To correct for a slow drift in the wavelength solution of Goodman, we took Ne arcs throughout the night in addition to standard calibration data during the daytime. 

We perform bias subtraction, flat fielding, optimal extraction of the target spectrum, and map pixels to wavelengths using a 5th-order polynomial derived from the nearest Ne arc set. We stacked the five extracted spectra using a robust weighted mean. The stacked spectrum had peak SNR $>100$ for all targets. We placed each star in their rest wavelength using radial velocity standards taken with the same setup. 

\subsubsection{Las Cumbres Observatory -- NRES}
\label{nres}

The LCO Network of Robotic Echelle Spectrographs (NRES; \citealt{Siverdetal2018}) was used to obtain a high-resolution ($R\sim53,000$), optical spectrum (3800--8600\AA) of a candidate co-moving companion to HD 110082, \gaia\ DR2 6348657448190304512 (see Section \ref{neighbors}). This spectrum is used to measure the source radial velocitiy to assess its kinematic similarity to HD 110082. All data are reduced and wavelength calibrated using the standard LCO pipeline\footnote{\url{https://lco.global/documentation/data/nres-pipeline/}}. Only one source was observed with NRES due to the very low declination of HD 110082 and its neighbors, which proved difficult to acquire with NRES.

\subsection{Literature Photometry}
\label{lit_phot}

To aid in our characterization of HD 110082's stellar properties, we compile photometry from the literature, which we provide in Table \ref{tab:stellar_props}. Our analysis also includes photometry from the SkyMapper survey \citep{Wolfetal2018}, which we exclude from Table \ref{tab:stellar_props} for brevity.

\subsection{\gaia\ Astrometry and Limits on Close Neighbors from \gaia\ DR2}
\label{gaia}

We use astrometry from \gaia\ \citep{GAIA2016,GAIAdr2} to determine the space motion, and thereby, the potential membership of HD 110082 to known young moving groups and associations. The position and proper motion, and the parallax distance derived by \citet{Bailer-Jonesetal2018}, are presented in Table \ref{tab:stellar_props}. We used these values and our radial velocity (Section \ref{rv}) to compute the $UVW$ space velocity \citep[e.g.,][]{Johnson&Soderblom1987}.

Our null detection for close companions from speckle interferometry (Section~\ref{zorro}) is consistent with the limits set by the lack of {\it Gaia} excess noise, as indicated by the Renormalized Unit Weight Error \citep[$RUWE$;][]{Lindegrenetal2018}\footnote{\url{https://gea.esac.esa.int/archive/documentation/GDR2/Gaia_archive/chap_datamodel/sec_dm_main_tables/ssec_dm_ruwe.html}}. HD 110082 has $RUWE = 1.08$, consistent with the distribution of values seen for single stars. Based on a calibration of the companion parameter space that would induce excess noise (\citealt{Rizzutoetal2018}; \citealt{Belokurovetal2020}; Kraus et al., in prep), this corresponds to contrast limits of $\Delta G \sim 0$ mag at $\rho = 30$ mas, $\Delta G \sim 4$ mag at $\rho = 80$ mas, and $\Delta G \sim 5$ mag at $\rho \ge 200$ mas. For an assumed age of $\tau = 300$ Myr at $D = 105$ pc, the evolutionary models of \citet{Baraffeetal2015} indicate corresponding physical limits for equal-mass companions at $\rho \sim 3$ AU; $M \sim 0.61 M_{\odot}$ at $\rho \sim 8$ AU; and $M \sim 0.50 M_{\odot}$ at $\rho > 21$ AU.

\citet{Ziegleretal2018a} and \citet{Brandeker&Cataldi2019} mapped the completeness limit close to bright stars to be $\Delta G \sim 6$ mag at $\rho = 2\arcsec$, $\Delta G \sim 8$ mag at $\rho = 3\arcsec$, and $\Delta G \sim 10$ mag at $\rho = 6\arcsec$. For an  age of $\tau = 300$ Myr at $D = 105$ pc, the evolutionary models of \citet{Baraffeetal2015} indicate corresponding physical limits of $M \sim 0.37 M_{\odot}$ at $\rho = 210$ AU; $M \sim 0.17 M_{\odot}$ at $\rho = 315$ AU; and $M \sim 0.09 M_{\odot}$ at $\rho = 630$ AU. At wider separations, the completeness limit of the Gaia catalog ($G \sim 20.5$ mag at moderate Galactic latitudes; \citealt{GAIAdr2}) corresponds to a mass limit $M \sim 0.068 M_{\odot}$. 

The Gaia DR2 catalog does not report any comoving, codistant neighbors in the immediate vicinity of HD 110082 (within $\sim$10s of arcsconds), though as we discuss in Section~\ref{wide_binary}, there appears to be at least one very wide neighbor that is comoving and codistant.

Finally, we used the \gaia\ DR2 catalog to identify co-moving, co-distance sources that may be coeval neighbors to HD 110082, a process that we describe in more detail in Section \ref{Age} and Appendix A.

\subsubsection{A Wide Binary Companion}
\label{wide_binary}

Using \gaia\ DR2 we search for wide binary companions to HD 110082. We find a co-moving, co-distant companion 59.3\arcsec\ to the northeast, TIC 383400718 (\gaia\ DR2 5765748511163760640). This companion has a parallax difference consistent with zero and a projected physical separation of $\rho = 6240\pm20$AU, assuming the primary's distance. Its $RUWE$ value is 1.08, indicating its astrometric measurements are well-fit by a single-star model. This pair was previously reported as a likely wide binary by \citet{Tianetal2020}.

The relative proper motion between the pair is $\Delta \mu = 0.3\pm0.1$ mas yr$^{-1}$, corresponding to relative tangential velocity of $\Delta v_{tan} = 0.16\pm0.06$ \kms\ at the primary's distance. These values are less than the relative motion of a face-on circular orbit with a semi-major axis of $a = 1.26\rho$ \citep{Fischer&Marcy1992} and a combined mass of $\sim$1.5 $M_\odot$ (0.4 \kms). We therefore consider the likelihood that the pair are gravitationally bound is high, and explore the possible orbital parameters in Section \ref{binarity}. In Table \ref{tab:comp} we summarize the properties of the companion, providing \gaia\ astrometry, literature photometry, and derived stellar parameters. 

\begin{deluxetable}{l c c}
\tablecaption{Stellar Properties of the HD 110082 Binary Companion
\label{tab:comp}}
\tablewidth{0pt}
\tabletypesize{\footnotesize}
\tablecolumns{3}
\phd
\tablehead{
  \colhead{Parameter} &
  \colhead{Value} &
  \colhead{Source}
}
\startdata
\multicolumn{3}{c}{\textbf{Identifiers}} \\
2MASS     & J12514562-8806328    & 2MASS \\
\gaia\ DR2& 5765748511163760640  & \gaia\ DR2 \\
TIC       & 383400718            & \citet{Stassunetal2018} \\
\hline
\multicolumn{3}{c}{\textbf{Astrometry}} \\
$\alpha$ RA (J2000)          & 12:51:45.509 & \gaia\ DR2\\
$\delta$ Dec (J2000)         & $-$88:06:33.009 & \gaia\ DR2\\
$\mu_\alpha$ (mas yr$^{-1}$) & $-$18.486 $\pm$ 0.109 & \gaia\ DR2\\
$\mu_\delta$ (mas yr$^{-1}$) & $-$17.988 $\pm$ 0.095 & \gaia\ DR2\\
$\pi$ (mas)                  & 9.4309 $\pm$ 0.0643   & \gaia\ DR2\\
\hline
\multicolumn{3}{c}{\textbf{Photometry}} \\
{\it B} (mag)   & 18.923 $\pm$ 0.169 & USNO A2.0 \\
{\it G$_{\rm BP}$} (mag)  & 17.9899 $\pm$ 0.018276 & \gaia\ DR2 \\
{\it G} (mag)   & 16.4009 $\pm$ 0.001101 & \gaia\ DR2\\
{\it G$_{\rm RP}$} (mag)  & 15.156 $\pm$ 0.001917 & \gaia\ DR2\\
\tess\ (mag)    & 15.074 $\pm$ 0.008 & TIC\\
{\it J} (mag)   & 13.376 $\pm$ 0.026 & 2MASS\\
{\it H} (mag)   & 12.792 $\pm$ 0.027 & 2MASS\\
{\it K$_s$} (mag)   & 12.522 $\pm$ 0.027 & 2MASS\\
{\it W}1 (mag)  & 12.378 $\pm$ 0.023 & {\it WISE}\\
{\it W}2 (mag)  & 12.225 $\pm$ 0.022 & {\it WISE}\\
{\it W}3 (mag)  & 12.095 $\pm$ 0.245 & {\it WISE}\\
\hline
\multicolumn{3}{c}{\textbf{Positions}} \\
X (pc)                 & 52.00 $\pm$ 0.36 & This Work \\
Y (pc)                 & $-$80.27 $\pm$ 0.55 & This Work \\
Z (pc)                 & $-$45.08 $\pm$ 0.31 & This Work \\
Distance (pc)          & $105.72^{+0.73}_{-0.72}$ & \citet{Bailer-Jonesetal2018} \\
Separation, $\rho$ (\arcsec) & 59.3 & \gaia\ DR2\\
\hline
\multicolumn{3}{c}{\textbf{Physical Parameters}} \\
$P_{\rm rot}$ (d)      & 0.80 $\pm$ 0.01 & This Work\\
$F_{\rm Bol}$ (ergs\,s$^{-1}$\,cm$^{-2}$) & (1.96 $\pm$ 0.05) $\times$ 10$^{-11}$ & This Work \\
$T_{\rm eff}$ (K)      & 3250 $\pm$ 75 & This Work\\
$M_\star$ (\msun)      & 0.26 $\pm$ 0.01 & This Work \\
$R_\star$ (\rsun)      & 0.26 $\pm$ 0.02 & This Work\\
$L_\star$ (\lsun)      & (6.28 $\pm$ 0.29) $\times$ 10$^{-3}$ & This Work\\
Spectral Type          & M4V $\pm$ 1 & This Work \\
\enddata
\end{deluxetable}

\section{Measurements}
\label{meas}

\subsection{Fundamental Stellar Properties}
\label{stellar_props}

\subsubsection{Luminosity, Effective Temperature, and Radius}
\label{sed}
We derived fundamental parameters of \targ\ by fitting its spectral-energy-distribution (SED) using literature photometry and spectral templates of nearby young stars following \citet{Mannetal2015a}. In cases of low reddening (as expected for a star within $\simeq$100\,pc), the method yields precise (1-5\%) estimates of \fbol\ from the integral of the absolutely calibrated spectrum, $L_\star$ from \fbol\ and the \gaia\ distance, and \teff\ from comparing the calibrated spectrum to atmospheric models \citep{Baraffeetal2015}. Our fitting procedure also provides an estimate of $R_\star$ from the ratio of the absolutely calibrated spectrum to the model \citep[equal to $R_*^2/D^2$, ][]{Blackwell1977}.

We use optical and NIR photometry from the Two-Micron All-Sky Survey \citep[2MASS; ][]{Skrutskie2006}, the Wide-field Infrared Survey Explorer \citep[\textit{WISE}; ][]{allwise}, \gaia\ data release 2 \citep[DR2; ][]{Evansetal2018, GAIAdr2}, and Tycho-2 \citep{Hog2000}. To account for variability of the source, we assume all photometry had an addition error of 0.02\,mag (for optical) or 0.01\,mag (for near-infrared). We compare photometry to synthetic values derived from the template spectra. In addition to the overall scale of the spectrum, there are four free parameters of the fit that account for imperfect (relative) flux calibration of the spectra and both the model and template spectra used.

\begin{figure*}[!t]
    \centering
    \includegraphics[width=0.49\textwidth]{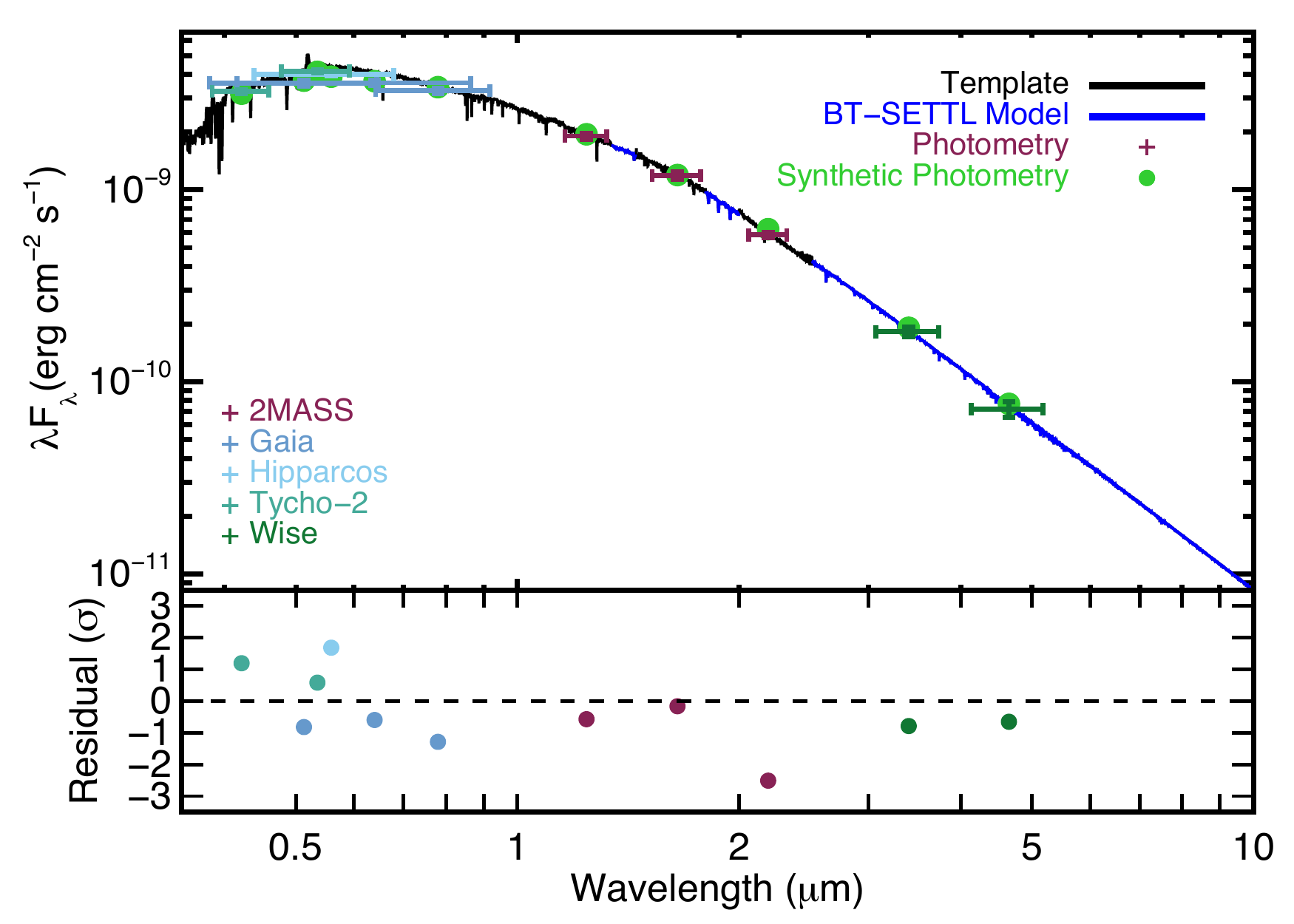}
    \includegraphics[width=0.49\textwidth]{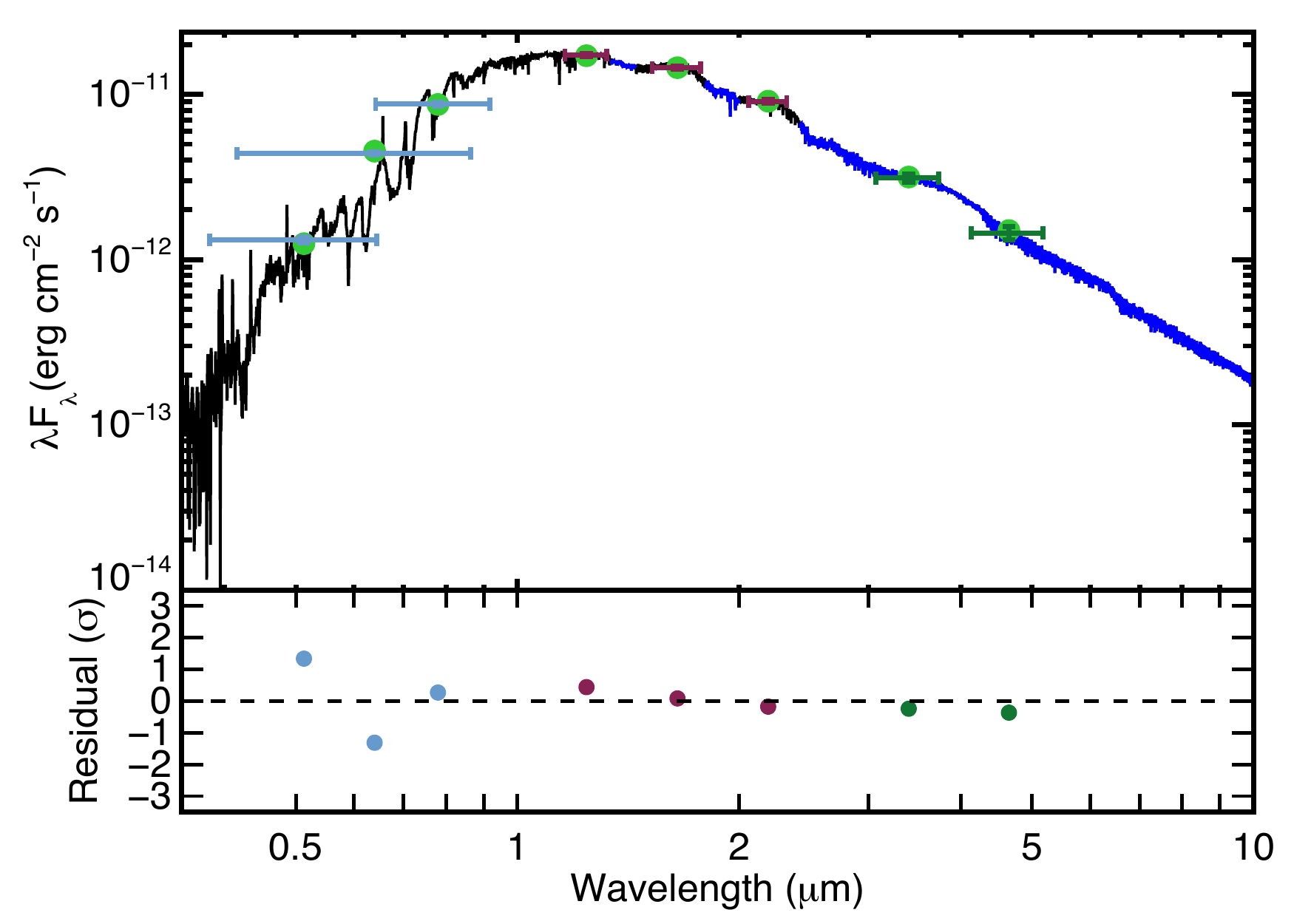}
    \caption{Best-fitting spectral template to the broadband photometry of HD 110082 (left) and its companion 2MASS J12514562-8806328 (right). Photometry are color-coded by the source, with vertical errors representing the uncertainty in the literature flux measurement and horizontal errors represent the filter width. Green points are the synthetic photometry of the best-fitting spectral template. Regions heavily affected by telluric contamination or beyond the edge of our template spectra are filled in with a BT-SETTL model \citep{Baraffeetal2015} shown in blue. Bottom panels for each star show the residuals of the fit in standard deviations.}
    \label{fig:sed}
\end{figure*}

We show the best-fit spectrum and photometry in Figure~\ref{fig:sed} (left). From our fit we estimate $T_{\rm eff}=6200\pm100$\,K, \fbol = $5.56\pm0.28\times 10^{-9}$\,erg\,cm$^{-2}$\,s$^{-1}$, $L_\star=1.91\pm0.03\,L_\odot$, and $R_\star=1.19\pm0.06\,R_\odot$. The best-fit template has a spectral type of F8 with templates within one spectral type having similar $\chi^2$ values. These values are included in Table \ref{tab:stellar_props}.

We perform an identical analysis for the M dwarf companion, using optical and near-infrared spectral templates from \citet{Gaidos2014} and available photometry from \gaia\ DR2, 2MASS, and {\it WISE}. The resulting fit is shown in Figure~\ref{fig:sed} (right), and the adopted stellar parameters are presented in Table~\ref{tab:comp}.

\subsubsection{Metallicity and Element Abundances}
\label{metal}

The atmospheric parameters of the star, including the \teff, $\log{g}$, and [Fe/H] are determined from the combined CHIRON spectrum using the LoneStar Stellar Parameter and Abundance Pipeline (Nelson et al, in preparation). LoneStar is a python-based code, which uses Bayesian inference to determine the \teff, $\log{g}$, [Fe/H], microturblent velocity, and the rotation velocity (\vsini) and propagates these posterior distributions to abundance measurements. We use the Markov Chain Monte Carlo (MCMC) algorithm implemented in {\tt emcee} \citep{Foreman-Mackeyetal2013} to sample the log-likelihood efficiently, therefore, we require a rapid, on-the-fly generation of synthetic spectra. 
 
In order to facilitate this, we linearly interpolate a grid of 9400 precomputed synthetic spectra using the linear ND interpolation implemented in \cite{scipy2}. The parameters in this grid range as follows: $3800 \le \teff \mathrm{(K)} \le 8000$ with steps of 100~K ; $3.5 \le \log{g} \mathrm{(dex)}\le 5.0$ with steps of 0.25~dex; $-3.0 \le \mathrm{[Fe/H]} \mathrm{(dex)}\le 1.0$ with steps 0.5~dex. The $v_{\mathrm{micro}}$\ value was synthesized with values of 0, 1, 2, and 4~\kms. Additionally, the \vsini\ convolution is computed at runtime using the functional form from \cite{gray_book}, as implemented in \citet{starfish_paper, starfish_code}. These pre-computed spectra were constructed from MARCs model atmospheres \citep{MARCs} using TURBOSPECTRUM \citep{turbospectrum} for radiative transfer. These models were generated with solar abundances from \cite{grevesse_solar_abunds}. We use Gaia-ESO line list version 5 \citep{Heiter2019} for atomic transitions. We also include molecular data for CH \citep{CH_data_98, CH_data_14}, C$_2$\citep{12C2_data, 12C13C_data}, and CN \citep{CN_data_93, CN_data_14}. The interpolator and grid enables us to generate theoretical spectra with a median per pixel error $<10^{-4}$. For comparison, it takes 0.18 seconds to interpolate through our grid compared to 90 seconds to generate a spectrum using TURBOSPECTRUM. 
 
The stellar parameters are determined by maximizing the log-likelihood function. We include a systematic error term which is added in quadrature with the normalized flux error vector. This attempts to account for underestimation in the error propagation in the spectral data reduction. This error term is fit simultaneously with the other parameters. The best fit stellar parameters are included in Table \ref{tab:stellar_props}, except for the $T_{eff}$, which is adopted from our SED-based approached (Section \ref{sed}). The value determined here is 6230$^{+90}_{-10}$ K, which is in good agreement with the other independently determined value. 

Once the stellar parameters are derived, the Lithium abundance is determined by generating synthetic spectra with A(Li) = [2.5, 2.6, 2.7, 2.8, 2.9, 3.0, 3.1, 3.2]. These synthetic spectra are compared against the observation using a $\chi^2_{\nu}$ minimization approach. Once an internal minimum is found, the best fit abundance is determined by interpolating the $\chi^2_{\nu}$ points with a cubic spline (to enforce smoothness) and resampling. The minimum of the resampled $\chi^2$ curve gives the best fit abundance. The Magnesium abundance is measured in a similar fashion to Lithium. The synthetic spectra are generated with [Mg/Fe] at $\pm 0.6, \pm 0.3, \mathrm{and} \ 0.0$ dex relative to the solar composition. 
 
Abundance errors for individual lines are calculated using the inverse Fisher matrix added in quadrature with the propagated uncertainty from \teff, $\log{g}$, [Fe/H], and v$_{\mathrm{micro}}$ (see Nelson et al. in prep.). If there are multiple lines for an element, the measured abundance is taken as average weighted by the inverse variance. 

The results of our detailed abundance measurements are listed in Table \ref{tab:stellar_props}. [Fe/H] and Magnesium and Lithium abundances point to a young thin-disk star. A(Li) is the most constraining, placing HD 110082 within the overlapping spread of Pleiades \citep{Bouvieretal2018} and Hyades \citep{Takedaetal2013} abundances at this temperature, and above the field-age Lithium-abundance sequence \citep{Ramirezetal2012}. 

To facilitate comparison to a broader range of known clusters and moving groups (specifically young groups), we also measure the equivalent width (EW) of \lii\ 6707.8 \AA. For many young groups, \lii\ EWs are more readily available than detailed abundances in the literature. In our measurement, we consider a 50 \AA\ region around the rest wavelength and use a rotationally-broadened, synthetic model \citep{Husseretal2013} to define adjacent line-free continuum regions. We fit the continuum with a linear slope, which is well-suited to the blaze-corrected, combined spectrum. The fit is made with an MCMC approach (using {\tt emcee}), providing a posterior distribution for the slope and y-intercept. We sample these posteriors 5000 times, where for each iteration, we normalize the spectrum and measure the EW through a numerical integration 10 separate times, adding random Gaussian noise determined by the RMS of the continuum. At each iteration, we also vary the integration bounds randomly over a normal distribution with a standard deviation of three resolution elements. The integration bounds are set initially at $\pm$10 \kms\ beyond the point where line absorption is expected from our rotational broadening measurement (see Section \ref{rv}).  This exercise results in a distribution of 50,000 EW measurement where the continuum fit, noise, and integration bounds are varied. From the mean and standard deviation of this distribution we measure a \lii\ EW of $0.09\pm0.02$. We note that our measurement includes an adjacent iron line (Fe {\scshape i} 6707.4\AA), which cannot be deconvolved given the SNR and rotational broadening of our spectrum. This is also generally the case for measurements of young moving group members (see Section \ref{liew_compare}).

\subsubsection{Mass and Density}
\label{mass_density}

To determine the mass of HD 110082, we use the empirical mass and radius relations derived in \citet{Torresetal2010}. The relations are a function of the stellar \teff, log $g$, and metallicity, and are calibrated by a large sample of detached eclipsing binaries. We determine the mass, radius, and luminosity within an MCMC formalism, using {\tt emcee}, that is fit to the \teff, log $g$, and metallicity values derived above. We placing a Gaussian prior on the luminosity determined from the SED-based approach (Section \ref{sed}), and employ 30 walkers in the fit. The convergence of the fit is determined by measuring the auto-correlation timescale, $\tau$, of all chains, ending the run when our measure of $\tau$ converges (fractional change less than 5\%) and the chain length is greater than 100 $\tau$. We discard the first five auto-correlation times as burn in. 

The mass and radius posteriors provide values of 1.21$\pm$0.02 M$_\odot$ and 1.20$\pm$0.04 R$_\odot$, respectively (uncertainties correspond to the 68\% confidence interval). The radius is in good agreement with the adopted value above. The formal mass uncertainty, $\sim$2\%, is less than the scatter in the empirical relation for solar type stars quoted in \citet{Torresetal2010}, $\sim$5\%, so we conservatively adopt the empirical uncertainty, corresponding to a mass measurement of 1.21$\pm$0.06 M$_\odot$. With the radius derived in Section \ref{sed}, we compute a stellar density of 0.7$\pm$0.1 $\rho_\odot$. This mass and density are provided in Table \ref{tab:stellar_props}.

\subsection{Stellar Rotation Period}
\label{rotp}

To measure the stellar rotation period in the presence of a rapidly evolving spot pattern (Figure \ref{fig:tess}), we model the \tess\ light curve with a scalable Gaussian process from the {\tt celerite} package \citep{Foreman-Mackeyetal2017}. Our covariance kernel consists of two damped, driven, simple harmonic oscillators, one at the stellar rotation period and the other at half the rotation period. The kernel is described by five terms: $P$, the primary period, $A$, the primary amplitude, $Q_1$, the damping timescale (or quality factor) of the primary period, $Mix$, the ratio of the primary to secondary amplitude ($A_2/A_1$), and $Q_2$, the damping timescale of the secondary period ($P/2$). In practice, we fit the $Mix$ term as $m$, where $A_1/A_2 = 1+e^m$ to avoid placing a prior on its bounds. We also fit the $Q1$ term as $\Delta Q$, where $\Delta Q = Q_1 - Q_2$, ensuring that the oscillator at the primary period has the largest quality factor. We also include a jitter term, $\sigma_{GP}$, to account for stellar variability not associated with rotational star-spot modulation. 

Before fitting these parameters to the \tess\ light curve, we mask six-hour windows centered on the transit ephemerides predicted by the TPS pipeline. (An independent search for transit events described in Section \ref{inject-recover} did not return any additional signals that would merit masking.) To remove sections of the light curve contaminated by flares, we iteratively fit the light curve with the GP using least-squares minimization, clipping 3$\sigma$ outliers. This process converged after two iterations. With flares and transits removed, we use the best fitting parameters as initial guesses in a Markov Chain Monte Carlo (MCMC) fit implemented with the {\tt emcee}, where all parameters are fit in natural logarithmic space. Our fit employed 50 walkers and used the convergence assessment scheme described in Section \ref{mass_density}. 

We fit the rotation period to each \tess\ Sector individually measuring periods of 2.28$\pm$0.05, 2.29$\pm$0.03, and 2.43$\pm$0.03 days for Sectors 12, 13, and 27, respectively. The small ($\sim$6\%) change in the rotation period from Sector 13 to 27 is statistically significant ($>$3$\sigma$), and may be the result of an evolving latitudinal spot pattern in the presence of differential rotation. The shift in the rotation period is well within the range of differential rotation measured from active stars with \kepler\ light curves \citep[e.g.,][]{Reinholdetal2013,Lanzaetal2014}. We adopt an average value, weighted by the individual measurement uncertainty, 2.34$\pm$0.07, with the standard deviation as the adopted uncertainty (provided in Table \ref{tab:stellar_props}). The remainder of our analysis makes use of the flare-masked light curve.

\subsection{Radial \& Rotational Velocities}
\label{rv}

We derive stellar radial velocities (RVs) and projected rotational velocities (\vsini) from the CHIRON spectra by computing spectral-line broadening functions \citep[BFs;][]{Rucinski1992,Tofflemireetal2019}. The BF is computed from a linear inversion of the observed spectrum with a narrow-lined template and represents a reconstruction of the average photospheric absorption-line profile (intensity vs radial velocity). From a best-fit line profile model, we measure the stellar RVs and \vsini.

We compute the BF for individual CHIRON echelle orders that span wavelengths from 4600--7200\AA. Removing those with heavy telluric contamination results in a total of 38 orders. The BFs are initially computed over a grid of PHOENIX model spectra \citep{Husseretal2013} with 100 K spacing in temperature and 0.5 dex in log $g$. The best-fitting, narrow-lined template is selected as that returning the most stable BF shape (lowest RMS) from order-to-order. For HD 110082, we find a best-fitting template with a temperature of 6200 K and a log $g$ of 4.5, which agrees well with the stellar parameters derived above. With a template selected, a visual inspection determines whether the star is single- or double-lined, single in this case, indicating the signal originates from a single star in the CHIRON fiber. 

The BFs from individual orders are then combined into a single, high signal-to-noise BF, weighted by the standard deviation of the BF baselines at high and low velocities where there is no stellar signal. The uncertainty on the BF profile is determined with a boot-strap Monte-Carlo (MC) approach that creates 10,000 combined BFs made from random draws with replacement of the 38 individual orders. The uncertainty at each velocity is set by the RMS of the MC BF samples. A line-profile model that includes instrumental broadening, rotational broadening (\vsini; \citealt{gray_book}), broadening from macroturbulent velocity ($v_{macro}$), an RV, and amplitude is fit to the BF with {\tt emcee}. The \vsini\ and $v_{macro}$ terms are explored in natural logarithmic space to avoid placing a strict prior at zero. The fit uses 32 walkers and follows the same convergence assessment methodology described in Section \ref{mass_density}. 

Over a time baseline of $\sim$58 days, we measure a systemic, barycentric RV of 3.63 \kms\ (weighted mean) that is constant within our measurement precision, with a standard error of 0.06 \kms. This measurement agrees with the \gaia\ DR2 value (the only literature RV available). We adopt our measured value due to its higher precision. From our four highest SNR observations, we measure a \vsini\ of $25.3\pm0.3$ \kms\ (Table \ref{tab:stellar_props}). The low SNR of the spectra do not allow us to place tight constraints on the macroturbulent velocity; all the measurements are consistent with zero, and we place a 95\% confidence upper limit of 3.5 \kms\ on its value. 

\begin{deluxetable}{l c c c}
\tablecaption{HD 110082 Radial Velocity Measurements
\label{tab:rv}}
\tablewidth{0pt}
\tabletypesize{\footnotesize}
\tablecolumns{4}
\phd
\tablehead{
  \colhead{BJD} &
  \colhead{RV} & 
  \colhead{$\sigma_{\rm RV}$} &
  \colhead{S/N} \\
  \colhead{} &
  \colhead{(\kms)} & 
  \colhead{(\kms)} &
  \colhead{}
}
\startdata
2458822.85502273 & 3.47 & 0.15 & 13 \\
2458866.87923667 & 3.77 & 0.52 & 6 \\
2458871.85450011 & 3.84 & 0.17 & 13 \\
2458879.86515306 & 3.75 & 0.16 & 14 \\
2458880.87977273 & 3.52 & 0.14 & 17 \\
\hline
\multicolumn{2}{l}{Weighted Mean: 3.63 (\kms)}\\
\multicolumn{2}{l}{RMS: 0.14 (\kms)}\\
\multicolumn{2}{l}{Std. Error: 0.06 (\kms)}\\
\enddata
\tablecomments{S/N evaluated in continuum region near $\sim$6600 \AA.}
\end{deluxetable}

\subsection{Stellar Inclination}

We measure the inclination of HD 110082's rotation axis following the Bayesian inference method presented in \citet{Masuda&Winn2020}. The projected rotational velocity (\vsini\ = $25.3\pm0.3$ \kms) is formally consistent with the derived equatorial rotation velocity ($v = 2 \pi R_\star/P_{\rm{rot}} = 25.6\pm1.3$ \kms), and corresponds to an inclination constraint of $i > 63$\degs\ at 99\% confidence. This measurement assumes the $i<90$\degs\ convention, due to the $i < 90$\degs\ and $i>90$\degs\ degeneracy inherent in this approach.

\subsection{Wide Binary Companion}
\label{binarity}

In this subsection we present measurements for the wide binary companion to HD 110082 and place constraints on the binary orbit of the pair. Table \ref{tab:comp} compiles derived stellar parameters following the methodology described in the subsections above where relevant, given only data from photometric and astrometric surveys, and \tess\ time-series photometry. 

Two aspects of our analysis vary between HD 110082 and its companion. First is the \tess\ light curve and the associated rotation period measurement. The companion's light curve is extracted from 30-minute full-frame images (FFIs) in a 1-pixel aperture with a systematics corrections and a regression against the HD 110082 rotation period (which is present in the raw light curve). The corrected light curve exhibits stable sinusoidal variability. We measure a rotation period of 0.80 days with a Lomb-Scargle periodogram \citep{Scargle1982} applied to data from Sectors 12 and 13. An uncertainty of 0.01 days is determined from the standard deviation of the rotation period measurements from 10 discrete regions of the light curve. 

The second difference is in the mass measurement. Without a log $g$ measurement, and with a mass that falls below the \citet{Torresetal2010} mass relation, we use the \citet{Mannetal2015a} empirical mass relationship for low-mass stars. For the companion's absolute $K_s$-band magnitude, we calculate a mass of 0.26 M$_\odot$ and adopt a 3\% uncertainty.

In summary, the companion is a 0.26 M$_\odot$, M4 star with an effective temperature of \teff\ = 3250 K and a short rotation period (0.8 d), consistent with a generally young age. 

\subsubsection{Binary Orbital Parameters}
\begin{figure}[!t]
    \centering
    \includegraphics[width=0.47\textwidth]{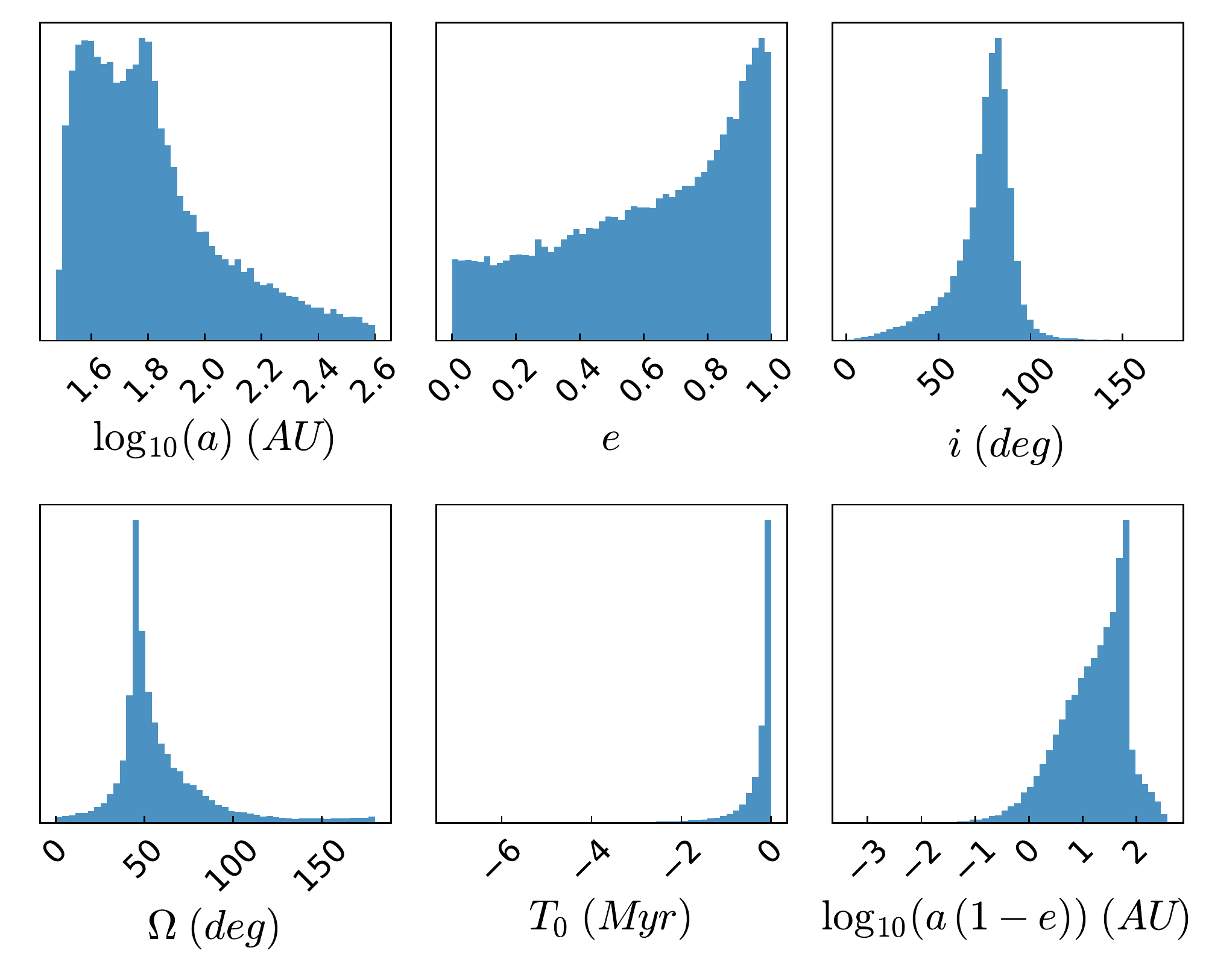}
    \caption{Posterior distributions of relevant orbital parameters of the HD 110082 wide binary pair. The semi-major axis posterior is truncated at log($a$) = 2.6 for display purposes only and does not reflect a real cutoff in the posterior. We exclude the posterior for $\omega$, the argument of periastron, for brevity as it is generally unconstrained by the observations.}
    \label{fig:LOFTI}
\end{figure}

We constrain the orbital parameters of the wide binary pair from \gaia\ proper motions and parallaxes. We use Linear Orbits for the Impatient ({\tt lofti\_gaiaDR2}; \citealt{Pearceetal2019,Pearceetal2020}), which given mass estimates and \gaia\ DR2 IDs for each component, queries the DR2 measurements and fits for the system's orbital elements using the rejection-sampling algorithm developed by \citet[Orbits for the Impatient]{Bluntetal2017}. Posteriors are returned for the six orbital elements: semi-major axis ($a$), eccentricity ($e$), inclination ($i$), longitude of ascending nodes ($\Omega$), argument of periastron ($\omega$), and the most recent epoch of periastron ($T_0$). Figure \ref{fig:LOFTI} presents posterior distributions for five of the binary orbital elements that are constrained by \gaia\ astrometry, as well as the posterior for the separation at closest approach (bottom right panel; the argument of periastron posterior is unconstrained and excluded from the figure).

The inclination posterior has a median and 68\% confidence interval of $i=78^\circ\,^{+17}_{-9}$, and has a mode of $82^\circ$. This result favors an edge on alignment, consistent with the interpretation that the binary and planetary orbital plans are near alignment. However, the position angle of the ascending nodes ($\Omega$) for the planet's orbit remains unknown. 

\section{Age Determination}
\label{Age}

HD 110082 was identified as likely member of the $\sim$40 Myr Octans moving group (Gagn{\'e}, J; private communication), motivating our initial investigation. In the following subsection we assess the membership of HD 110082 to Octans, ultimately finding it is not a member. With the goal of placing a more precise age estimate than is typically possible for a single star, we search for coeval phase-space neighbors, or ``siblings'', to the HD 110082 binary pair. We compare these stars to empirical cluster sequences and use gyrochonology and \lii\ EW-based age relationships to determine the system age.

We adopt the following ages for the clusters/associations used in our comparison based on the general consensus of various results and approaches in the literature: 
Octans $\sim$ 40 Myr \citep{TorresCetal2003,Murphyetal2015,Gagneetal2018},
Tuc-Hor $\sim$ 40 Myr \citep{daSilvaetal2009,Krausetal2014,Herczeg&Hillenbrand2015},
Columba, Carina, and Argus $\sim$ 40 Myr \citep{daSilvaetal2009,Elliottetal2016,Gagneetal2018},
Pleiades $\sim$ 125 Myr \citep{Staufferetal1998,Dahm2015,Bouvieretal2018},
Praesepe $\sim$ 700 Myr \citep{Kraus&Hillenbrand2007,Delormeetal2011,Brandt&Huang2015},
Hydaes $\sim$ 700 Myr \citep{Perrymanetal1998,deBruijneetal2001,Gossageetal2018}.

\subsection{Assessing Octans Membership}
With the addition of an RV from our high-resolution spectra, HD 110082 has a 99\% membership probability to the Octans moving group according to the BANYAN $\Sigma$ algorithm \citep{Gagneetal2018}\footnote{\url{http://www.exoplanetes.umontreal.ca/banyan/}}. Its wide binary companion also has a 99\% membership probability based on its astrometric measurements (an RV has not been measured).

\begin{figure}[!t]
    \centering
    \includegraphics[width=0.47\textwidth]{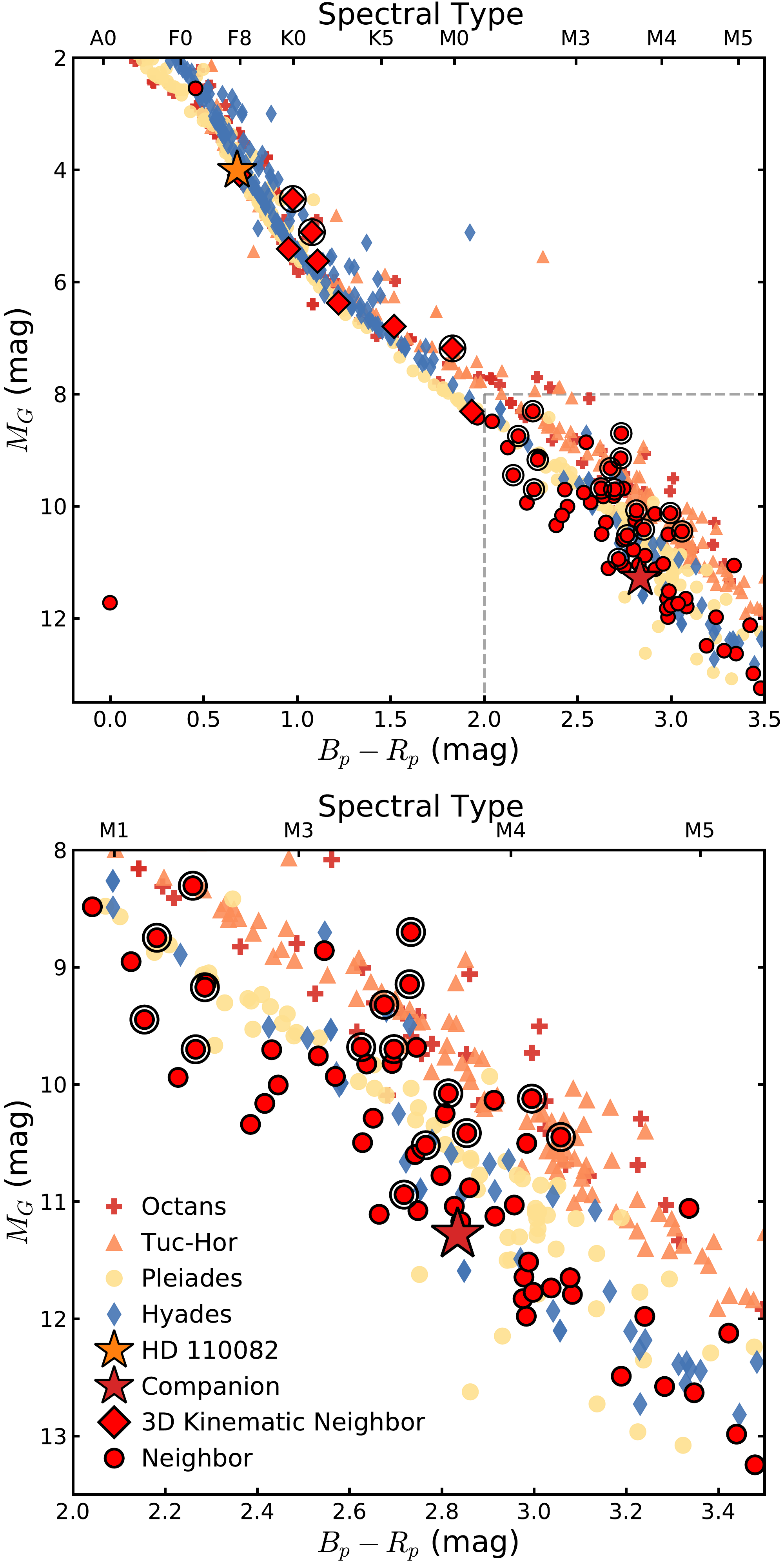}
    \caption{Color-magnitude diagram of young associations. In order of age: Octans is presented in red plus symbols \citep[$\sim$40 Myr;][]{Murphyetal2015,Gagneetal2018}, Tuc-Hor in orange triangles \citep[$\sim$40 Myr;][]{Krausetal2014}, the Pleiades ($\sim$125 Myr) in tan circles, and the Hyades ($\sim$700 Myr) in narrow blue diamonds \citep{Gagneetal2018}. HD 110082 and its wide binary companion are shown in the gold and red stars, respectively. HD 110082 sibling candidates are shown in bold, red circles. Candidates with consistent proper motions and RVs are show in bold, red diamonds. Candidates with \gaia\ $RUWE >$ 1.2 (likely wide binaries) are encircled. The bottom panel expands the region highlighted in grey above. The two elevated encircled diamonds near $B_p-R_p \sim 1$ are known Octans members. The HD 110082 companion falls significantly below the Octans and Tuc-Hor sequences.
    }
\label{fig:cmd}
\end{figure}

To investigate Octans membership, we compare the locations of HD 110082 and its binary companion in the color-magnitude diagram (CMD) to canonical Octans members \citep{Murphyetal2015, Gagneetal2018}. Figure \ref{fig:cmd} presents this CMD alongside members of Tuc-Hor \citep[$\sim$40 Myr;][]{Krausetal2014}, the Pleiades ($\sim$125 Myr), and the Hyades ($\sim$700 Myr); the latter two derive from \citet{Gagneetal2018}. (A further discussion of Figure \ref{fig:cmd} is provided in Section \ref{cmd}; the HD 110082 siblings are discussed in Section \ref{neighbors}.)

The CMD location of HD 110082 does not provide leverage to determine whether its age is consistent with known Octans members, or any of the older associations plotted. Its wide binary companion, however, falls at a color (mass) where the cluster sequences diverge substantially with age. In the bottom panel of Figure \ref{fig:cmd}, the binary companion falls well below the roughly coeval Octans and Tuc-Hor sequences, and below the core Pleiades sequence. The CMD location of the companion is secure; the \gaia\ photometric measurements do not have a ``$B_p/R_p$ flux error excess'', which can lead to erroneous colors (\citealt{Evansetal2018}\footnote{\url{https://gea.esac.esa.int/archive/documentation/GDR2/Data_processing/chap_cu5pho/sec_cu5pho_qa/ssec_cu5pho_excessflux.html}}), and the behavior is persistent in other color-bands. 

This large discrepancy with an age of $\sim$40 Myr is supported by additional lines of evidence that are explored below. Specifically, the \lii\ EW of the HD 110082 falls below Octans members of the same color, suggesting an age $\gtrsim$40 Myr (Section \ref{liew_compare}); and the rotation period of HD 110082 is slower than Pleiads of the same color, suggesting an age $\gtrsim$120 Myr (Section \ref{gyro}). Given this evidence, we conclude that HD 110082 is not a member of the Octans moving group and is, in fact, more evolved. Although both HD 110082 and its companion are high-confidence kinematic members, Octans is the dynamically ``loosest'' of the groups included in BANYAN $\Sigma$ and therefore the most likely to produce field interlopers from kinematics alone. And indeed, HD 110082 falls in the outskirts of the $X, Y, Z$ and $U, V, W$ distributions of canonical Octans members. 

\subsection{A Search for Comoving, Codistant, Coeval Stars}
\label{neighbors}

Precise age measurements of star clusters and associations are made possible by their coeval ensemble that spans a range in stellar mass (or empirically, color). Depending on the age and age-sensitive diagnostic in question (e.g., CMD location, Lithium abundance, rotation period), stars within certain mass ranges provide stronger constraints than others. For instance, higher-mass stars are the first to converge onto a rotation sequence (i.e., gyrochronology; \citealt{Barnesetal2015}), while lower-mass stars retain elevated CMD locations longer than their higher-mass siblings \citep{Hayashi1961}. Together they provide a level of precision that is not generally obtainable for a single star. In the case of HD 110082, the lack of known and well-studied siblings is particularly acute, given that its mass ($\sim$1.2 $M_\odot$; F8 spectral type) falls in a regime that provides very little age-distinguishing leverage. 

To better constrain the age of HD 110082, we search for a population of siblings. A thorough discussion of our approach and its motivation is provided in Appendix \ref{appendix}. In short, we rely on the fact that stars form in clustered environments \citep{Lada&Lada2003}, where even low-mass associations can retain the common phase-space characteristics (three-dimensional positions and velocities) they inherit from their natal cloud \citep[e.g.,][]{Maloetal2013}. This motivates our search for candidate siblings in the comoving, codistant (phase-space) neighborhood of HD 110082.

First, we query the \gaia\ DR2 source catalog within a 25 pc volume of HD 110082, and compute the difference between the observed tangential (plane of the sky) velocity ($v_t$) and the projected tangential  velocity that would correspond with comotion at source's location ($v_{t,cand}$). Sources with absolute tangential velocity offsets, $v_{off} = |v_t-v_{t,cand}| < 5$ \kms\ comprise our HD 110082 neighborhood sample of 134 candidate siblings. Second, we query the {\it GALEX} \citep{Bianchietal2017} and {\it WISE} \citep{Cutrietal2012} archives for each candidate to search for youth signatures in elevated chromospheric activity levels \citep[far-UV flux excess;  e.g.,][]{Findeisen&Hillenbrand2010} and presence of disk material \citep[infrared excess; e.g.,][]{Krausetal2014}, respectively. (This query only acts to compile youth indicators and is not required for our sample selection.) Finally, an initial cut on the \gaia\ ``$Bp-Rp$ excess error'' \citep[see][]{Evansetal2018} reduces the sample to 96 stars. We refer to this initial sample as 2D kinematic neighbors. (A python package, {\tt FriendFinder}, performing these queries and calculations has been made publicly available\footnote{\url{https://github.com/adamkraus/Comove}}.)

We additionally use RVs to identify comoving sources where possible. We measure RVs from reconnaissance spectra (1 target) and compile RV measurements from the literature (19 targets) to assess whether the sources are comoving with HD 110082 in three dimensions (see Appendix \ref{app:neighborRV}). Nine of the 20 stars with available RV measurements meet the 3$\sigma$ agreement we require with the projected comoving RV, $v_{r,cand}$ (allowing a 2 \kms\ uncertainty on the predicted comoving RV, motivated by the velocity dispersion observed in young moving groups; e.g., \citealt{Gagneetal2018}). We refer to these as 3D kinematic neighbors. 

To further clean the sample, we remove sources with very large uncertainties in their \gaia\ astrometric solutions (larger than the excess error induced by an unresolved binary companion; Section \ref{gaia}), specifically, removing sources with $RUWE$ values $>$ 10. We note sources with $RUWE$ values greater than 1.2, which signifies a likely unresolved binary companion (Section \ref{gaia}). This is particularly relevant for distinguishing young stars from binaries in  the \gaia\ CMD. In total, 82 sources survive these cuts: 9 3D kinematic neighbors (3 of which have $RUWE>1.2$) and 73 2D kinematic neighbors (16 of which have $RUWE>1.2$). We refer to these as candidate siblings. An inclusive table of all 134 phase-space neighbors along with their compiled and measured properties is provided in Table \ref{tab:neightborhood}; 3D kinematic neighbors are given the ``RV-comoving'' note. 

The candidate-sibling sample contains two bonafide members of the Octans moving group \citep{Gagneetal2018}. Bonafide in this context signifies that they are part of the canonical sample of members that is used to define the group's kinematics. Both have independent evidence for an age near 40 Myr based on their \lii\ EWs \citep{Murphyetal2015}. The stars, CPD-82 784 (\gaia\ DR2 6347492932234149120) and CD-87 121	(\gaia\ DR2 6343364815827362688) are both 3D kinematic neighbors to HD 110082 and have $RUWE$ values $>$ 1.2 (likely binary). The presence of Octans members is not surprising given that HD 110082 is listed as a high-probability Octans member. These stars stand out as being younger than HD 110082, its companion, and the majority of the sibling-candidate sample. As such, we exclude them from our analysis below and label them with the ``Octans M'' note in Table \ref{tab:neightborhood}.  

We also note that six candidate-siblings are listed as members of the Theia 786 moving group identified by \citet{Kounkeletal2020}. These stars are labeled with the ``Theia 786'' note in Table \ref{tab:neightborhood}. Theia groups are derived from a hierarchical clustering of \gaia\ DR2 astrometry ($\mu_\alpha$, $\mu_\delta$, $\pi$) and position in galactic coordinates ($l$, $b$). The age of Theia 786 is estimated to be 280$^{+80}_{-60}$ Myr based on the \gaia\ CMD locations of group members, as interpreted by a machine-learning algorithm. None of the six overlapping stars have a literature RV, or a reconnaissance spectrum from our follow up that would more directly tie HD 110082 to Theia 786, either with 3D kinematic agreement or a consistent \lii\ or gyrochonology based age. Although this Theia-estimated age provides a better match to the characteristics of HD 110082, confirming Theia 786 as a cohesive group (e.g., 3D kinematics, coherent \lii\ EW and/or stellar rotation sequences), and assessing HD 110082's membership to it, is beyond the scope of this work. 

It is likely that other yet-unidentified members of Octans and/or Theia 786 reside in this sample, as well as older field interlopers. The approach we take here to constrain the age of HD 110082 does not rely on coevality of the entire sample, only that some number of stars are coeval, and that they trace out a discernible sequence in age-dependent parameters that can be compared to other empirical sequences of known age. 

\subsubsection{A Comoving, Codistant White Dwarf}

The sample of sibling-candidates includes a white dwarf, \gaia\ DR2 5766091009035511680. If this white dwarf is indeed coeval with HD 110082, it is likely massive and may supply a useful constraint on cooling ages and the initial-to-final mass relation. Without an RV measurement to confirm its kinematic association, we do not analyze this source further but note it as an interesting target for future study. 

\

\subsection{Comparison with Empirical Cluster Sequences}
\label{sequences}

We compare the age-dependent properties of HD 110082, its wide binary companion, and its candidate siblings to the CMD, \lii\ EW, and rotation period sequences of other young associations. 

Comparisons to two additional age-dependent cluster sequences are presented in Appendix \ref{appendix} (Figure \ref{fig:neighborhood3}), which we summarize here. First, the \textit{WISE} $W1-W3$, $B_p-R_p$ color-color diagram does not show evidence for IR excesses that would be indicative of a very young population ($\lesssim 20$ Myr). Second, the ratio of \textit{GALEX} NUV to $J$-band flux as a function of $B_p-R_p$ color, though sparsely sampled, shows a spread of stars above the Hyades sequence, indicating an age less than $\sim$700 Myr. These comparisons suggests the group's age is between 20 and 700 Myr.

\subsubsection{Color-Magnitude Diagram}
\label{cmd}
\begin{figure}[!t]
    \centering
    \includegraphics[width=0.47\textwidth]{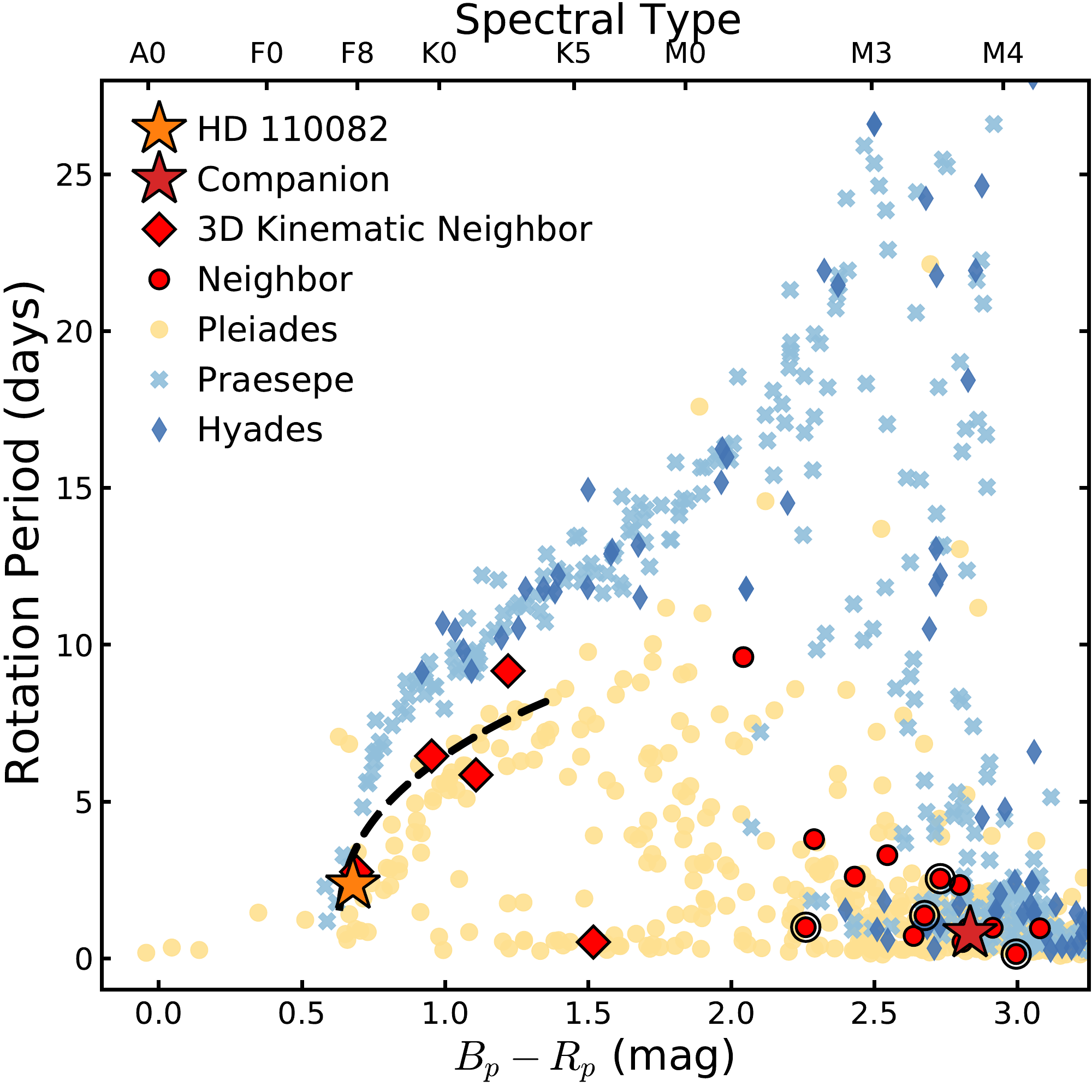}
    \caption{Rotation period distribution of young clusters. In order of age: the Pleiades is presented in tan circles \citep[$\sim$125 Myr;][]{Rebulletal2016}, Praesepe ($\sim$700 Myr) in light blue crosses, and the Hyades ($\sim$700 Myr) in narrow blue diamonds \citep[the latter two from][]{Douglasetal2019}. HD 110082 and its wide binary companion are shown in the gold and red stars, respectively. Sibling candidates to HD 110082 are shown in bold, red circles. Neighbors with consistent proper motions and RVs are show in bold, red diamonds. Neighbors with high \gaia\ $RUWE$ values ($RUWE > 1.2$; likely wide binaries) are encircled. The dashed line represents the 50th percentile of the ensemble age distribution, 250 Myr, using the \citet{Mamajek&Hillenbrand2008} gyrochronology relation.}
    \label{fig:p_rot}
\end{figure}

The CMD of sibling candidates is presented in Figure \ref{fig:cmd}. 2D kinematic neighbors are shown with bold, red circles; 3D kinematic neighbors are shown with bold, red diamonds.  Encircled points denote those that have \gaia\ $RUWE$ values $>$1.2, indicating they are likely binary. The top axis provides corresponding spectral types based on the updated (2019) \citet{Pecaut&Mamajek2013}\footnote{\url{https://www.pas.rochester.edu/~emamajek/EEM_dwarf_UBVIJHK_colors_Teff.txt}} analysis of main sequence standards in the solar neighborhood. The same plotting scheme is used in Figures \ref{fig:LiEW} and \ref{fig:p_rot}.

For context, high-likelihood members of Octans, Tuc-Hor, the Pleiades, and the Hyades are over-plotted. All absolute magnitudes are computed using distances inferred by \citet{Bailer-Jonesetal2018} and Pleiades members are corrected for 0.1054 mags of visual extinction \citep{Taylor2008} using the \citet{Cardellietal1989} extinction law. All Pleiades bonafide members provided in \citet{Gagneetal2018} are plotted to $B_p-R_p=2.5$, beyond which a random draw of 50 likely members are plotted as to not obscure the other cluster sequences. 

Most, but not all, of the presumed single ($RUWE<1.2$) neighbors fall well below the Octans/Tuc-Hor sequences. Given that HD 110082 itself is listed as a 99\% kinematic member of Octans, it is likely that true Octans members reside in our neighbor sample, but do not appear to dominate. The two bonafide Octans members described above are the encircles diamonds at a $B_p-R_p$ color of $\sim$1.0 and are likely elevated in the CMD due to the presence of a close binary companion ($RUWE>1.2$).

\subsubsection{\lii\ Equivalent Width}
\label{liew_compare}

\begin{figure}[!t]
    \centering
    \includegraphics[width=0.47\textwidth]{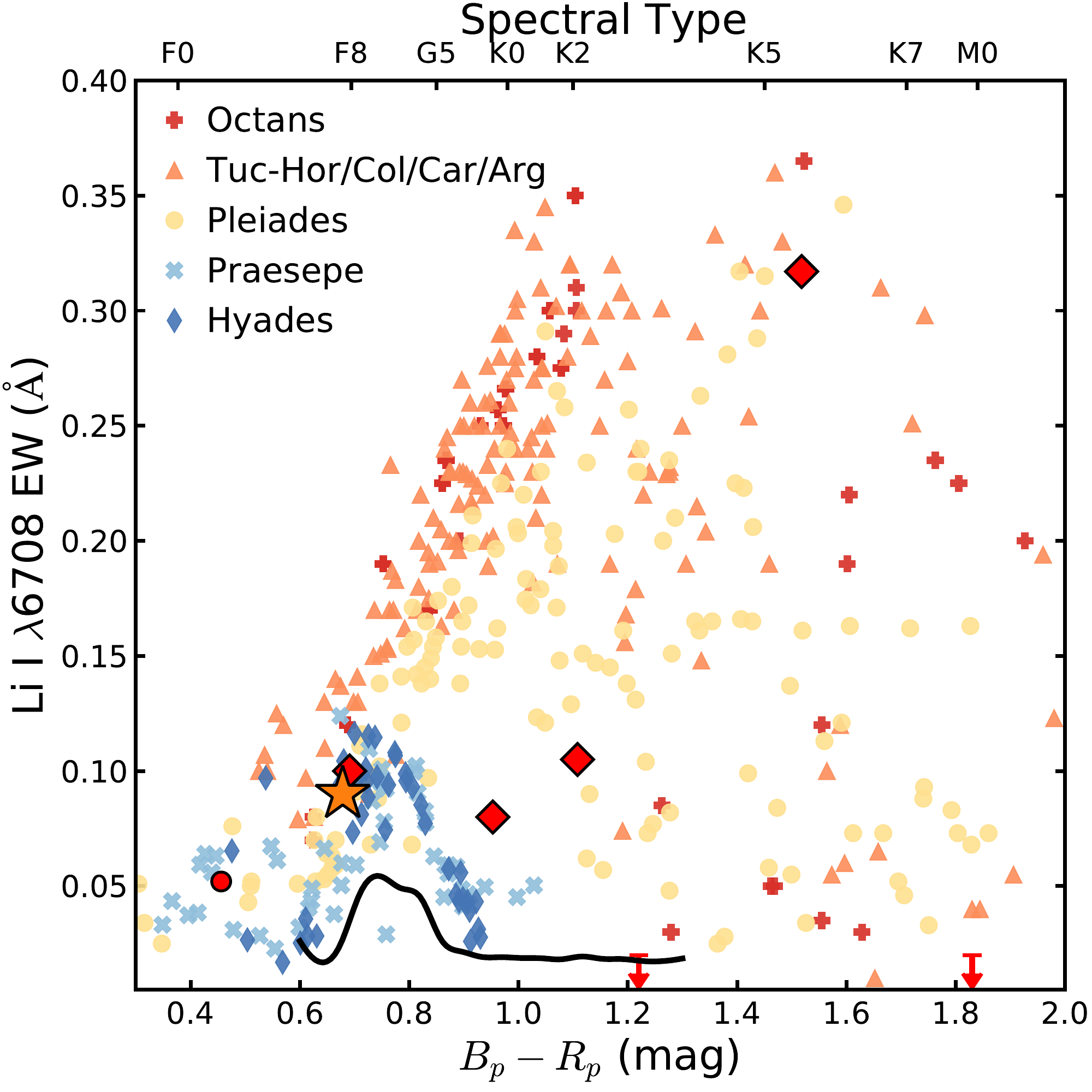}
    \caption{\lii\ EW of young moving group or cluster members as a function of their $B_p-R_p$ color. In order of group/cluster age: Octans \citep[$\sim$40 Myr;][]{Murphyetal2015} is presented in red plus symbols, members of similar age moving groups ($\sim$40 Myrs; Tuc-Hor, Columba, Carina, Argus; \citealt{daSilvaetal2009}) are presented in orange triangles, the Pleiades are shown in tan circles \citep[$\sim$125 Myr;][]{Bouvieretal2018}, and Praesepe and the Hyades in light blue crosses and narrow blue diamonds, respectively \citep[$\sim$700 Myr;][]{Cummingsetal2017}. The black line represents the median \lii\ EW of field-age systems from \citet{Bergeretal2018}. The \lii\ EW of HD 110082, represented with the gold star (its uncertainty is roughly the size of the symbol), falls at a $B_p-R_p$ color that provides little leverage to distinguish its age.} 
    \label{fig:LiEW}
\end{figure}

The presence of \lii\ 6707.8 \AA\ in a stellar atmosphere is an indication of youth ($<$1 Gyr) and can provide a strong age constraint depending on the stellar mass. Figure \ref{fig:LiEW} presents the \lii\ EW for HD 110082 and a handful of sibling candidates for which we have obtained reconnaissance spectra (see Appendix \ref{app:li}). The symbols for sibling candidates follows that in the CMD (Figure \ref{fig:cmd}). Also plotted are sequences from  Octans \citep{Murphyetal2015}, Tuc-Hor, Columba, Carina and Argus \citep{daSilvaetal2009}, the Pleiades \citep{Bouvieretal2018}, Praesepe and the Hyades \citep{Cummingsetal2017} for reference. We also plot the median \lii\ EW of field-age stars analyzed by \citet{Bergeretal2018} as the black line. In the case of Octans, Tuc-Hor, Columba, Carina, and Argus, the \lii\ EW measurement includes the contribution from an adjacent iron line (Fe {\scshape i} 6707.4\AA), as do our measurements. For the Pleiades, Praesepe, Hyades, and the field-star average, the literature EW does not include the contribution from this iron line. We add an average offset of 0.013\AA\ based on the moving group metallicities and temperature range plotted to makes these sequences more directly comparable (see \citealt{Bouvieretal2018}).

As with the CMD, HD 110082 has a $B_p-R_p$ color that provides little leverage to precisely determine its age. Still, with a \lii\ EW of 0.09$\pm$0.02 \AA, it falls below the Octans sequence, and below the average EW of 40 Myr stars with similar $B_p-R_p$ colors, 0.14$\pm$0.04 \AA\ (mean and standard deviation). It does, however, appear fully consistent with the $\sim$125 and $\sim$700 Myr populations. The sibling candidates, however, provide strong evidence for an intermediate age between 125 and 700 Myr. Particularly, the two, 3D-comoving, late-G/early-K stars whose \lii\ EWs reside between the Pleiades and Hyades/Praesepe sequences. We also note the presence of a high-\lii\ EW star at a $B_p-R_p\sim1.6$ (UCAC2 43064; \gaia\ DR2 6341894558326196480), which likely a unidentified Octans member. This star is removed from our analysis below and labled with the ``Octans LM" in Table \ref{tab:neightborhood}. 

Two \lii\ non-detections are also plotted as upper limits. Both stars (\gaia\ DR2 5775111230632453248 and \gaia\ DR2 6346649808677390464) are 3D kinematic neighbors and have rotation periods less than 10 days (Section \ref{rotp} and Table \ref{tab:neightborhood}), indicative of an age $<$700 Myr for their $Bp-Rp$ colors. We suspect these non-detections are indicative of recently exhausted \lii\ supplies rather than field interlopers.

To place a more quantitative age estimate based on our \lii\ EWs measurements, we use {\tt BAFFLES} \citep{Stanford-Mooreetal2020}. This Bayesian inference tool produces age posteriors for given $B-V$ colors and \lii\ EWs, calibrated by benchmark clusters with well determined ages. In the calculation we truncate the flat age prior at 1 Gyr. Johnson $B-V$ colors are compiled from APASS DR9 \citep{Hendenetal2015} and the Guide Star Catalog v2.3 \citep{Laskeretal2008}. Individual age posteriors are presented for our sample of five stars in the top panel of Figure \ref{fig:age_post}, color-coded by their $B-V$ color. HD 110082 is displayed as a solid, colored line. The ensemble posterior, shown in black, corresponds to an age of 260$^{+60}_{-50}$ Myr (68\% confidence interval). Table \ref{tab:age} summarizes the properties of stars used in the \lii\ age ensemble. While the sharpness of the distribution is driven by two stars in the ensemble, those between $B_p-R_p$ of 1.0 and 1.2, the distribution of median ages from a 10$^5$ iteration random-sampling-with-replacement of the five-star sample is even more centrally peaked, indicating that the result is not heavily impacted by individual stars. 

\subsubsection{Rotation Period}
\label{gyro}

After the dispersal of circumstellar material and the end of pre-MS contraction, solar type stars lose angular momentum as their magnetic fields interact with stellar winds (i.e., magnetic breaking; \citealt{Weber&Davis1967}). Once stars converge to the slow-rotator sequence (I-sequence), gyrochronology provides a means to age-date systems. Figure \ref{fig:p_rot} presents the rotation period distribution for HD 110082 and its sibling candidates, in relation to the Pleiades \citep{Rebulletal2016}, Praesepe, and the Hyades \citep{Douglasetal2019}. The rotation periods of sibling candidates are broadly consistent with Pleiades members, but appear younger than a Praesepe/Hyades age population.

\begin{deluxetable*}{l c c c c c c c c c c}
\tablecaption{HD 110082 Age Determination Ensemble
\label{tab:age}}
\tablewidth{0pt}
\tabletypesize{\footnotesize}
\tablecolumns{11}
\phd
\tablehead{
  \colhead{\gaia\ DR2} &
  \colhead{$M_G$} & 
  \colhead{$B_p-R_p$} &
  \colhead{$B-V$} &
  \colhead{$RUWE$} &
  \colhead{$RV_{\rm Comoving}$} &
  \colhead{\lii\ EW} & 
  \colhead{Li Age} &
  \colhead{$P_{\rm rot}$} &
  \colhead{Gyro Age} &
  \colhead{Note} \\
  \colhead{} &
  \colhead{(mag)} &
  \colhead{(mag)} &
  \colhead{(mag)} &
  \colhead{} &
  \colhead{} &
  \colhead{(\AA)} &
  \colhead{(Myr)} &
  \colhead{(d)} &
  \colhead{(Myr)} &
  \colhead{}
}
\startdata
5765748511163751936 & 4.01 & 0.68 & 0.57 & 1.079 & Yes & 0.09  & $480^{+330}_{-270}$ & 2.3 & $100^{+50}_{-40}$  & HD 110082 \\
5196316421300498944 & 5.63 & 1.11 & 0.78 & 0.953 & Yes & 0.105 & $250^{+150}_{-100}$ & 5.9 & $230^{+120}_{-80}$ & \\
5196824739270050560 & 5.41 & 0.95 & 0.77 & 0.942 & Yes & 0.08  & $320^{+200}_{-110}$ & 6.5 & $270^{+80}_{-70}$  & \\
5210764416405839488 & 4.09 & 0.69 & 0.53 & 0.962 & Yes & 0.1   & $390^{+380}_{-250}$ & 2.8 & $210^{+120}_{-80}$ & \\
5775111230632453248 & 6.37 & 1.22 & 1.01 & 0.982 & Yes &       &                     & 9.2 & $360^{+70}_{-70}$  & Li Non-detection  \\
5195793466083385344 & 2.55 & 0.46 & 0.36 & 0.952 & N/A & 0.052 & $480^{+350}_{-300}$ &     &                    & Pulsator \\
\enddata
\end{deluxetable*}

\begin{figure}[!h]
    \centering
    \includegraphics[width=0.47\textwidth]{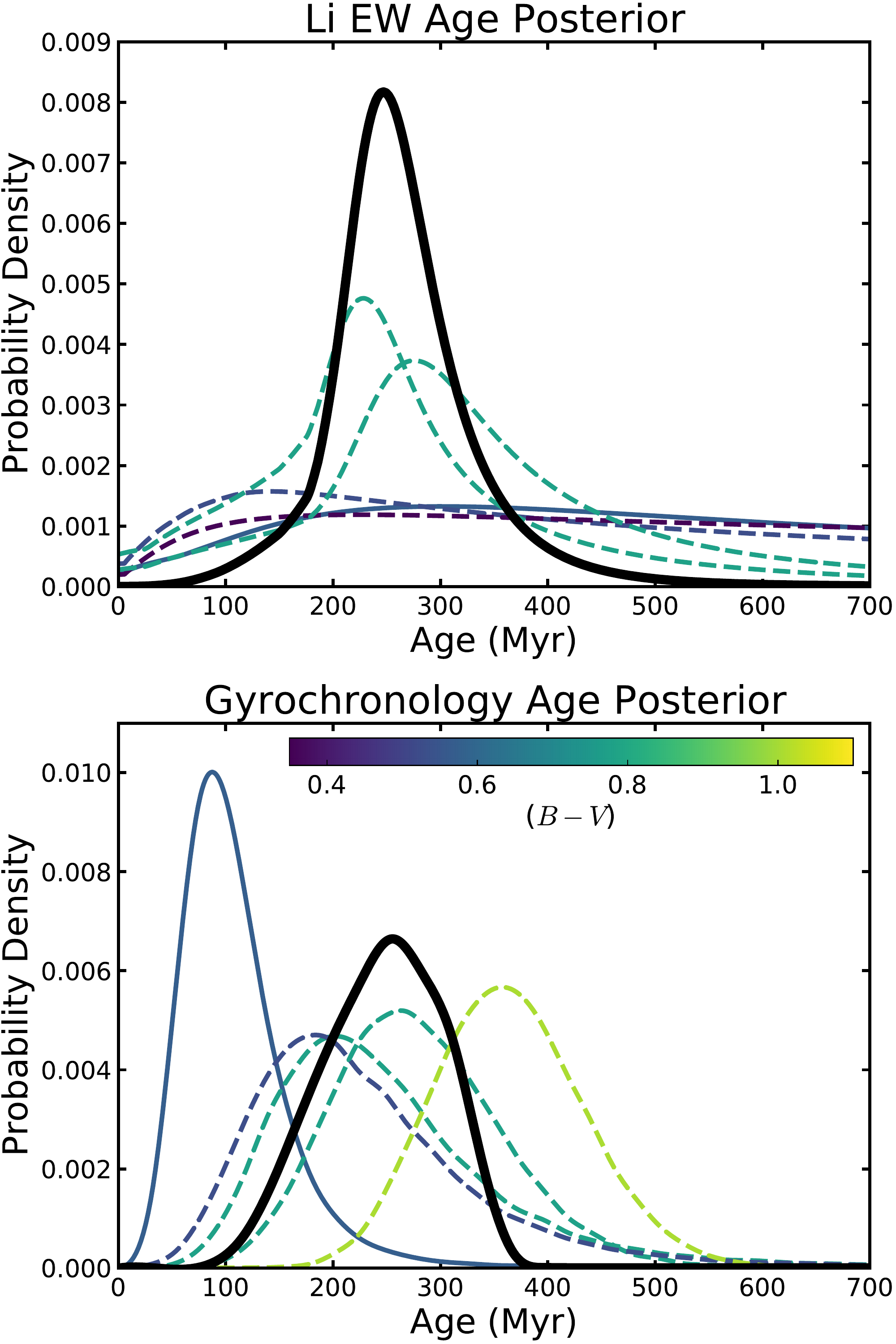}
    \caption{Individual and ensemble age posteriors from our \lii\ EW (top) and gyrochronology (bottom) analysis. Posteriors for individual stars are dashed lines, color coded by their $B-V$ color. The HD 110082 posterior is a solid, colored line. Ensemble posteriors are shown in black. The \lii\ EW analysis uses {\tt BAFFLES} \citep{Stanford-Mooreetal2020}; the ensemble posterior is a multiplication of the individual posteriors. The gyrochronology analysis uses the \citet{Mamajek&Hillenbrand2008} relation; the adopted ensemble posterior is the distribution of ages from a bootstrap simulation.}
    \label{fig:age_post}
\end{figure}

The rotation periods of sibling candidates are determined from \tess\ light curves via a Lomb-Scargle periodogram \citep{Scargle1982} and, in two cases, using the Gaussian process approach described above (Section \ref{rotp}). A full description of our methodology can be found in Appendix \ref{app:rotation}. There are two 3D kinematic neighbors for which we do not have a measured rotation period. The first (\gaia\ DR2 4612569033640835584; $B_p-R_P = 1.93$) is a slow rotator with a $>$20 day period, and is likely a field interloper. We label this star with the ``Field" note in Table \ref{tab:neightborhood}. The other (\gaia\ DR2 6346649808677390464; $B_p-R_p = 1.83$) hosts two distinct periods (0.49 and 0.76 days). With a high $RUWE$ value (5.01), this is likely the combination from two stars, and we exclude them from our gyrochronology analysis. Lastly, there is one star from the \lii\ age ensemble for which we do not have a measured rotation period, \gaia\ DR2 5195793466083385344, which is a pulsating F-star.

We quantitatively assess the age of HD 110082 using gyrochronology. The $B-V$ color-period gyrochronology relationship originally proposed in \citet{Barns2007} has been calibrated by multiple groups using different sets of empirical benchmarks (e.g., open clusters, the Sun, field stars with asteroseismic ages; \citealt{Meibometal2011,Angusetal2015}). We choose to adopt the \citet{Mamajek&Hillenbrand2008} calibration which uses younger benchmarks than the works referenced above, specifically, $\alpha$ Per ($\sim$85 Myr) and Pleiades ($\sim$125 Myr) members on the  I-sequence. For HD 110082, the \citet{Mamajek&Hillenbrand2008} relationships predicts an age of $100^{+50}_{-35}$ Myr (68\% confidence interval), which includes $B-V$ uncertainty and a 20\% uncertainty on the rotation period. This posterior and those for the four additional stars with measured rotation periods in the calibrated $B-V$ color range (0.5--1.4) are presented in the bottom panel of Figure \ref{fig:age_post}. Each dashed line is color-coded by the star's $B-V$ colors, with the HD 110082 posterior presented as a solid line. Table \ref{tab:age} summarizes the properties of these five stars.

\citet{Angusetal2019} point out that the functional form of the \citet{Barns2007} gyrochonology relationship systematically under-predicts the ages of stars more massive than the Sun and over-predicts the ages of lower mass stars. This agrees with the increasing age with $B-V$ color seen in Figure \ref{fig:age_post}, and slight tension between the HD 110082 rotational age and its companion's CMD location. This short-coming motivated \citet{Angusetal2019} and others \citep[e.g.,][]{Spada&Lanzafame2020} to develop a more flexible gyrochonology models. Unfortunately, these models are not calibrated well for young ages and do not capture the early rotational evolution relevant for HD 110082. 

To determine a more robust age for the system, we fit the five-star ensemble to with the the \citet{Mamajek&Hillenbrand2008} relation using {\tt emcee}, including color and period uncertainties. Our fit uses 10 walkers and convergence is assessed with the method described in Section \ref{mass_density}. The result is an ensemble age of 250$\pm$40 Myr. The black dashed line in Figure \ref{fig:p_rot} presents the 250 Myr rotational isochrone. Given the strong color dependence in the individual age posteriors, we perform a bootstrap sample with replacement of the five-star sample for 10$^5$ iterations. The distribution of these ages is more broad than the MCMC ensemble fit, with a median and 68\% confidence interval of 250$^{+50}_{-70}$. Because the bootstrap age distribution more readily accounts for the significant impact each star has on the ensemble age, we adopt it as our ensemble gyrochronology age posterior and present it as the black distribution in the bottom panel of Figure \ref{fig:age_post}.

\subsection{A New Stellar Association and its Age}
\label{age_final}

The stunning agreement between the independently derived ensemble ages suggests that our approach has indeed revealed a collection of coeval stars, and highlights their ability to place precise and robust age constraints. Although both age-dating methods ultimately rely on the same underlying age calibration (MS turnoffs of benchmark clusters), the agreement is still very encouraging given the only partial overlap of benchmark clusters used to calibrate each method, and the partial overlap of sibling candidates that make up our gyrochronology and \lii\ EW ensembles (Table \ref{tab:age}). We adopt the result of our gyrochronology analysis as our final age determination, 250$^{+50}_{-70}$ Myr, as it is the most common approach used in the literature for measuring the ages of individual young stars. This result agrees well with the comparisons to cluster sequences, which independently bracket the system age at greater than $\sim$125 Myr (Pleiades age; CMD and Li EW comparison), and less than $\sim$700 Myr (Hyades/Praesepe age; rotation-period comparison). 

Our effort to age date HD 110082 has revealed a new young stellar association. Our approach amounts to determining the group's \underline{m}embership and \underline{e}volution by \underline{l}everaging \underline{a}djacent \underline{n}eighbors in a \underline{g}enuine \underline{e}nsemble, so we accordingly name the association MELANGE-1. The consistent ages and partial phase-space overlap between MELANGE-1 and Theia 786 \citep{Kounkeletal2020} is compelling and may signify they are part of the same extended structure. Assessing their relation will be the subject of future work.

\section{Analysis of Transit Signal}
\label{analysis}

We derive the properties of HD 110082 b by modeling the \tess\ and \spitzer\ transits simultaneously with {\tt misttborn} \citep{Mannetal2016a,Johnsonetal2018}. This package uses {\tt batman} \citep{Kreidberg2015} to generate analytic transit models following \citet{Mandel&Agol2002}, combined with a Gaussian process describing the out-of-transit variability with {\tt celerite} \citep{Foreman-Mackeyetal2017}. The simultaneous modeling of the transits and stellar variability are fit with {\tt emcee} \citep{Foreman-Mackeyetal2013}. 

Transits are modeled by the time of mid-transit at a reference epoch ($T_0$), period ($P$), ratio of the planet and host star radii ($R_p/R_\star$), an impact parameter ($b$), stellar density ($\rho_\star$), two parameters combining the orbital eccentricity and argument of periastron ($\sqrt{e}sin\omega$ and $\sqrt{e}cos\omega$), and a quadratic limb-darkening law ($q_1$ and $q_2$ for each photometric band) using the \citet{Kipping2013} sampling method. Gaussian priors are placed on the stellar density, informed by the stellar parameters derived above. Quadratic limb-darkening coefficients specific to HD 110082's stellar parameters are taken from \citet{Claret2017} for the \tess\ band pass and from the \citet{Eastmanetal2013} interpolation of \citet{Claret&Bloemen2011} values for IRAC Channel 2, using Gaussian prior widths of 0.1. The impact parameter is allowed to vary between $b<\pm1+R_p/R_\star$ to allow for grazing transits. MCMC walkers for the remaining parameters are initialized at the values from the fit from the SPOC Transiting Planet Search, or at zero in the case of $\sqrt{e}sin\omega$ and $\sqrt{e}cos\omega$. 

A description of the Gaussian process, {\tt celerite}, modeling out-of-transit variability is provided above in Section \ref{rotp} where it is used to determine the stellar rotation period. The kernel consists of two damped, drive simple harmonic oscillators, the first at the stellar rotation period and the second at half the rotation period. The same implementation is applied here, with six parameters: the stellar rotation period ($P_{{\rm rot}, GP}$), the amplitude of primary rotation kernel ($A_{1, GP}$) and its decay timescale ($Q_{1, GP}$, fit as $\Delta Q_{GP} = Q_{1, GP} - Q_{2, GP}$), the ratio of the secondary to primary kernel amplitude (Mix$_{GP}$, fit as $m$ where $A1/A2 = 1+e^m$), the decay timescale of the secondary kernel ($Q_{2, GP}$), and a jitter term ($\sigma_{GP}$). These parameters are explored in natural logarithmic space. The same Gaussian process is applied to both the \tess\ and \spitzer\ light curves for computational ease, despite the expected difference in the spot variability amplitude in the IR. The separation of the \spitzer\ observations from \tess\ Cycles 1 and 3 are larger than the decay timescale of the kernels informed by the \tess\ data, and their durations are sufficiently short ($<P_{\rm {rot}}/2$) that the out-of-transit variability is smooth and easily modeled by the flexible Gaussian process. Walkers for these parameters are initialized at the location of the best fit found in Section \ref{rotp}. 

\begin{figure*}[!h]
    \centering
    \includegraphics[width=1.0\textwidth]{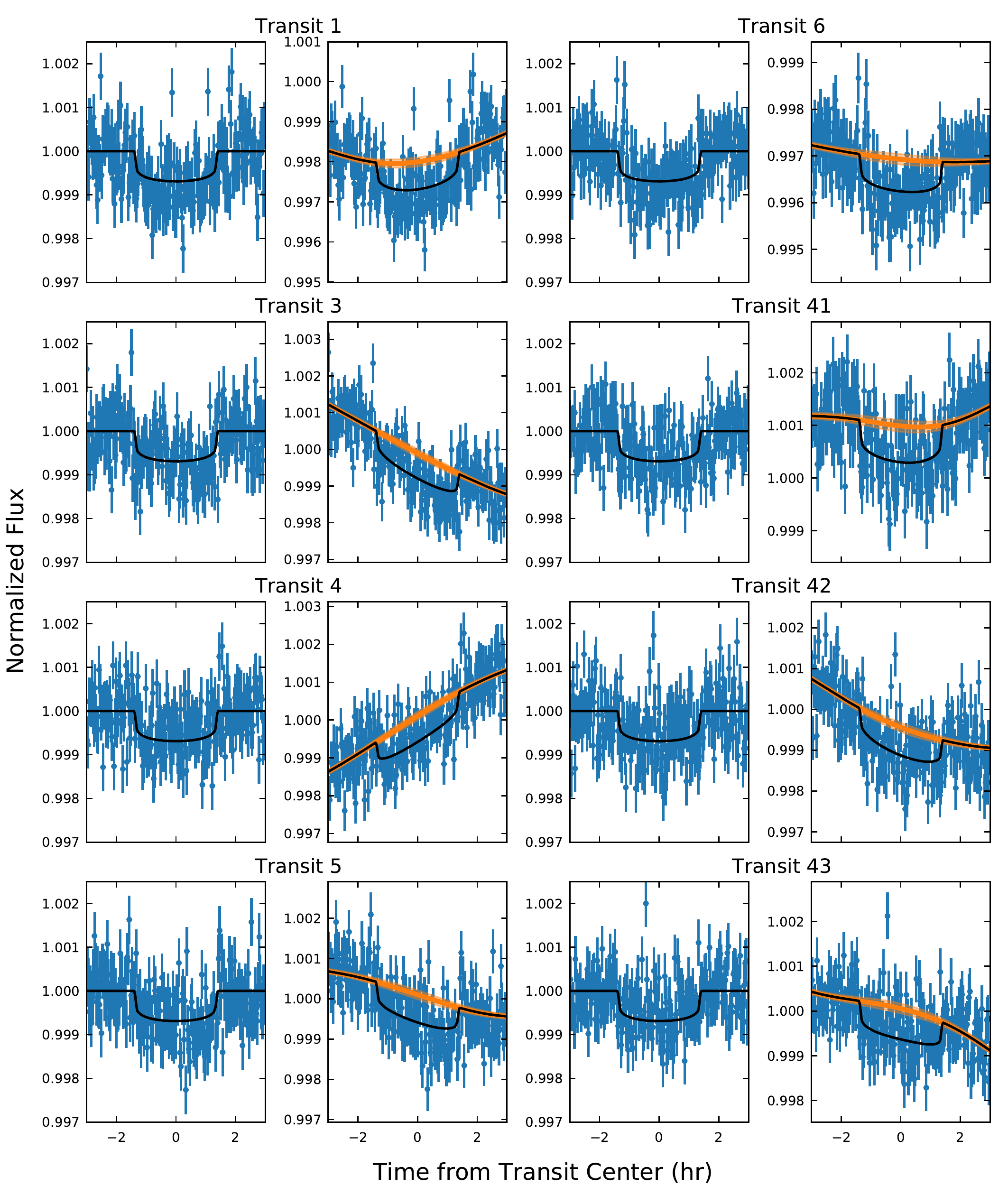}
    \caption{\tess\ light curves for eight HD 110082 b transits. Each transit is presented twice: detrended and with stellar variability included on the left and right of each panel pair, respectively. The orange line represents the Gaussian-proccess stellar variability model. Labels at the top of each panel pair correspond to the vertical ticks in Figure \ref{fig:tess}.}
    \label{fig:tess_trans}
\end{figure*}

\begin{figure*}[!t]
    \centering
    \includegraphics[width=0.9\textwidth]{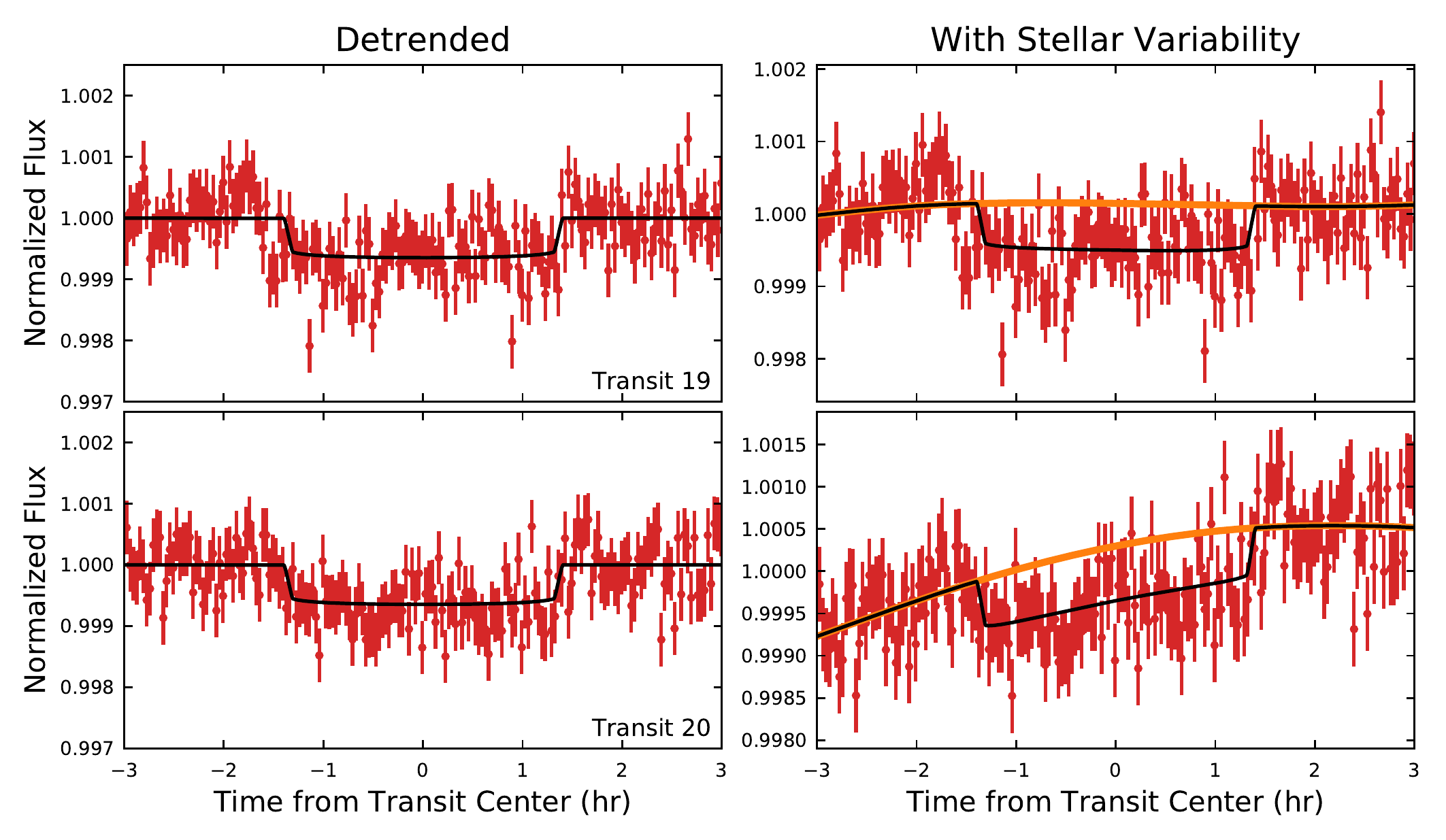}
    \caption{\spitzer\ light curve of two HD 110082 b transit events. Light curves have been binned to a 2-m cadence. Transits are shown sequentially from top to bottom with the best-fit transit model in black. The label in the bottom left corner is consistent with the numbering scheme in Figure \ref{fig:tess_trans}. Left panels present light curves where stellar variability has been removed by our Gaussian-process model. In the right panels stellar variability is left in, highlighting our model with the orange line. }
    \label{fig:spitzer}
\end{figure*}

\begin{figure}[!ht]
    \centering
    \includegraphics[width=0.47\textwidth]{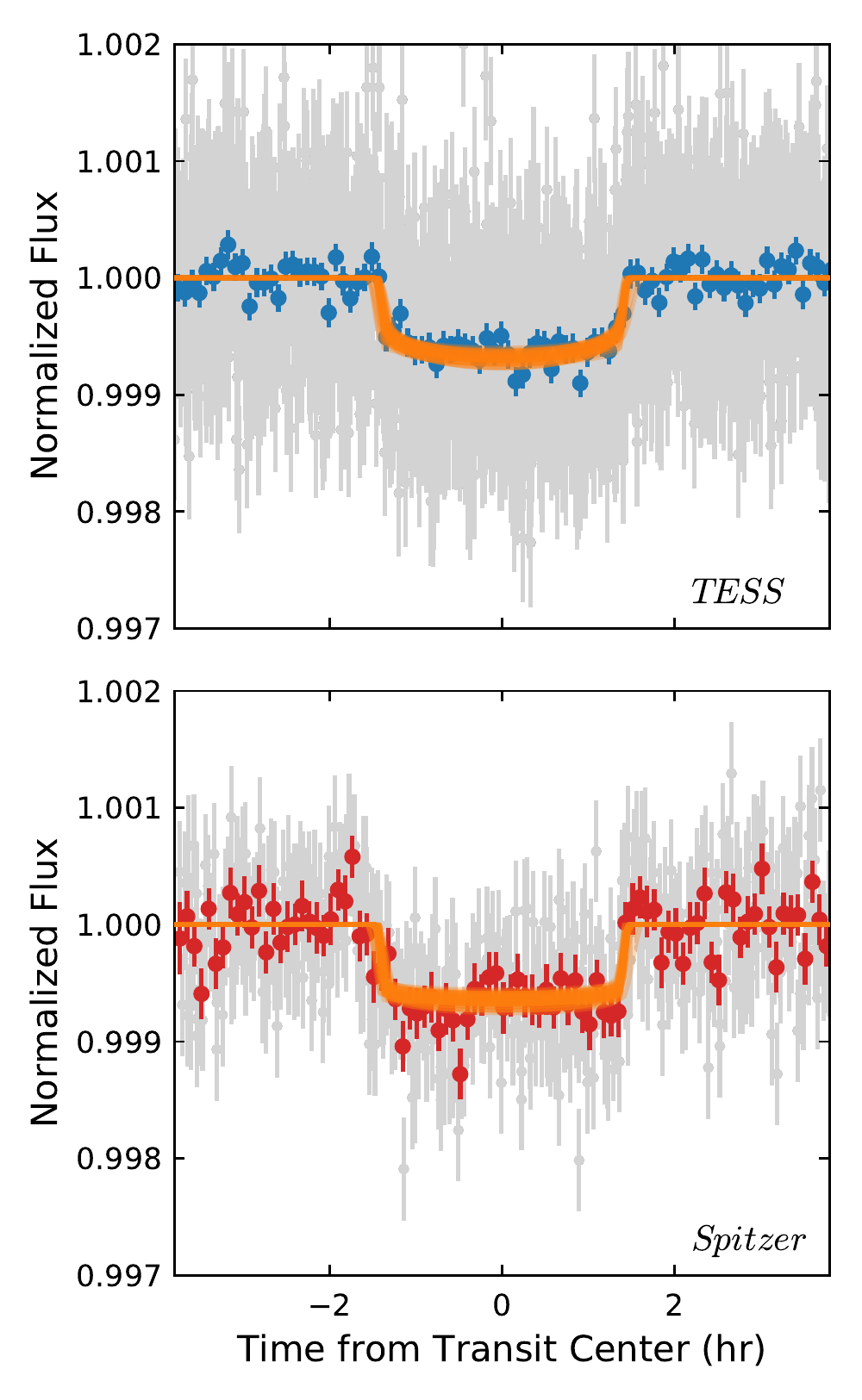}
    \caption{Detrended, phase folded, \tess\ (top) and \spitzer\ (bottom) light curves. The \spitzer\ light curve in grey has been binned to a 2-m cadence. Colored points in each panel present a 5-m binned light curve of the phase-folded data. Orange curves represent transit models constructed from 50 random draws of parameters in the MCMC posteriors. }
    \label{fig:phase}
\end{figure}

Our fit is made with 144 walkers, each taking a total of 240,000 steps. We estimate the fit convergence by measuring the auto-correlation time, which converges (fractional change $<$ 5\%) at a value of $\sim$16,000 steps. We discard the first 12 auto-correlation times  as burn in (200,000 steps), using the distribution of walkers from the last 40,000 steps to infer the values of fit parameters. The 50th percentile and 68\% confidence interval for each fit parameter are presented in Table \ref{tab:planet_props}, as well as a collection of derived parameters. Figures \ref{fig:tess_trans} and \ref{fig:spitzer} present the \tess\ and \spitzer\ transits, respectively, with the best-fit transit model. The right side of each panel pair includes out-of-transit variability, modeled by our Gaussian process in orange, which is removed in the left panel. Transits are numbered sequentially since the beginning of \tess\ Sector 12 (Transit 2 occurred during a \tess\ data downlink; see Figure \ref{fig:tess}). Finally, a phase folded light curve from each data set is provided in Figure \ref{fig:phase} where orange lines display transit model realizations for 50 random draws of the parameter posteriors. 

\begin{figure*}[!ht]
    \centering
    \includegraphics[width=1.0\textwidth]{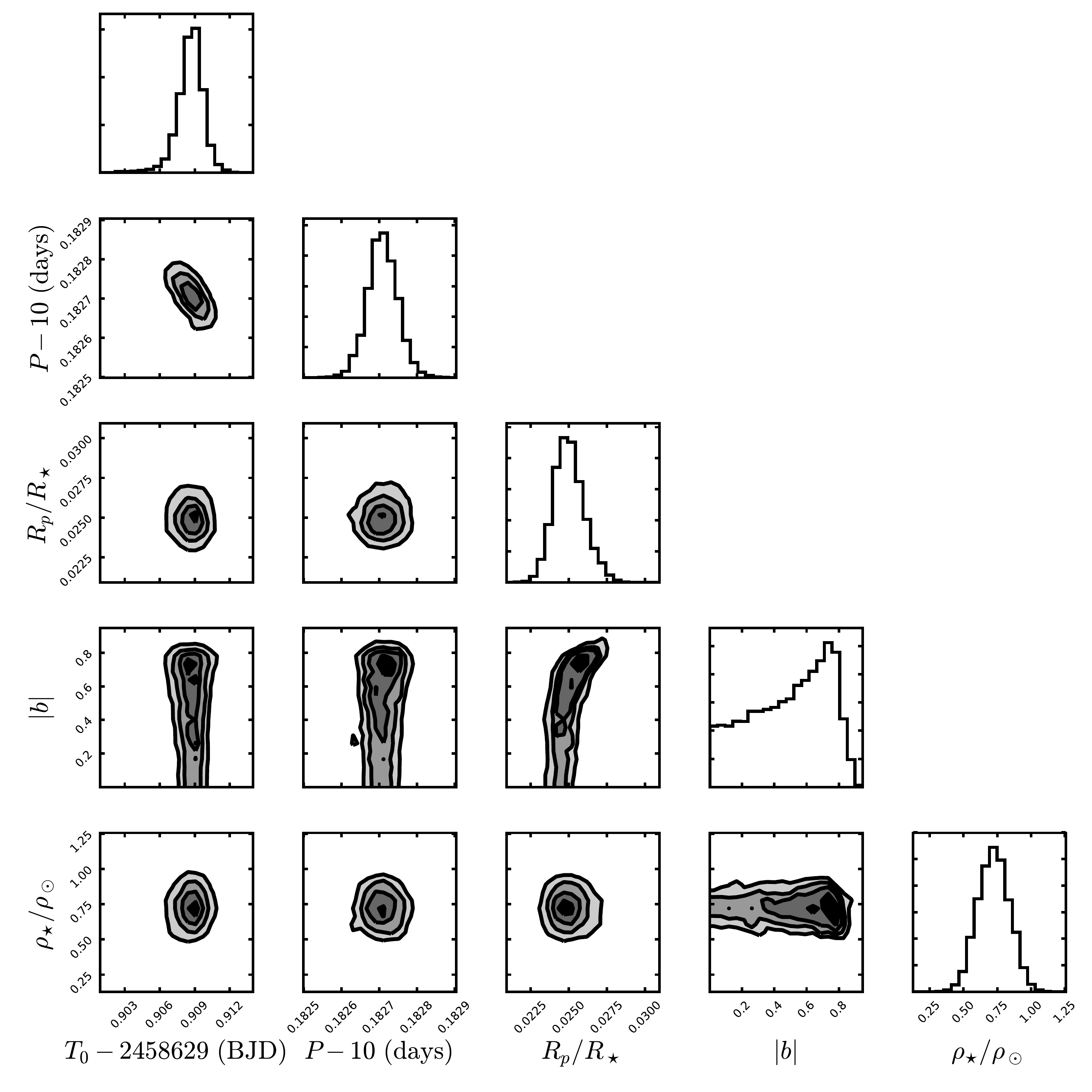}
    \caption{Corner plot of orbital parameters fit to HD 110082 b transits. The absolute value of the impact parameter, $b$, is presented due to the degeneracy between the hemisphere the planet transits. }
    \label{fig:corner}
\end{figure*}

\begin{figure}[!ht]
    \centering
    \includegraphics[width=0.47\textwidth]{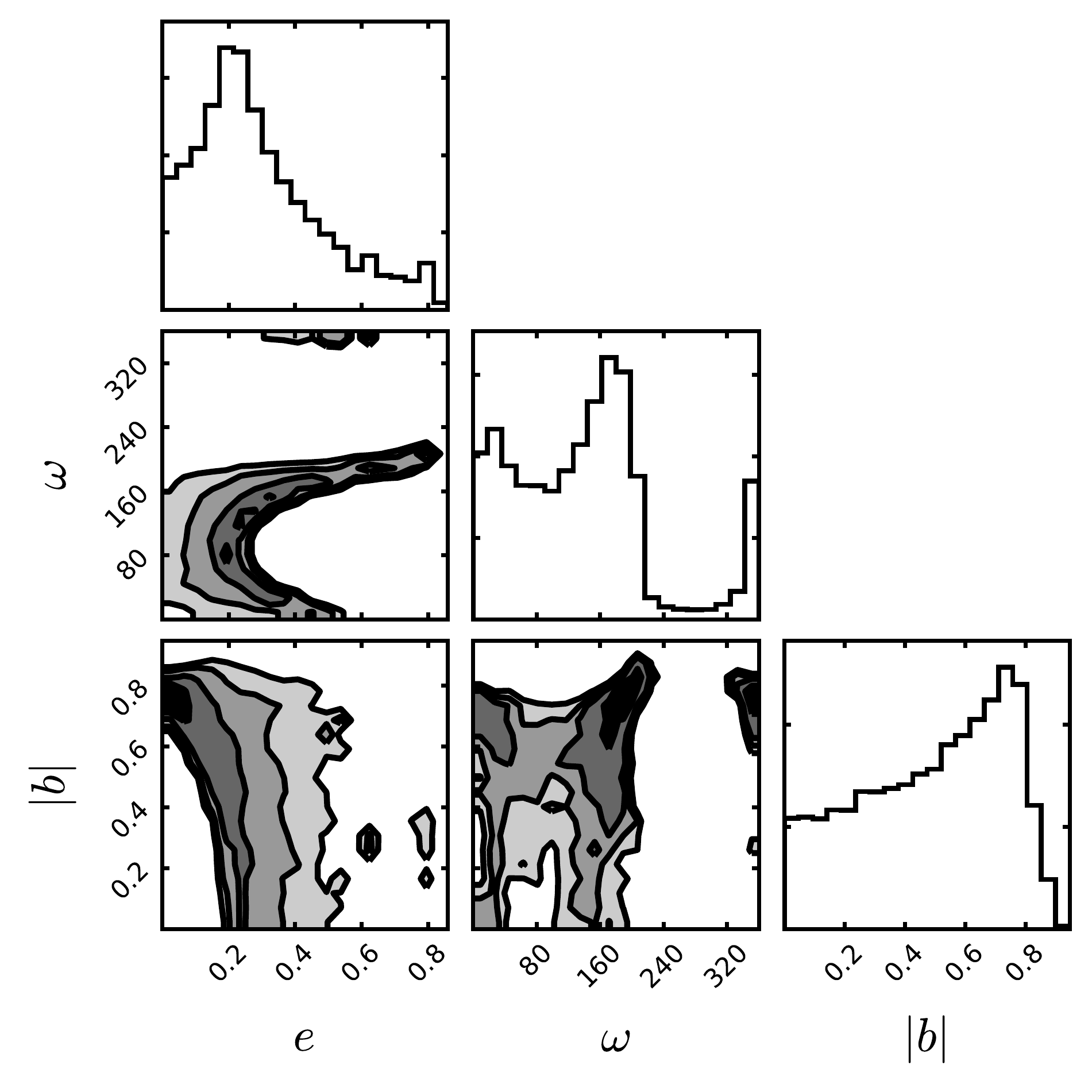}
    \caption{Corner plot for derived orbital parameters, eccentricity ($e$) and the longitude of periastron ($\omega$).}
    \label{fig:corner_derived}
\end{figure}

Figure \ref{fig:corner} presents a corner plot of the transit parameters. The absolute value of the impact parameter, $b$, is shown, given the degeneracy in the projected hemisphere the planet transits. Figure \ref{fig:corner_derived} presents the corner plot for the derived eccentricity and argument of periastron, as well as the impact parameter. 

There is some evidence that the SPOC pipeline may overestimate the background level when performing photometry in some cases, particularly in crowded fields or for dim stars. Due to the large angular size of \tess\ pixels, there are typically no completely dark pixels in the 2m postage stamps, and if the background level is overestimated, the transit depth will be correspondingly inflated. To test this, we run a separate MCMC fit including a dilution term for \tess\ transits. With no bright nearby sources, we assume the \spitzer\ transits are not diluted. The fit returned a dilution of $0.05^{+0.15}_{-0.14}$ (68\% confidence interval), with all other parameters agreeing within confidence intervals listed in Table \ref{tab:planet_props}. Given the similarities between the two fits, and that the dilution term is consistent with zero, we adopt the values derived from the dilution-free fit as fiducial. 
 
With 10 observed transits spanning 43 orbital periods, we also search for TTVs by fitting each transit individually. For \tess\ data, we fit light curves detrended by the {\tt celerite}, Gaussian process model described in Section \ref{rotp}, where transits are masked in 6 hr windows centered on the expected transit center. Fits are made to 10 hr regions of the light curves centered on each transit. For \spitzer\ data, we fit the full $\sim$8.5 hr time series, using the BLISS, ramp-corrected (detrended) light curve. For each transit we fit a limited set of parameters: $T_0$, $R_p/R_\star$, $b$, $\rho_\star$, and limb darkening coefficients, following the approach described above. We do not find evidence for the evidence for TTVs with amplitudes $>$8 minutes (standard deviation), which agrees with our $T_0$ measurement precision at 95\% confidence level.

\begin{deluxetable}{l c}
\tablecaption{Planetary Properties of HD 110082 b
\label{tab:planet_props}}
\tablewidth{0pt}
\tabletypesize{\footnotesize}
\tablecolumns{2}
\phd
\tablehead{
  \colhead{Parameter} &
  \colhead{Value}
}
\startdata
\multicolumn{2}{c}{\textbf{Fit Parameters}} \\
$T_0$ (BJD)             & 2458629.909$\pm0.001$ \\
$P$ (days)              & 10.18271$\pm0.00004$ \\
$R_p/R_\star$           & 0.025$\pm0.001$ \\
$b$                     & 0.5$^{+0.2}_{-0.3}$ \\
$\rho_\star/\rho_\odot$ & 0.7$\pm0.1$ \\
$\sqrt{e}sin\omega$     & 0.2$^{+0.2}_{-0.3}$ \\
$\sqrt{e}cos\omega$     & -0.1$^{+0.6}_{-0.5}$ \\
$q_{1,\tess}$           & 0.31$\pm0.09$ \\
$q_{2,\tess}$           & 0.23$^{+0.09}_{-0.10}$ \\
$q_{1,\spitzer}$        & 0.09$^{+0.08}_{-0.06}$ \\
$q_{2,\spitzer}$        & 0.13$^{+0.09}_{-0.08}$ \\
\hline
\multicolumn{2}{c}{\textbf{GP Parameters}} \\
$P_{{\rm rot}, GP}$ (d) & 2.33$\pm0.03$ \\
ln $A_{1, GP}$          & -13.6$\pm0.1$ \\
ln $Q1_{1, GP}$         & 3.0$\pm0.4$ \\
Mix$_{GP}$              & 1.00$^{+0.00}_{-0.01}$ \\
ln $Q_{2, GP}$          & 1.1$^{+0.2}_{-0.1}$ \\
ln $\sigma_{GP}$        & -9.0$\pm0.1$ \\
\hline
\multicolumn{2}{c}{\textbf{Derived Parameters}} \\
$R_p (R_\earth)$        & 3.2$\pm0.1$ \\
$a/R_\star$             & 20$\pm2$ \\
$a$ (AU)                & 0.113$^{+0.009}_{-0.013}$ \\
$t_{14}$ (hr)           & 2.91$^{+0.06}_{-0.04}$ \\
$i$ ($^\circ$)          & 88.2$^{+1.1}_{-0.7}$ \\
$e$                     & 0.2$^{+0.2}_{-0.1}$ \\
$\omega$ ($^\circ$)     & 138$^{+60}_{-100}$ \\
\enddata
\end{deluxetable}

\subsection{Analysis of False Positives}
\label{false_positive}

In this section we address some possible scenarios that could give rise to a false positive in our detection of HD 110082 b. 

\begin{enumerate}
    \item \textit{Transit signal originates from instrumental artifacts:} Five periodic transits of equal duration are detected in the \tess\ light curve. The period does not coincide with any know periodicities in the \tess\ satellite system (e.g., momentum dumps). Finally, the detection of transits with \spitzer\ rules out instrumental false positives associated with either system.
    
    \item \textit{Transit signal originates from stellar variability:} While the amplitude of stellar variability is much larger than the transit signal, the \tess\ transit signal occurs on a period longer than, and not an integer multiple of, the stellar rotation period. Additionally, the transit is detected at IR wavelengths with the same depth, whereas the amplitude of stellar variability is expected to be reduced in the IR. For these reasons, we rule out stellar variability as a false positive. 
    
    \item \textit{HD 110082 b is an eclipsing binary or brown dwarf:} The low RV variability, single-peaked broadening function (Section \ref{rv}), and flat-bottomed transits (Figure \ref{fig:phase}) rule out most stellar companions. To test whether the transit signal could arise from the grazing eclipse of a low-mass stellar, or brown dwarf companion, we simulate 1,000,000 binary systems to compare with our time-series RV measurements. Binary systems are made from a random (uniform) draw of mass ratio (0 to 1), eccentricity (0 to 0.99), and argument of periastron (0 to 2$\pi$). Periods are set to 10.1827 days and the inclinations are limited to $i>70^\circ$ to ensure an eclipse is feasible. The RVs associated with these orbits are generated at the phases of our observations and compared against our RV measurements. A jitter term of 100 m s$^{-1}$ is added in quadrature to the simulated RVs to account for stellar activity, and both data sets are offset to have a mean RV of zero. All generated binaries with companion masses above 8 $M_J$ were rejected at $>5\sigma$, and greater than 95\% were rejected down to 2 $M_J$. This exercise rules out virtually all stellar and brown dwarf companions, but allows for the grazing eclipse of a Jupiter-size planet. This scenario, however, is heavily disfavored by the transit fit, which measures short ingress and egress times compared to the transit duration. 
    
    \item \textit{Light from a physically associated eclipsing binary or planet-hosting companion is blended with HD 110082:} 
    In this scenario, the eclipsing/transiting pair would have to reside within $\sim$10 AU of HD 110082 based on the LCO time-series imaging (clearing companions to beyond 2.5$\arcsec$; Section \ref{lco}), the \gaia\ DR2 source catalog (clearing companions beyond $\sim$0.2$\arcsec$; Section \ref{gaia}), and the Zorro speckle imaging (clearing companions beyond $\sim$0.1$\arcsec$; Section \ref{zorro}). Because grazing eclipse/transit are confidently ruled out by our fit, we can place three constrains on potential companions. 
    
    First, we use the ratio of the ingress time, $t_{12}$, to the time between the first and third contact, $t_{13}$, to constrain the radius ratio of a potential companion pair, irrespective of any diluting flux. The companion would require a magnitude difference of:
    \begin{equation}
    \label{eqn:V19}
        \Delta m \lesssim 2.5 log_{10} \left( \frac{t^2_{12}}{t^2_{13}\delta} \right),
    \end{equation}
    compared to HD 110082 to in order to recreate the observed transit depth, $\delta$ \citep[see][]{Vanderburgetal2019}. From a separate MCMC transit fit of the \tess\ and \spitzer\ light curves removing priors on stellar parameters, we determine $\Delta m_{\tess} < 2.0$ and $\Delta m_{4.5\mu m} < 1.5$ (to 95\% confidence). For the \tess\ light curve, this corresponds to a maximum (undiluted) transit depth of $\sim$0.004, which would corresponds to a planetary-size object for any feasible host star in this scenario. As such, we can ignore any flux contribution from the third (transiting) body.
    
    Second, we can use the independent transit depths measured in the \tess\  ($\delta_T; \ \lambda_{eff}\sim0.75\mu m$) and \spitzer\ IRAC Channel 2 ($\delta_S; \ \lambda_{eff}\sim4.5 \mu m$) bandpasses to constrain the \tess--\spitzer\ color of a potential companion, $C_{TS_2}$. With a reduced contrast ratio from optical to IR wavelengths, a transit on a companion star should appear deeper in the \spitzer\ light curve. Following \citet{Desertetal2015}, the observed transit depth in either band is, $\delta = \delta_{true}F_2/(F_1+F_2)$, where $F_1$, and $F_2$ are the primary (HD 110082) and secondary flux, respectively. The ratio of the two transit depths is then:
    \begin{equation}
    \label{eqn:CST1}
    \begin{aligned}
        \frac{\delta_S}{\delta_T} &= \frac{(F_{2,S}/F_{2,T})}{(F_{1,S}+F_{2,S})/(F_{1,T}+F_{2,T})} \\ 
        &= \left(\frac{10^{-0.4M_{1,T}}+10^{-0.4M_{2,T}}}{10^{-0.4M_{1,S}}+10^{-0.4M_{2,S}}}\right)\frac{10^{-0.4M_{2,S}}}{10^{-0.4M_{2,T}}},
    \end{aligned}
    \end{equation}
    where $S$ and $T$ subscripts correspond to \spitzer\ and \tess\ fluxes (or magnitudes), respectively. We can then use this expression to rewrite the combined (observed) \tess--\spitzer\ color of the primary and (theoretical) secondary, $C_{TS_{1,2}}$ in terms of the observed transit depths and secondary color:
    \begin{equation}
    \label{eqn:CST2}
    \begin{aligned}
    C_{TS_{1,2}} &= -2.5\rm{log} \frac{10^{-0.4M_{1,T}}+10^{-0.4M_{2,T}}}{10^{-0.4M_{1,S}}+10^{-0.4M_{2,S}}} \\
    &=-2.5\rm{log} \left(\frac{\delta_S}{\delta_T}\frac{10^{-0.4M_{2,T}}}{10^{-0.4M_{2,S}}}\right) \\
    &=-2.5\rm{log} \left(\frac{\delta_S}{\delta_T}\right) + C_{TS_{2}}.
    \end{aligned}
    \end{equation}
    While both transit depths are consistent, the \spitzer\ values skews to larger depths. Taking the 95\% percentile of the $\delta_S/\delta_T$ distribution, we can place the limit:
    \begin{equation}
    \label{eqn:CST3}
    C_{TS_{1,2}} \leq C_{TS_2} \leq C_{TS_{1,2}} + 2.5\rm{log}\left(\frac{\delta_S}{\delta_T}\bigg\rvert_{95\%}\right),
    \end{equation}
    corresponding to allowed \tess--\spitzer\ colors for the companion between 0.69 and 1.17.
    
    Finally, with precise astrometric measurements from \gaia\ we demand that the combined light of HD 110082 and any companion match the observed apparent magnitudes at the measured distance. Conservatively, we require agreement within 0.1 mag in the \tess\ and \spitzer\ bandpasses, which is much larger than the measurement uncertainty (or derived uncertainty in the case of the \tess\ magnitude; \citealt{Stassunetal2019}), or the propagated uncertainty from the distance measurement. 
    
    With the three constraints above, we simulate binary pairs from MIST isochrones (provided in the \tess\ and IRAC Channel 2 bandpasses) at an age of 200 Myr. Primary masses are sampled from 0.4 to 2.0 $M_\odot$ in steps of 0.01 $M_\odot$, each with an array of companions from 0.1 $M_\odot$ to the primary mass (also sampled in 0.01 $M_\odot$ steps). No binary pairs jointly meet the constraints above. Specifically, meeting the observed apparent magnitudes requires primary masses between 1.20 and 1.11 $M_\odot$ and companion masses $\leq0.73 M_\odot$. For the same primary mass range, Equation \ref{eqn:V19} requires companion masses $\geq0.75 M_\odot$, and Equation \ref{eqn:CST3} requires a companions masses: $0.86 M_\odot \leq M_c \leq 1.20M_\odot$. With no companion pairs meeting the observed requirements, we can confidently rule out this scenario as a false positive. 
    
    \item \textit{Light from a background eclipsing binary or planet-hosting system is blended with HD 110082:} \\ The proper motion of HD 110082 is too small to rely on archival plate images to determine the presence of any currently aligned stars. With one epoch of speckle imaging, however, we can rule out companions within a 0.1\arcsec\ radius. Without the means to conclusively rule out an aligned source, we calculate the probability of this scenario by measuring the projected density of sources that meet the constraint from Equation \ref{eqn:V19} (which holds whether the source of the false positive is physically associated or a chance alignment). First, we query the \gaia\ DR2 source catalog for stars within one square degree of HD 110082, and cross-match them with the \tess\ Input Catalog (TIC) to compile \tess\ bandpass magnitudes, $T$, for each source. Of the 17,858 stars found within the DR2 search radius, all but 9 were successfully cross matched with the TIC. Unmatched sources are faint ($>$6 mags fainter than HD 110082 in $G$) and would not meet the $\Delta m_{\tess}$ requirement. Only 36 stars within one square degree meet the criterion from Equation \ref{eqn:V19}: $8.6 < T < 10.6$. Given this source density, we can expect $\simeq 9\times10^{-8}$ stars within a 0.1\arcsec\ radius of HD 110082. To compute the false-positive probability (FPP), we assume that 1 in 100 sources will transit/eclipse, which multiplied by the source density, corresponds to a FPP $\simeq 9\times10^{-10}$. With this vanishing small probability, which would be made even smaller with the inclusion of color constraints, we can confidently exclude this false positive scenario. 

\end{enumerate}

\subsection{Injection/Recovery Analysis for Additional Planets}
\label{inject-recover}

\begin{figure}
    \centering
    \includegraphics[width=0.48\textwidth]{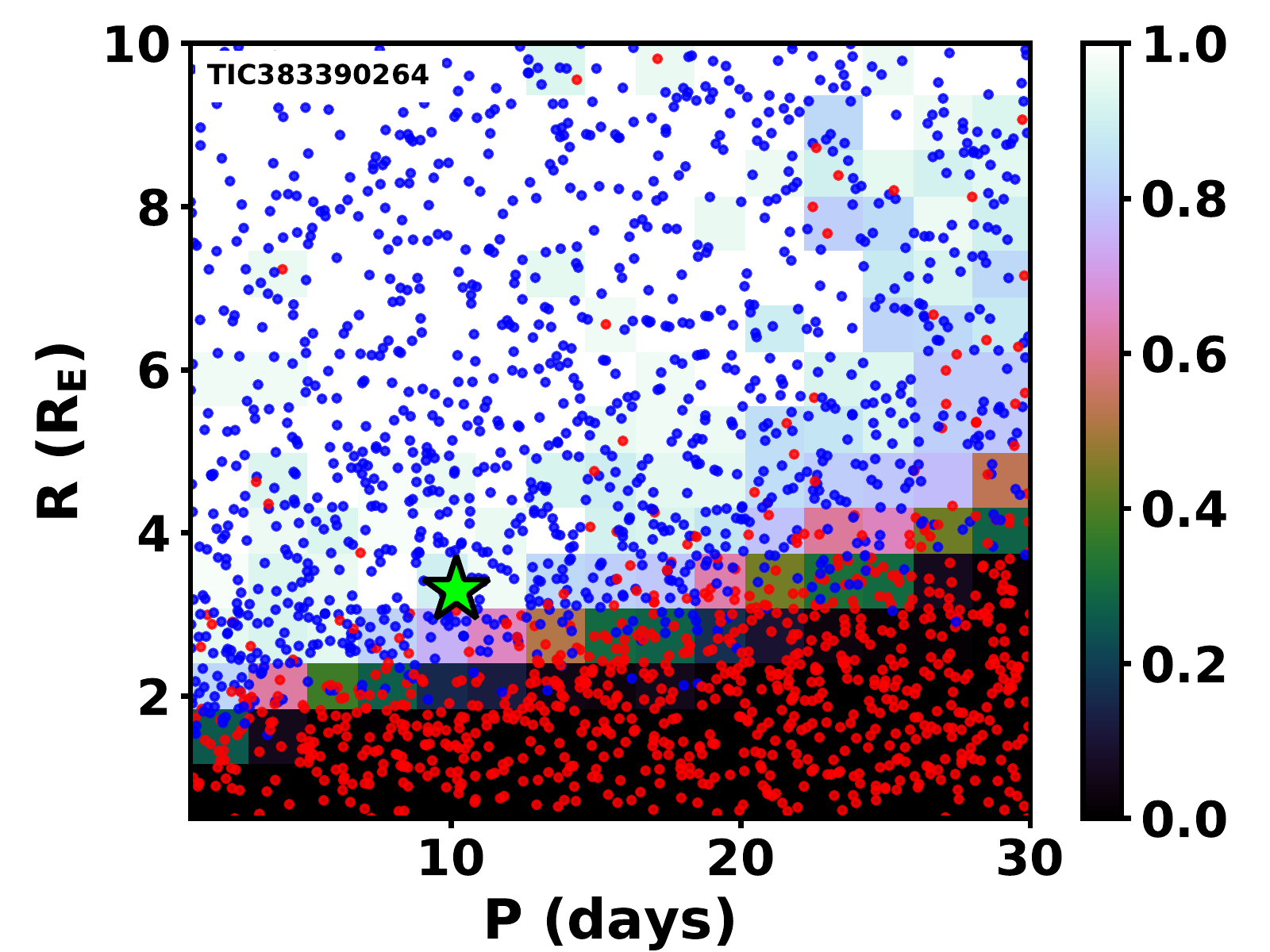}
    \caption{Injection recovery test results for the \tess~ light curve of HD 110082 using the notch filter pipeline and injection recovery formalism of \citet{Rizzutoetal2017}. Blue/red points are recovered/missed injected planet signals, and the shading indicates recovery completeness. The green star indicates the period and radius of HD 110082 b.}
    \label{fig:inject}
\end{figure}

We search for additional planets using a notch filter, as described in \citet{Rizzutoetal2017}. We recover the planet identified above and find no additional planet signals. We set limits on the existence of additional planets using an injection/recovery test, again following \citet{Rizzutoetal2017}. In summary, we generate planets using the \texttt{BATMAN} package following a uniform distribution in period, $b$, and orbital phase. Half of the planet radii are drawn from a uniform distribution spanning $0.5-10$\,R$_\oplus$, and the other half from a $\beta$ distribution with coefficients $\alpha=2$ and $\beta=6$ on the same range. We use this mixed distribution to ensure higher sampling around smaller and more common planets, where the transition between non-detection and detection expected to occur. We then detrend the light curve using the notch filter and search for planets in the detrended curve using a Box-Least Squares (BLS) algorithm \citep{Kovacs2002}, requiring at least two transits for recovery. We apply 5000 such random trials, the results of which are summarized in Figure~\ref{fig:inject}. We find that our search would be sensitive to $R_P\simeq 2 R_\oplus$ planets at periods of $<$10\,d and $R_P\simeq 3R_\oplus$ out to 20\,d. 

\section{Discussion \& Summary}
\label{discussion}

\begin{figure*}[!ht]
    \centering
    \includegraphics[width=1.0\textwidth]{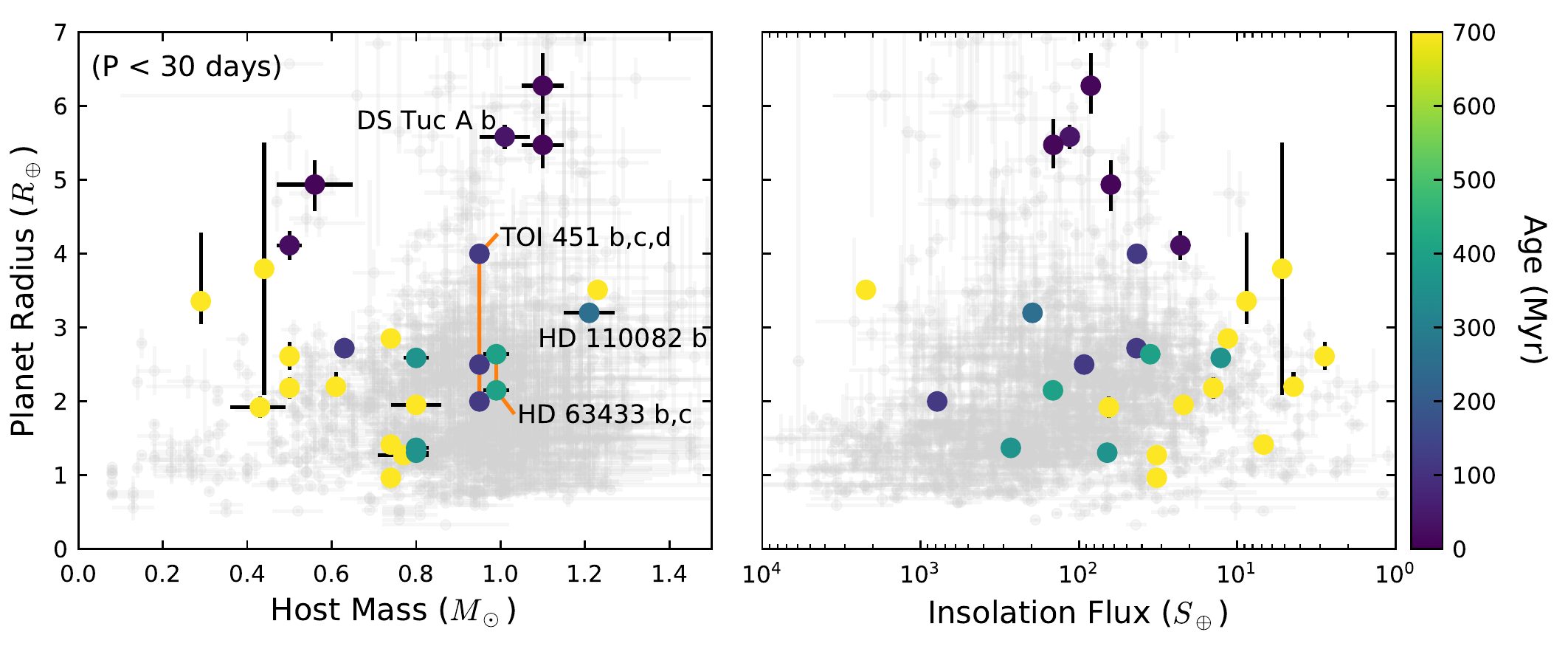}
    \caption{Distribution of young planets compared to field-age systems. Planet radii as a function of host mass and insolation flux are presented in the left and right panels, respectively. Young planets are shown as large circles, color-coded by their age. Field-age planets are shown as small grey circles. In addition to HD 110082 b, small planet from the THYME survey are highlighted (DS Tuc A b; Tuc-Hor, $\sim$40 Myr; \citealt{Newtonetal2019}, HD63433 b, c; Ursa Major, $\sim$400 Myr;
    \citealt{Mannetal2020}), TOI 451 b, c, d; Pisc-Eri Stream, $\sim$120 Myr; \citealt{Newtonetal2021}).}
    \label{fig:compare}
\end{figure*}

\begin{figure}[!ht]
    \centering
    \includegraphics[width=0.49\textwidth]{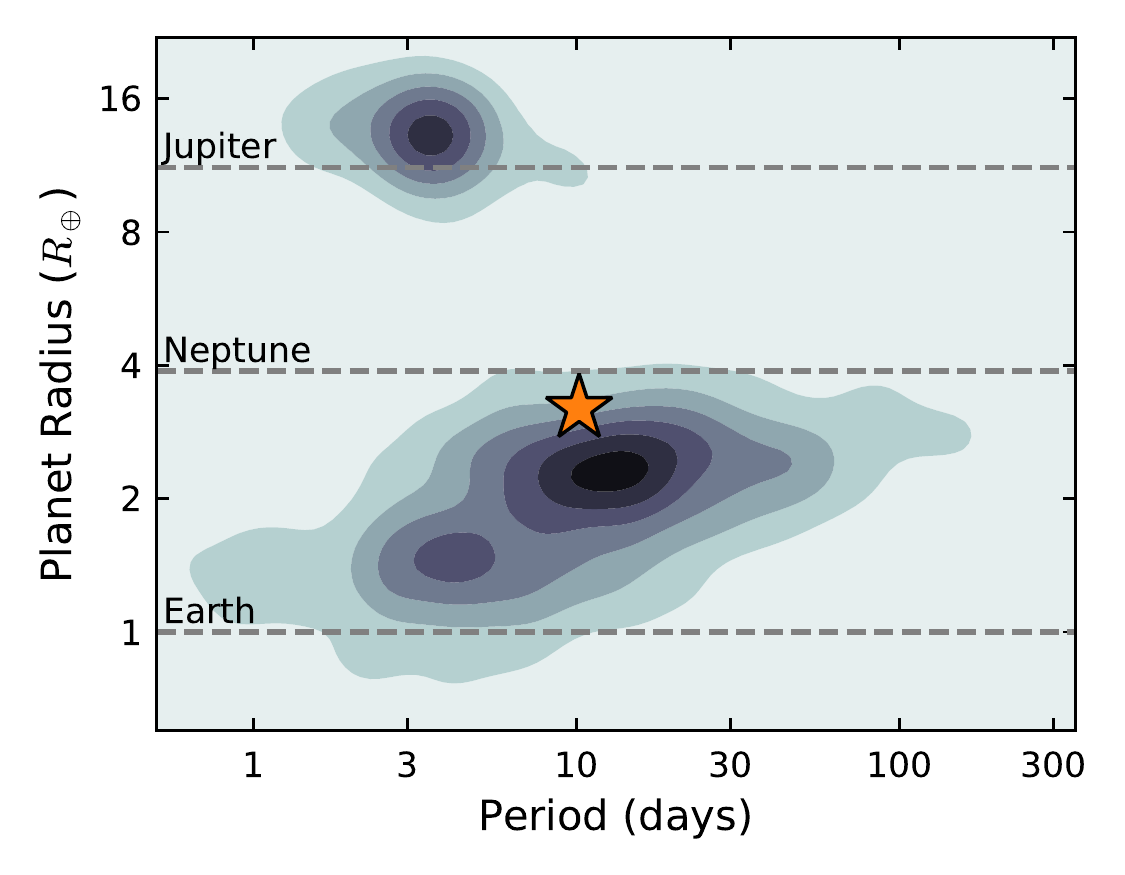}
    \caption{Distribution of planet radii as a function of orbital period. The population of planets without known ages, which is dominated by field-age systems, is represented as contours of a Gaussian kernel density estimate. HD 110082 b is represented by the gold star.}
    \label{fig:compare_dist}
\end{figure}

We report on the detection, characterization, and validation of a sub-Neptune-size planet transiting a young field star in a 10.1827 day orbit. In addition to the initial discovery with the \tess\ satellite, follow-up observations including space-based transit monitoring, high-contrast imaging, and high-resolution spectroscopy, are able to rule out sources of false positives and validate the planetary origin of the transit signal. 

The 1.21 $M_\odot$ planet host, HD 110082, is also host to a wide binary companion separated by $\sim$1\arcmin\ ($\sim$6200 AU in projection). With precise astrometry from \gaia\ DR2, we are able to not only conclude the companion is gravitationally bound, but also place constraints on the orbital parameters. While loose, our constraint on the orbital inclination favors those that are edge on, agreeing with limits on the HD 110082 rotational inclination, allowing for the mutual alignment of the system as a whole.

HD 110082 has kinematics consistent with membership to the $\sim$40 Myr, Octans moving group, however, its age-dependent characteristics (rotation period, Li abundance) are in tension with an age this young. This tension is made worse when considering the host's wide binary companion, whose lower mass (0.26 $M_\odot$) places it in a regime where the spread in evolutionary sequences is more pronounced. Without membership to a known moving group or association, we search for phase-space neighbors to HD 110082 to compile a population of coeval siblings. We use these candidate siblings in comparison with well-characterized young clusters and gyrochronology and \lii\ EW-based age relationships to place an age constraint on the system. In doing so, we have discovered a new young stellar association we designate MELANGE-1 with an age of $250^{+50}_{-70}$ Myr. This approach, relying primarily on archival data, offers an accessible and reliable means to age-date young field stars that are not members known associations. 

HD 110082 b solidly resides at an age where planet evolution is expected to be ongoing. Photoevaporation models predict that atmospheric losses driven by X-ray and UV photons primarily take place over the first few 100 Myr, but can extend throughout a Gyr depending on the planetary properties (planetary mass, core/envelope mass fraction, stellar radiation field, etc; \citealt{Owen&Jackson2012,Lopez&Fortney2013}). The effect on planetary radii is expected to be the largest for sub-Neptune-size planets, which HD 110082 b is, and has been suggested as an explanation for the reduced occurrence rate of 1.5--2.0 $R_\oplus$ planets \citep{Owen&Wu2017,Fulton&Petigura2018}. Alternatively (or jointly), core-powered mass loss of gaseous envelopes may also play an important role in the radius evolution of sub-Neptune-size planets over a similar timescale \citep{Ginzburgetal2018}. 

Empirically, young planets are found to be statistically larger than their field-age counterparts \citep{Mannetal2018,Rizzutoetal2018,Bergeretal2018}. This trend appears to continue in the case of HD 110082 b. Figure \ref{fig:compare} presents the radius distribution for short-period planets ($P<30$ days) as a function of the host mass and insolation flux. Systems with well-constrained ages $\lesssim$700 Myr are plotted as points, color-coded by their age. Small planets from the THYME survey are labeled. Field-age systems are shown as small grey circles (compiled from the exoplanet archive\footnote{\url{exoplanetarchive.ipac.caltech.edu/}}). In relation to systems with similar host masses or insolation fluxes, HD 110082 b's radius is in the 88th percentile of systems with masses and insolation fluxes within 20\% of the HD 110082's values (including orbital periods less than 30 days and planet radii less than 7 $R_\earth$). 

In Figure \ref{fig:compare_dist} we compare HD 110082 b to the larger exoplanet population in the period-radius plane. Contours represent a Gaussian kernel density estimate of the field-age planet population. HD 110082 b resides above the observed sub-Neptune density peak. If still in the process of radius evolution, it could eventually fall comfortably among the most commonly observed exoplanets. 

Lastly, given our stellar density prior, the HD 110082 b transit duration suggests low orbital eccentricities, 0.2$^{+0.2}_{-0.1}$, with a 95\% confidence upper limit of 0.7. 
Given the long tidal circularization timescale for this system ($\gtrsim$ 10 Gyr following \citealt{Goldreich&Soter1966}), any eccentricity gained in reaching its current orbit should remain. As such, from transit timing alone, we do not find strong evidence to suggest HD 110082 b reached its current orbit through a dynamically violent process that would result in large eccentricities (i.e. planet-planet scattering; \citealt{Chatterjeeetal2008,Kennedy&Kenyon2008}).

\subsection{Prospects for Follow-up}

HD 110082 is both nearby ($\sim$105 pc) and bright ($V$ = 9.2; $K$ = 8.0), making it a potentially attractive target for detailed follow-up observations. The primary challenge for ground-based follow-up is its very high southern declination ($\sim-$88$^\circ$), and correspondingly low elevation ($\sim$30$^\circ$; minimum airmass $\sim$ 1.8) at even the most southern observatories (excluding Antarctica). 

A mass estimate would allow for a direct comparison with the planet evolution scenarios described above. Based on its radius, HD 110082 b has a predicted mass of 11$^{+9}_{-5}$ $M_\oplus$ \citep{Chen&Kipping2017}. These masses correspond to RV semi-major amplitudes between 1.5 and 5 m s$^{-1}$. While within reach of current instruments, they are well below the expected RV jitter for this host star. Based on its amplitude of photometric variability in the \tess\ bandpass, $\sigma_{\rm{phot}}$, and the measured \vsini, we predict an optical RV jitter (RV dispersion) of $RV_{\rm{jitter}}\sim\sigma_{\rm{phot}} v$sin$i\sim$100 m s$^{-1}$, which generally agrees with the RV dispersion we measure in our CHIRON spectra ($\sim$140 m s$^{-1}$). However, the stellar and planetary signals are on different timescales (2 days versus 10 days) and may be possible to disentangle with a dedicated photometric and spectroscopic campaign.

More readily obtainable for this system is a measure of the planet's obliquity through the Rossiter-McLaughlin (RM) effect. For HD 110082 b's parameters, we predict an RM, RV amplitude of $\sim$16 m s$^{-1}$, following \citet{Gaudi&Winn2007}.

\

\noindent In summary, HD 110082 b contributes to the growing population of young planetary systems capable of testing planet formation and evolution theory, and provides a demonstration of how phase-space neighbors can provide age constraints for young systems. 

\acknowledgements
The authors would like to acknowledge Quinn Konopacky and Erik Petigura for useful discussions regarding the selection of coeval neighbor populations for young field stars. BMT acknowledges helpful discussions with David French and Benjamin Rosenwasser on spectral continuum normalization. 
This work was supported by the \tess\ Guest Investigator programs 80NSSC19K0636 and G03130 (awarded to AWM), and G03141 (awarded to ERN). 
Funding for the TESS mission is provided by NASA's Science Mission directorate. We acknowledge the use of public TESS Alert data from pipelines at the TESS Science Office and at the TESS Science Processing Operations Center. This paper includes data collected by the TESS mission, which are publicly available from the Mikulski Archive for Space Telescopes (MAST). Resources supporting this work were provided by the NASA High-End Computing (HEC) Program through the NASA Advanced Supercomputing (NAS) Division at Ames Research Center for the production of the SPOC data products.
Observations with the SMARTS 1.5m CHIRON spectrograph were obtained through the NOAO.
Some of the Observations in the paper made use of the High-Resolution Imaging instrument Zorro. Zorro was funded by the NASA Exoplanet Exploration Program and built at the NASA Ames Research Center by Steve B. Howell, Nic Scott, Elliott P. Horch, and Emmett Quigley. Zorro was mounted on the Gemini South telescope of the international Gemini Observatory, a program of NSF’s OIR Lab, which is managed by the Association of Universities for Research in Astronomy (AURA) under a cooperative agreement with the NSF on behalf of the Gemini partnership. 
This work makes use of observations from the LCOGT network.
This work is based in part on observations obtained at the Southern Astrophysical Research (SOAR) telescope, which is a joint project of the Minist\'{e}rio da Ci\^{e}ncia, Tecnologia e Inova\c{c}\~{o}es (MCTI/LNA) do Brasil, the US National Science Foundation’s NOIRLab, the University of North Carolina at Chapel Hill (UNC), and Michigan State University (MSU).
The authors acknowledge the Texas Advanced Computing Center (TACC) at The University of Texas at Austin for providing HPC resources that have contributed to the research results reported within this paper\footnote{\url{http://www.tacc.utexas.edu}}.
The national facility capability for SkyMapper has been funded through ARC LIEF grant LE130100104 from the Australian Research Council, awarded to the University of Sydney, the Australian National University, Swinburne University of Technology, the University of Queensland, the University of Western Australia, the University of Melbourne, Curtin University of Technology, Monash University and the Australian Astronomical Observatory. SkyMapper is owned and operated by The Australian National University's Research School of Astronomy and Astrophysics. The survey data were processed and provided by the SkyMapper Team at ANU. The SkyMapper node of the All-Sky Virtual Observatory (ASVO) is hosted at the National Computational Infrastructure (NCI). Development and support the SkyMapper node of the ASVO has been funded in part by Astronomy Australia Limited (AAL) and the Australian Government through the Commonwealth's Education Investment Fund (EIF) and National Collaborative Research Infrastructure Strategy (NCRIS), particularly the National eResearch Collaboration Tools and Resources (NeCTAR) and the Australian National Data Service Projects (ANDS). 
Figures in this paper made use of color-impaired-friendly schemes provided by ColorBrewer2.0 (Cynthia Brewer, Mark Harrower)\footnote{\url{https://colorbrewer2.org/}}.

\vspace{5mm}
\facilities{TESS, Spitzer, SMARTS 1.5m, Gemini South, SOAR, LCOGT (Sinistro; NRES), CDS, MAST, Simbad}

\software{{\tt astropy} \citep{astropy1,astropy2}, 
{\tt emcee} \citep{Foreman-Mackeyetal2013}, 
{\tt celerite} \citep{Foreman-Mackeyetal2017}, 
{\tt saphires} \citep{Tofflemireetal2019}, 
{\tt lofti\_gaiaDR2} \citep{Pearceetal2020}, 
{\tt misttborn} \citep{Johnsonetal2018},
{\tt lightkurve} \citep{lightkurve2018},
{\tt starspot} \citep{Angusetal2018},
{\tt eleanor} \citep{Feinsteinetal2019},
{\tt galpy} \citep{Bovy2015},
{\tt BALLFES} \citep{Stanford-Mooreetal2020}
}

\appendix

\section{The HD 110082 Neighborhood}
\label{appendix}

Most stars form in clusters \citep{Lada&Lada2003}, and while the majority of these groups may not remain bound past the dispersal of their natal molecular cloud \citep{Tutukov1978,Hills1980}, the stellar members will retain the common motions that they inherit from that cloud and will only gradually disperse into the Milky Way field population. If a group of siblings can be identified, the ensemble offers a more robust measure of their properties (such as age and metallicity) than any individual member. These populations can comprise recognizable over-densities in spatial density for tens of Myr, especially when clearly young stars (such as disk hosts) act as signposts for their existence \citep[e.g.,][]{Kastneretal1997,BarradoyNavascuesetal1999}. Even after the density drops below that of the field, these groups can still remain distinct in chemo-kinematic phase space for much longer \citep[e.g.,][]{Meingastetal2019,Tangetal2019}, potentially for billions of years \citep{Brownetal2010}. The precise astrometry and photometry of \gaia\ DR2 \citep{GAIAdr2} now enable a deeper search for stellar populations across numerous access of parameter space, but nonetheless, the Solar Neighborhood is a complex melange of star-forming events integrated over 10 Gyr of star formation. It remains a daunting task to disentangle the entire star formation history of the Milky Way disk.

It is more tractable to consider whether any one star (such as the host of a potentially young planet) remains detectably associated with its siblings. Even a simple hard-edged selection box in phase space can be successful in whittling the field population such that a visually obvious over-density emerges, or in selecting a feasible number of candidates for further down-selection based on follow-up observations. This strategy has already been adopted by the ACRONYM survey \citep{Krausetal2014,Shkolniketal2017}, which achieved very high yields in member searches for nearby young moving groups. Once a population has been identified, the candidates can be inspected for additional signatures of youth, and eventually the phase-space distribution can be characterized more rigorously by either fitting it directly within phase space (as is common for open clusters; \citealt{Sanders1971}; \citealt{Francic1989}) or by using confirmed members to estimate posterior distributions \citep{Maloetal2013,Gagneetal2018}. We therefore have developed a formalism to conduct this preliminary reconnaissance, searching for comoving neighbors to a specified target star using data from \gaia\ DR2 and other all-sky surveys. We have used this tool to search for comoving neighbors to HD 110082, in the hopes of finding a populations of siblings.

Our initial candidate list was drawn from \gaia\ DR2 \citep{GAIAdr2}, using the \gaia\ positions and parallaxes to identify all sources within a 3D volume of radius $R$ centered on the target star's own position and distance. We then used the \gaia\ proper motion of the target star, combined with an externally measured radial velocity $v_{r,targ}$, to compute the target star's $UVW$ space velocity \citep[e.g.,][]{Johnson&Soderblom1987}. Once computed, we reproject the target star's $UVW$ space velocity at the position and distance of each candidate, estimating the proper motion $\mu_{cand}$ (and hence tangential velocity $v_{t,cand}$) and radial velocity $v_{r,cand}$ that the candidate would have if it were exactly comoving at the same $UVW$. Finally, we select or reject each candidate based on a comparison of its observed tangential velocity $v_{t}$ to the value $v_{t,cand}$ that it would have if comoving, retaining those candidates that agree to within a threshold velocity difference.\footnote{All velocity conversions were computed using the Python package galpy \citep{Bovy2015}}. For our reconnaissance around HD 110082, we adopted a search radius of $R = 25$ pc and required a tangential velocity difference of $V_{\rm {off}}=|v_{t}-v_{t,cand}| \le 5$ km/s, yielding 134 \gaia\ sources that are candidate comoving siblings. From this sample, we make an initial cut to exclude sources with unreliable \gaia\ photometry (as judged from the phot\_bp\_rp\_excess\_factor value in the \gaia\ DR2 catalog; \citealt{GAIAdr2}; \citealt{Evansetal2018}), leaving us with 96 candidate comoving siblings.

\begin{figure}[!ht]
    \centering
    \includegraphics[width=0.6\textwidth]{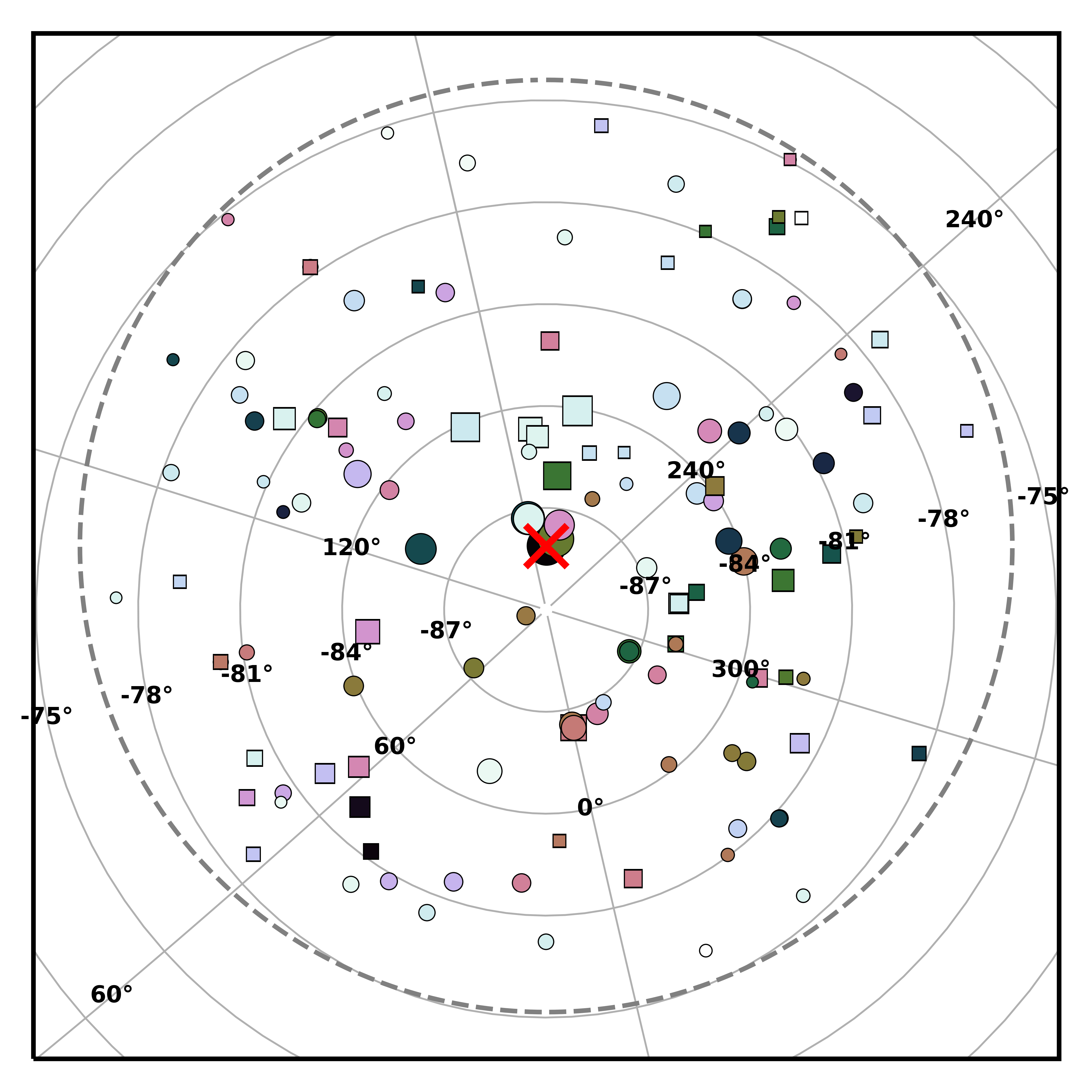}
    \caption{Candidate siblings in the neighborhood of HD 110082, as identified with {\tt FriendFinder}. Sky map of HD 110082 (red ``$\times$'') and the 96 candidate siblings that are located within a spherical volume of $R < 25$ pc. The symbol sizes are scaled linearly with 3D distance and the symbol colors denote the magnitude of the tangential velocity offset from comovement (see color scale in Figure \ref{fig:neighborhood2}), with circles used for objects with $RUWE < 1.2$ (likely single stars) and squares used for objects with $RUWE \ge 1.2$ (likely binary systems). There is not clear evidence of a spatial over-density, though several close neighbors also appear to be closely comoving in $v_{tan}$.}
    \label{fig:neighborhood1}
\end{figure}

If the siblings of HD 110082 remain sufficiently concentrated, their presence should be apparent as a recognizable spatial over-density on the sky, as can be seen for the kinematic candidate members of nearby young moving groups like Tucana-Horologium \citep{Krausetal2014}. In Figure~\ref{fig:neighborhood1} we plot the locations of the 96 candidate companions on the sky relative to HD 110082 (red ``$\times$'' symbol). For each candidate, the symbol size is scaled linearly with the 3D distance from HD 110082, while the point color is set by the absolute difference in tangential velocity $|v_{t}-v_{t,cand}|$. Sources with $RUWE > 1.2$ (denoting likely binarity; Section \ref{gaia}) are shown with squares, while all others are shown with circles.

\begin{figure}[!ht]
    \centering
    \includegraphics[width=0.495\textwidth]{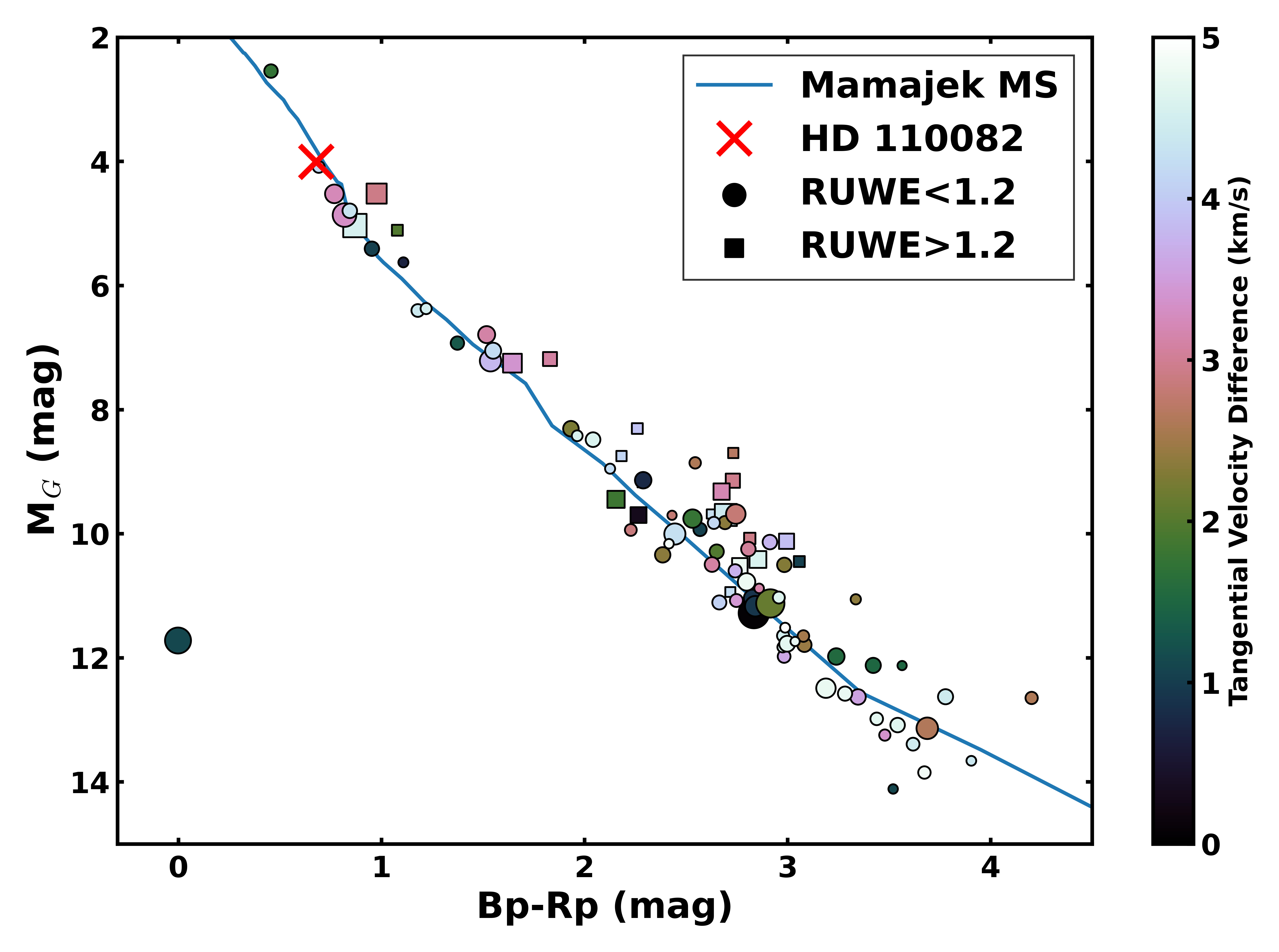}
    \includegraphics[width=0.495\textwidth]{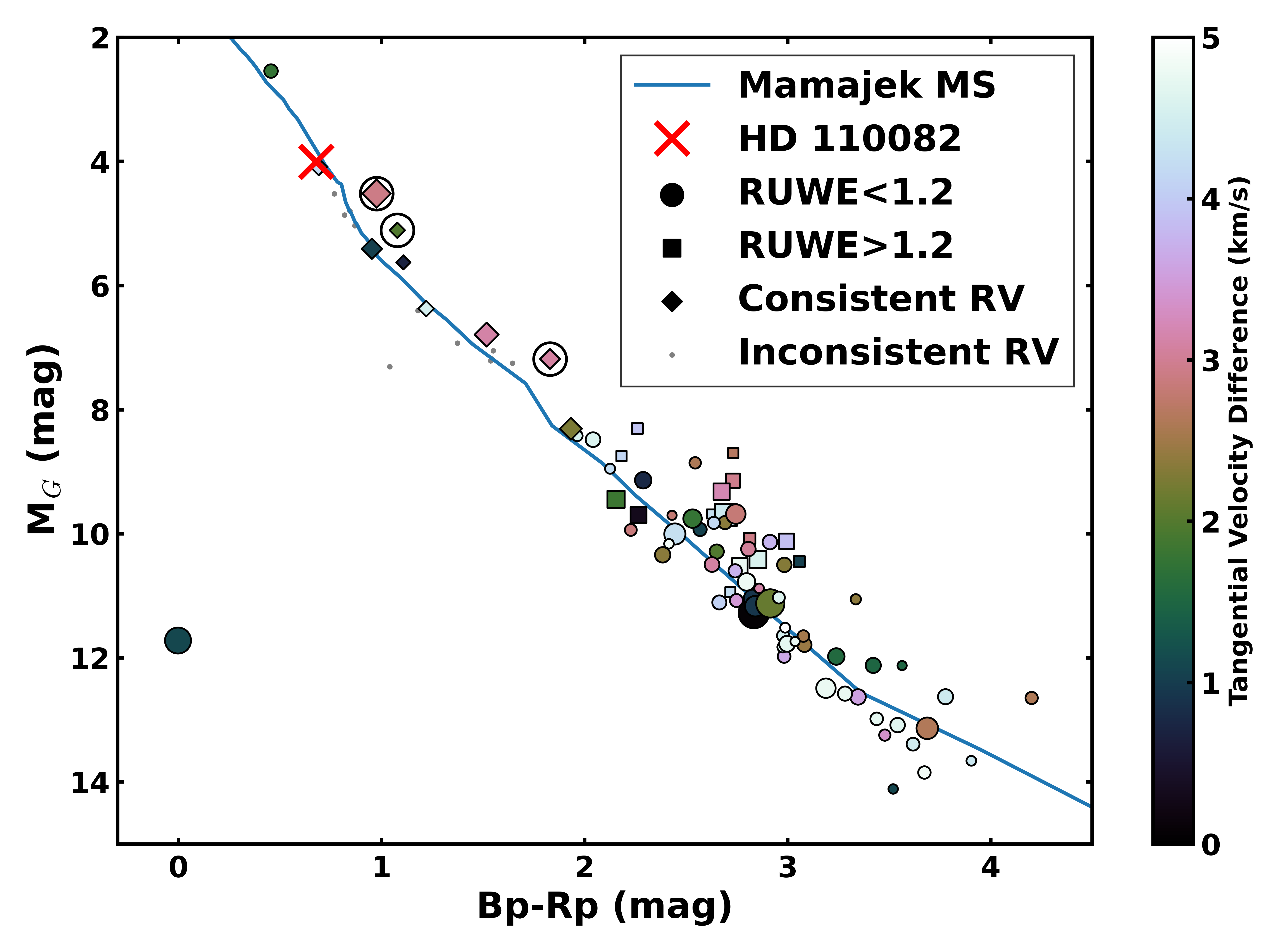}
    \caption{Candidate siblings in the neighborhood of HD 110082, as identified with {\tt FriendFinder}. {\bf Left:} \gaia\ CMD for HD 110082 and the candidate siblings. Symbols are as defined in Figure \ref{fig:neighborhood1}. There is also no clear evidence for a pre-main sequence population, but we note the existence of a likely wide binary companion and a potential sibling white dwarf. {\bf Right:} The same CMD, cleaned with RV information. Neighbors comoving in three dimensions are displayed as diamonds (color and size reflecting the same conventions above). Neighbors with inconsistent RVs are shown as small grey circles. RV-comoving neighbors with $RUWE > 1.2$ are encircled.}
    \label{fig:neighborhood2}
\end{figure}

If HD 110082 and its siblings are young enough for some stars to have not yet reached the zero-age main sequence, then they might also be apparent as a recognizable pre-MS sequence in the CMD. In Figure~\ref{fig:neighborhood2} (left), we plot the \gaia\ CMD ($M_G$ vs $B_p-R_p$) for the 96 candidate companions with the same symbol scheme as the sky map. To define the locus expected for typical Milky Way field stars, we also plot the typical main sequence for the Solar Neighborhood as defined by \citet{Pecaut&Mamajek2013} and updated by E. Mamajek on 2019 03 22\footnote{\url{http://www.pas.rochester.edu/~emamajek/EEM_dwarf_UBVIJHK_colors_Teff.txt}}. To aid in subsequent interpretation, we have also used the same color-magnitude sequence to infer an approximate spectral type for each source.

\begin{figure}[!ht]
    \centering
    \includegraphics[width=0.495\textwidth]{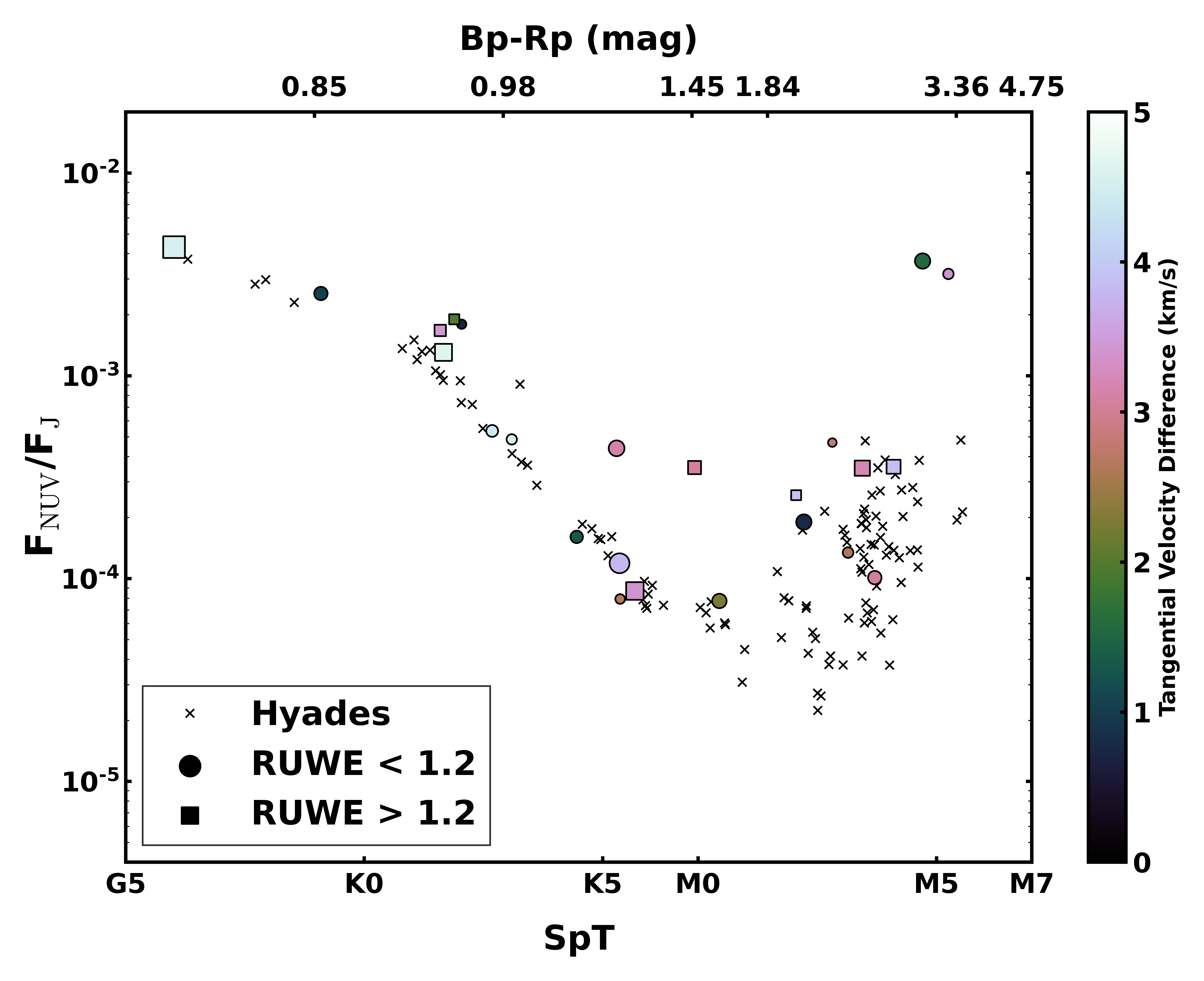}
    \includegraphics[width=0.495\textwidth]{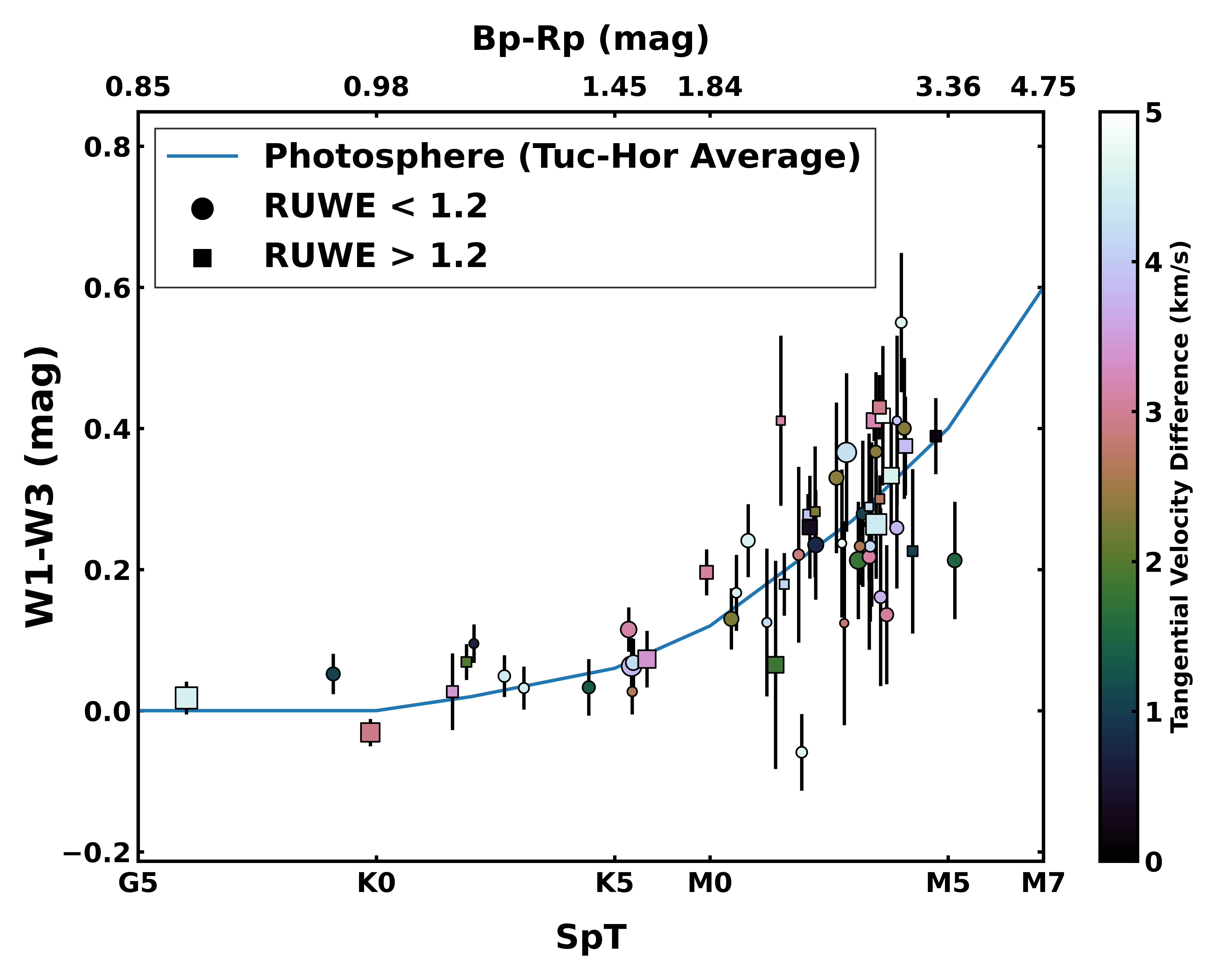}
    \caption{Candidate siblings in the neighborhood of HD 110082, as identified with {\tt FriendFinder}. The symbol sizes are scaled linearly with 3D distance and the symbol colors denote the magnitude of the tangential velocity offset from comovement, with circles used for objects with $RUWE < 1.2$ (likely single stars) and squares used for objects with $RUWE \ge 1.2$ (likely binary systems). {\bf Left:} {\it GALEX}/2MASS UV/NIR flux ratio as a function of $B_p-R_p$ color (or equivalently spectral type) for the 32 sources detected by {\it GALEX}. Symbols are as defined in the sky map, and we also show the corresponding sequence for the Hyades (as measured by Newton et al. 2020) to outline the threshold for stars to appear younger than the Milky Way field. There is no evidence of a young, active population among the candidate siblings. {\bf Right:} {\it WISE} $W1-W3$ color as a function of $B_p-R_p$ color (or equivalently spectral type) for the 67 sources detected in {\it WISE} $W3$. Symbols are as defined in the sky map, and we also show the photospheric color-SpT sequence measured for disk-free young stars in the Tuc-Hor moving group (as measured by \citealt{Krausetal2014}). No candidate siblings show a mid-infrared excess denoting the existence of a protoplanetary disk that would indicate youth. In summary, we conclude that while there are intriguing candidate siblings that should be followed up in more detail, identifying a cohort of siblings to HD 110082 will require follow-up observations to further whittle the candidate list along additional axes of parameter space. (Neither panel includes RV information.)}
    \label{fig:neighborhood3}
\end{figure}

Chromospheric activity is also a common feature for young stars, driven by rapid rotation and the resulting production of strong magnetic fields \citep{Skumanich1972}. X-ray emission (as detected by {\it ROSAT}; \citealt{Bolleretal2016}) is frequently interpreted as an indicator of youth, though an incomplete one since it is only universally detectable for the closest stars \citep{Shkolniketal2009}. UV excess emission (as detected by {\it GALEX}; \citealt{Bianchietal2017}) can also indicate youth \citep{Findeisen&Hillenbrand2010,Shkolniketal2011,Rodriguezetal2011}, and while more subtle than X-ray emission, it is also detectable for most young stars to a much larger distance (e.g., Newton et al.\ 2020). In Figure~\ref{fig:neighborhood3} (left), we plot the UV/NIR flux ratio, $F_{NUV}/F_J$, as a function of $B_p-R_p$ color (or inferred spectral type) for the 32 candidate companions that have GALEX counterparts. We also show the Hyades sequence as established by Newton et al.\ (2020), which defines the threshold distinguishing young stars from older stars that have spun down. We again define the symbols as in the sky map. We have also queried the 2nd ROSAT All-Sky Survey, finding that 12 candidate companions have an X-ray counterpart within $\le$1\arcmin.

Extremely young stars typically retain a gas-rich protoplanetary disk for $\sim$1--10 Myr \citep{Williams&Cieza2011}, with extremely long-lived disks seeming to survive for up to several tens of Myr \citep[e.g.,][]{Murphyetal2018}. The stellar luminosity that is absorbed and reprocessed by dust in the inner disk can produce an infrared excess in all-sky surveys such as 2MASS \citep{Skrutskie2006} and {\it WISE} \citep{Cutrietal2012} that acts as another robust youth indicator. In Figure~\ref{fig:neighborhood3} (right), we plot the {\it WISE} mid-infrared color ($W1 - W3$) as a function of $B_p-R_p$ color (or inferred spectral type) for the 67 sources that have a robust {\it WISE} $W3$ counterpart, again defining the symbols as in the sky map. We also plot the photospheric color-SpT sequence defined for disk-free members of the Tuc-Hor moving group by \citet{Krausetal2014}.

In Table \ref{tab:neightborhood}, we list all of these measured quantities for the 134 candidate companions that we identified to fall within $R < 25$ pc of HD 110082 and to comove in tangential velocity to $|v_{tan}-v_{t,cand}| \le 5$ km/s. To aid in such reconnaissance studies in the future, we also have created the flexible Python package {\tt Friendfinder} that can repeat all of these analyses for any other target that is in \gaia\ DR2\footnote{\url{https://github.com/adamkraus/Comove}}.

Inspection of Figure~\ref{fig:neighborhood1} reveals that there is no compelling evidence of a well-populated spatial over-density in the neighborhood of HD 110082, nor that a substantial number of pre-main sequence stars are present in the \gaia\ CMD (Figure \ref{fig:neighborhood2}). Similarly, while there are a few stars with UV flux excesses that fall above the Hyades sequence in the plot of $B_p-R_p$ vs $F_{NUV}/F_J$ (Figure \ref{fig:neighborhood3}; left), they do not fall closer than the other candidates in 3D distance (symbol size) or tangential velocity (symbol color), and none of the neighbors detected in {\it WISE} $W3$ demonstrate a clear excess (Figure \ref{fig:neighborhood3}; right). We therefore conclude that there is no compelling evidence of a comoving population, at odds with the classification of HD 110082 as a young ($\tau \sim 40$ Myr) member of the Octans moving group. However, this candidate list is quite tractable as a starting point for follow-up efforts to measure radial velocities and youth indicators, further winnowing the candidate list to see if a phase space over-density can be revealed (see subsections below).

Finally, we note that while there does not appear to be a large population of siblings in the vicinity of HD 110082, there are two intriguing candidates worthy of closer inspection. As we discuss in Section \ref{wide_binary}, the candidate sibling TIC 383400718 (\gaia\ DR2 5765748511163760640) is located only 59\arcsec\, northeast of HD 110082 and appears both comoving and codistant at high confidence, suggesting that it is likely a bound wide binary companion. We also identify the candidate sibling TIC 954359238 (\gaia\ DR2 5766091009035511680) which is located 1 degree northwest of HD 110082, differs in tangential velocity by only 1.1 km/s, and appears to sit on the white dwarf sequence in the \gaia\ CMD. TIC 954359238 was also identified as a candidate white dwarf by \citet{GentileFusilloetal2019}, who estimated a photometric surface temperature of $T_{eff} \sim 12000$ K. Follow-up spectroscopy would confirm its nature as a white dwarf, determine whether it is a hydrogen-dominated DA or helium-dominated DB, measure a radial velocity to further test its association with HD 110082, and measure its surface gravity and hence mass. With these measurements in hand, the cooling time for the white dwarf and the main-sequence lifetime of its progenitor could be estimated, with their sum offering an independent estimate of the age of HD 110082 and its transiting planet.

\subsection{Radial Velocities}
\label{app:neighborRV}

To strengthen the candidate-sibling status of neighborhood stars, while also thinning the neighbor sample, we obtain and compile RV measurements to test whether neighbors are indeed comoving in three dimensions. Follow-up observations with the LCO NRES echelle spectrograph are made for one bright neighbor. The spectrograph characteristics and RV measurement methodology are outlined in Sections \ref{nres} and \ref{rv}, respectively. We also query RV measurements derived from high-resolution spectra ($R>10000$) in the Vizier archive, adopting values from \gaia\ DR2 \citep{GAIAdr2,Cropperetal2018} and the SACY survey \citep{TorresCetal2006,Elliotetal2014}. In total, we are able to measure, or compile RVs for 20 stars in the neighbor sample. 

To assess whether neighbors are comoving radially as well as tangentially, we compare the measured RV to the predicted RV, $v_{r,cand}$, computed by {\tt FriendFinder}. We set the threshold for radial comotion as 3$\sigma$ agreement, assuming a 2 \kms\ error on the predicted RV, motivated by the velocity dispersion of known moving groups \citep[e.g.,][]{Gagneetal2018}. Nine sources meet this criterion. Unfortunately, none of the nine sources with comoving RVs come from our LCO/NRES follow-up that would allow for \lii\ EW measurements. 

The right panel of Figure \ref{fig:neighborhood2} presents the CMD cleaned with RV-information where available. The region of the CMD where RV information is present is not particularly sensitive to ages beyond $\sim$100 Myr, so the winnowing of candidate siblings in this setting does not further constrain the age of HD 110082. In the color-rotation period diagram (Figure \ref{fig:p_rot}), however, five of the RV members for which a stellar rotation period is measured appear to follow a distribution similar to the Pleiades. The fact that candidate siblings comoving in three dimensions appear to have similar age characteristics, supports the efficacy of the {\tt FriendFinder} approach and the use of the sibling-candidate sample in determining the age of the HD 110082 system. 

\subsection{\lii\ Equivalent Widths}
\label{app:li}

We estimated Li equivalent widths for each of the seven stars kinematically associated with HD 110082 using our Goodman spectra (Section~\ref{soar}). We simultaneously fit the width, depth, center, and the continuum line using a region around 6707.8\AA\ and assuming a Gaussian profile. The fit parameters were limited by physical and measurement constraints, e.g., the line center was allowed to deviate by 0.1\AA\ from 6707.8\AA\ (the error on our wavelength solution and rest-frame correction) and the width restricted by the spectral resolution and expected rotational broadening. The continuum was fit iteratively, rejecting points $>3\sigma$ below the continuum fit to handle nearby absorption lines. We measured the final equivalent width from the Gaussian fit to the region. These measurements include the contribution from the nearby iron line, Fe I 6707.4\AA. We consider EW measurements below 0.01\AA\ to be upper limits. 

We also tested removing the nearby absorption by dividing out spectral templates; this yielded consistent equivalent widths. We considered this template-fitting method less reliable because no template perfectly removed the absorption lines.

\subsection{Rotation Periods}
\label{app:rotation}

To test if the candidate sibling sample is able to reveal a coherent rotation-period sequence, we search for rotation periods in \tess\ 2-minute postage stamps and in  30-minute FFI data. We use PDCSAP light curves, accessed via {\tt lightkurve} \citep{lightkurve2018}, if available. Otherwise, we use full-frame image data reduced using {\tt eleanor} \citep{Feinsteinetal2019}. After an initial search for rotation periods and visual inspection of the results, we only consider stars with $T<15.5$. We note only one signal around a fainter target, TIC 1003702441 (\gaia\ DR2 5786574219173122560), which appears to be an eclipsing binary. This star is given the ``EB'' label in Table \ref{tab:neightborhood}.

We then search for rotation periods using both the autocorrelation function \cite[ACF;][]{McQuillanetal2013} and Lomb-Scargle periodograms \citep{Scargle1982}. We use the implementations of these functions in {\tt starspot}\footnote{\url{https://github.com/RuthAngus/starspot}} \citep{Angusetal2018} and consider periods between 0.05 and 25 days. We make our initial rotation period measurements by selecting the highest peak of the LS periodogram, finding that the ACF was more likely to be impacted by long-term systematics. We then visually inspect the results. If the ACF or LS periodogram indicate a signal at twice the initial period, we forced selection of the longer period peak, assuming the shorter period is a harmonic of the true period (period aliasing is unlikely to be an issue given the continuous sampling of \tess\ light curves). We flag stars with multiple distinct frequencies in the periodogram, for which the aperture can be assumed to contain multiple stars. 

Our final sample is selected by eye, and we report periods as determined from the LS periodogram. We distinguish between clear detections and candidate detections; the latter are either lower amplitude or potentially impacted by systematics. Excluding stars with multiple distinct frequencies, we detect periods in 21 stars and candidate periods in 4 stars, out of 78 searched. Our sensitivity diminishes beyond around 10 days, and as light curve systematics affect stars differently, so we do not interpret the {\it lack} of a detection.

Two candidate siblings comoving in three dimensions (\gaia\ DR2 6347492932234149120 and \gaia\ DR2 6343364815827362688) have rapidly evolving spot patterns that prove difficult to interpret with the method described above. For these 3D kinematic neighbors (of particular interest), we measure their rotation periods with the scalable Gaussian process tool, {\tt celerite} as described in Section \ref{rotp}. While scalable, this approach is not tractable for the entire sample and we reserve it for stars of interest where the above method fails.

\clearpage

\begin{longrotatetable}
\begin{deluxetable*}{lccccccccccccccccc}
\tablecaption{The HD 110082 Neighborhood -- MELANGE-1 Candidates
\label{tab:neightborhood}}
\tablewidth{0pt}
\tabletypesize{\scriptsize}
\tablecolumns{18}
\phd
\tablehead{
\colhead{\gaia\ DR2} &
\colhead{RA} &
\colhead{Dec} &
\colhead{\textit{G}} &
\colhead{\textit{B$_p$-R$_p$}} &
\colhead{RUWE} &
\colhead{$V_{\rm{off}}$} &
\colhead{$d_{3D}$} &
\colhead{$v_{r,cand}$} &
\colhead{$RV$} &
\colhead{$RV_{\rm{ref}}$} &
\colhead{Li EW} &
\colhead{Li EW$_{\rm{ref}}$} &
\colhead{P$_{\rm{rot}}$} &
\colhead{F$_{\rm NUV}$/F$_J$} &
\colhead{$W1-W3$} &
\colhead{Note} \\
\colhead{} &
\colhead{(deg)} &
\colhead{(deg)} &
\colhead{(mag)} &
\colhead{(mag)} &
\colhead{} &
\colhead{(km s$^{-1}$)} &
\colhead{(pc)} &
\colhead{(km s$^{-1}$)} &
\colhead{(km s$^{-1}$)} &
\colhead{} &
\colhead{(\AA)} &
\colhead{} &
\colhead{(d)} &
\colhead{(10$^{-3}$)} & 
\colhead{(mag)} &
\colhead{}
}
\startdata
5765748511163751936 & 192.5893 & -88.1211 & 9.12 & 0.68 & 1.079 & 0.0 & 0.0 & 3.6 & 3.63$\pm$0.06 & 1 & 0.09 &  1 & 2.3 &   & 0.02$\pm$0.02 & HD 110082 \\ 
5765748511163760640 & 192.9372 & -88.1092 & 16.4 & 2.83 & 1.082 & 0.06 & 0.6 & 3.6 &   &   &   &   & 0.8 &   &   & Wide Companion \\ 
4619368443610283136 & 48.1903 & -81.2354 & 15.62 & 3.25$^{a}$ & 3.535 & 0.14 & 22.1 & 5.8 &   &   &   &   &   &   & 0.39$\pm$0.05 & \\ 
4619046943834660864 & 55.6668 & -82.0326 & 14.74 & 2.27 & 1.246 & 0.29 & 17.4 & 5.7 &   &   &   &   &   &   & 0.26$\pm$0.07 & \\ 
5775823812950658432 & 247.2398 & -78.9273 & 20.43 & 2.38$^{a}$ & 1.026 & 0.5 & 19.4 & 1.4 &   &   &   &   &   &   &   & \\ 
5196316421300498944 & 122.9698 & -81.7354 & 11.08 & 1.11 & 0.953 & 0.7 & 23.9 & 4.7 & 5.06$\pm$0.52 & 2 & 0.105 &  1 & 5.9 & 1.8 & 0.09$\pm$0.03 & RV-comoving\\ 
5775319725524297728 & 254.7343 & -80.7611 & 14.33 & 2.29 & 1.14 & 0.78 & 16.5 & 1.9 &   &   &   &   & 3.8 & 0.19 & 0.24$\pm$0.08 & Theia 786 \\ 
5769040792575680384 & 240.0126 & -82.3 & 15.92 & 2.83 & 1.037 & 0.9 & 15.7 & 2.2 &   &   &   &   & 1.0 &   &   & \\ 
5767641114272601216 & 262.0139 & -84.2618 & 16.4 & 2.84 & 0.981 & 0.93 & 11.9 & 2.7 &   &   &   &   &   &   &   & Theia 786 \\ 
5196824739270050560 & 135.523 & -79.7764 & 10.65 & 0.95 & 0.942 & 1.04 & 18.9 & 4.4 & 3.64$\pm$0.9 & 2 & 0.08 &  1 & 6.5 & 2.54 & 0.05$\pm$0.03 & RV-comoving\\ 
6362272601894546816 & 304.024 & -78.2656 & 15.56 & 3.06 & 1.226 & 1.04 & 23.0 & 2.7 &   &   &   &   &   &   & 0.23$\pm$0.12 & \\ 
6348657177608778752 & 324.8093 & -80.8252 & 15.13 & 2.57 & 0.986 & 1.07 & 20.2 & 3.7 &   &   &   &   &   &   & 0.28$\pm$0.1 & \\ 
5215394391149157248 & 136.5243 & -76.7858 & 18.95 & 3.52 & 1.015 & 1.1 & 24.6 & 4.5 &   &   &   &   &   &   &   & \\ 
5766091009035511680 & 181.3996 & -87.2421 & 16.94 & -0.0 & 1.014 & 1.12 & 5.7 & 3.6 &   &   &   &   &   &   &   & White Dwarf \\ 
5199745003498944896 & 171.0696 & -79.773 & 19.5 & 1.67$^{a}$ & 3.777 & 1.12 & 24.6 & 3.0 &   &   &   &   &   &   &   & \\ 
5190133695618132992 & 128.4791 & -85.8922 & 20.61 & 3.03$^{a}$ & 1.098 & 1.15 & 7.8 & 4.2 &   &   &   &   &   &   &   & \\ 
6360516269508973696 & 271.6289 & -81.4359 & 19.1 & 3.11$^{a}$ & 3.551 & 1.26 & 19.2 & 2.3 &   &   &   &   &   &   &   & \\ 
6348657448190304512 & 324.694 & -80.8145 & 12.11 & 1.37 & 1.051 & 1.32 & 20.1 & 3.7 & -13.8$\pm$0.3 & 1 &   &   & 7.7 & 0.16 & 0.03$\pm$0.04 & \\ 
6345939597180530176 & 276.0333 & -85.5465 & 17.6 &   & 1.482 & 1.43 & 21.3 & 3.1 &   &   &   &   &   & 0.11 & 0.5$\pm$0.15 & \\ 
5791965987972663296 & 223.4406 & -76.8435 & 20.95 & 2.17$^{a}$ & 1.609 & 1.48 & 21.3 & 1.0 &   &   &   &   &   &   &   & \\ 
6346634076213813376 & 302.0372 & -83.5788 & 17.61 & 3.56 & 0.962 & 1.49 & 24.9 & 3.2 &   &   &   &   &   &   &   & \\ 
6342372197344151296 & 309.1032 & -87.2774 & 17.54 & 3.42 & 1.106 & 1.5 & 18.1 & 3.7 &   &   &   &   &   &   & 0.21$\pm$0.08 & \\ 
5773796558322153088 & 267.9736 & -82.8659 & 17.28 & 3.24 & 1.084 & 1.55 & 16.5 & 2.5 &   &   &   &   &   & 3.68 &   & Theia 786 \\ 
6342730019659961984 & 297.4382 & -86.0634 & 17.02 &   & 1.535 & 1.68 & 21.5 & 3.4 &   &   &   &   &   & 0.13 & 0.23$\pm$0.05 & \\ 
5195793466083385344 & 142.3744 & -81.2224 & 7.91 & 0.46 & 0.952 & 1.74 & 20.0 & 4.1 &   &   & 0.052 &  1 &   &   & 0.31$\pm$0.02 & \\ 
6342372197346011904 & 309.1002 & -87.2782 & 15.1 & 2.53 & 1.128 & 1.77 & 14.3 & 3.7 &   &   &   &   &   &   & 0.21$\pm$0.08 & \\ 
5786574219173122560 & 215.252 & -77.9258 & 20.59 & 2.46$^{a}$ & 1.956 & 1.79 & 24.8 & 1.4 &   &   &   &   &   &   &   & EB \\ 
5769271037181871104 & 197.2876 & -86.0412 & 19.56 & 2.17$^{a}$ & 1.399 & 1.82 & 10.8 & 3.2 &   &   &   &   &   &   &   & \\ 
6359400024688240512 & 275.605 & -82.9713 & 14.72 & 2.16 & 1.219 & 1.82 & 15.8 & 2.6 &   &   &   &   &   &   & 0.07$\pm$0.15 & Theia 786 \\ 
5195699732716798592 & 142.7282 & -81.2167 & 15.63 & 2.65 & 1.015 & 1.98 & 19.1 & 4.1 &   &   &   &   &   &   & 0.26$\pm$0.12 & \\ 
6347492932234149120 & 298.4877 & -82.6783 & 10.54 & 1.08 & 1.237 & 1.98 & 23.0 & 3.1 & -2.0$\pm$1.2 & 3 & 0.275 &  3 & 1.4 & 1.9 & 0.07$\pm$0.03 & Octans M \\ 
5765776686149621632 & 200.2258 & -87.8915 & 16.29 & 2.92 & 1.024 & 2.12 & 3.0 & 3.6 &   &   &   &   &   &   &   & \\ 
5792020168984546816 & 222.977 & -76.5819 & 18.91 & 2.84$^{a}$ & 4.832 & 2.17 & 24.5 & 0.9 &   &   &   &   &   &   &   & \\ 
4612569033640835584 & 63.8214 & -87.2734 & 13.06 & 1.93 & 1.162 & 2.25 & 17.6 & 4.5 & 4.32$\pm$1.45 & 2 &   &   &   & 0.08 & 0.13$\pm$0.04 & Field \\ 
5775905662142663040 & 269.3568 & -80.631 & 14.58 & 2.29 & 1.391 & 2.3 & 23.8 & 2.1 &   &   &   &   &   &   & 0.28$\pm$0.09 & Theia 786 \\ 
6348023721472455552 & 319.9424 & -82.624 & 15.36 & 2.98 & 1.18 & 2.32 & 18.9 & 3.6 &   &   &   &   &   &   & 0.4$\pm$0.1 & \\ 
6347977816861902976 & 320.451 & -83.1081 & 15.17 & 2.69 & 1.1 & 2.34 & 20.3 & 3.7 &   &   &   &   &   &   & 0.37$\pm$0.11 & \\ 
4614742321451898624 & 80.8928 & -83.9181 & 15.16 & 2.38 & 1.097 & 2.36 & 17.8 & 5.2 &   &   &   &   &   &   & 0.33$\pm$0.11 & \\ 
5768556526423021696 & 246.3388 & -83.8435 & 19.0 & 2.34$^{a}$ & 3.863 & 2.37 & 18.8 & 2.5 &   &   &   &   &   &   &   & \\ 
6347713968430779648 & 297.7706 & -82.1703 & 16.48 & 3.34 & 1.116 & 2.38 & 23.6 & 3.0 &   &   &   &   & 0.2 &   &   & Theia 786 \\ 
4611730823526295424 & 86.4126 & -89.3799 & 17.25 & 3.08 & 1.069 & 2.45 & 19.1 & 4.0 &   &   &   &   &   &   &   & \\ 
5766013802703686912 & 215.1249 & -86.4623 & 16.25 & 3.08 & 1.162 & 2.52 & 22.2 & 3.2 &   &   &   &   & 1.0 &   &   & \\ 
4616265759236854016 & 0.0148 & -86.5417 & 20.77 & 1.72$^{a}$ & 1.187 & 2.55 & 13.0 & 4.3 &   &   &   &   &   &   &   & \\ 
6342730019661547520 & 297.4324 & -86.0634 & 14.34 & 2.54 & 1.115 & 2.6 & 22.3 & 3.4 &   &   &   &   & 3.3 & 0.13 & 0.23$\pm$0.05 & \\ 
6344641859927264640 & 334.3651 & -84.1984 & 18.06 & 4.2 & 1.023 & 2.61 & 21.4 & 4.0 &   &   &   &   &   &   &   & \\ 
6351629290680888320 & 336.4267 & -81.0421 & 11.95 & 1.54 & 0.933 & 2.62 & 23.5 & 4.1 & 15.99$\pm$0.62 & 2 &   &   &   & 0.08 & 0.03$\pm$0.03 & \\ 
5773616646434379264 & 268.8401 & -84.016 & 18.26 & 3.69 & 1.005 & 2.63 & 10.8 & 2.7 &   &   &   &   &   &   &   & \\ 
4629766151413181312 & 9.3493 & -83.1858 & 14.12 & 2.73 & 1.42 & 2.69 & 23.7 & 4.9 &   &   &   &   &   &   & 0.3$\pm$0.03 & \\ 
5210764416404258304 & 93.2598 & -80.3093 & 14.25 &   & 3.658 & 2.72 & 22.6 & 5.7 &   &   &   &   & 2.8 & 11.41 & 0.13$\pm$0.02 & \\ 
5778812770888584320 & 241.5359 & -78.522 & 15.09 & 2.43 & 1.102 & 2.8 & 24.8 & 1.3 &   &   &   &   & 2.6 & 0.47 & 0.12$\pm$0.14 & \\ 
6343364712746486144 & 359.5301 & -86.4411 & 14.95 & 2.74 & 1.111 & 2.82 & 13.1 & 4.3 &   &   &   &   &   &   &   & \\ 
5207675785164300544 & 94.344 & -81.1086 & 15.29 & 2.23 & 1.157 & 2.86 & 21.9 & 5.5 &   &   &   &   &   &   & 0.22$\pm$0.12 & \\ 
6343364815827362688 & 359.5753 & -86.44 & 9.77 & 0.98 & 2.855 & 2.91 & 12.4 & 4.3 & -0.86$\pm$1.41 & 4 & 0.266 &  3 & 1.3 &   & -0.03$\pm$0.02 & Octans M \\ 
5203947169434827520 & 158.2246 & -77.7689 & 14.9 & 2.81 & 2.78 & 2.92 & 22.5 & 3.4 &   &   &   &   &   &   &   & \\ 
6350751231863421440 & 354.8266 & -81.6904 & 14.32 & 2.73 & 1.254 & 2.96 & 19.2 & 4.7 &   &   &   &   & 2.5 &   & 0.43$\pm$0.05 & \\ 
4630355111688690432 & 17.6717 & -81.9341 & 15.22 & 2.81 & 1.086 & 3.03 & 18.8 & 5.2 &   &   &   &   & 0.5 & 0.1 & 0.14$\pm$0.1 & \\ 
5783085576148686848 & 193.3898 & -82.0885 & 18.7 &   & 1.296 & 3.06 & 19.5 & 2.6 &   &   &   &   &   &   &   & \\ 
6346649808677390464 & 300.5504 & -83.4467 & 12.54 & 1.83 & 5.01 & 3.08 & 19.3 & 3.2 & 5.46$\pm$4.52 & 2 & $<$0.02 &  1 &   & 0.35 & 0.2$\pm$0.03 & RV-comoving\\ 
6342844678107493760 & 313.0195 & -86.2167 & 20.98 & 2.34$^{a}$ & 1.064 & 3.08 & 19.4 & 3.7 &   &   &   &   &   &   &   & \\ 
5191887519744419584 & 140.0097 & -84.1909 & 15.91 & 2.63 & 1.077 & 3.1 & 18.6 & 4.1 &   &   &   &   &   &   & 0.22$\pm$0.13 & \\ 
5793192007862350720 & 220.737 & -74.9447 & 16.3 & 2.17$^{a}$ & 1.518 & 3.12 & 24.8 & 0.6 &   &   &   &   &   &   & 0.41$\pm$0.12 & \\ 
6341894558326196480 & 346.4956 & -86.6 & 12.14 & 1.52 & 1.017 & 3.12 & 15.9 & 4.1 & -2.88$\pm$8.96 & 2 & 0.317 &  1 & 0.5 & 0.44 & 0.12$\pm$0.03 & Octans LM \\ 
5204855366041981184 & 153.6731 & -75.1995 & 15.95 & 2.86 & 1.094 & 3.14 & 24.4 & 3.5 &   &   &   &   &   &   &   & \\ 
5195525902504366720 & 143.7938 & -81.8532 & 19.32 & 3.29$^{a}$ & 3.482 & 3.17 & 18.7 & 4.1 &   &   &   &   &   &   &   & \\ 
4615952368358160512 & 62.3637 & -82.8152 & 14.27 & 2.68 & 1.346 & 3.19 & 16.8 & 5.5 &   &   &   &   & 1.4 & 0.35 & 0.41$\pm$0.03 & \\ 
5768542438930649088 & 234.9888 & -82.8646 & 9.42 & 0.77 & 0.922 & 3.24 & 14.1 & 2.3 & 34.65$\pm$0.37 & 2 &   &   &   & 7.88 & 0.04$\pm$0.02 & \\ 
5765852788673456512 & 201.2744 & -87.4697 & 10.14 & 0.82 & 1.135 & 3.34 & 8.4 & 3.5 & 15.89$\pm$0.25 & 2 &   &   & 6.0 & 6.03 & 0.03$\pm$0.03 & \\ 
5195109943510163968 & 141.2312 & -82.4654 & 18.7 & 3.48 & 1.008 & 3.37 & 22.5 & 4.1 &   &   &   &   &   & 3.18 &   & \\ 
5193876050946145664 & 95.4924 & -84.7167 & 12.52 & 1.64 & 1.379 & 3.39 & 13.8 & 4.9 & 56.28$\pm$0.89 & 2 &   &   &   & 0.09 & 0.07$\pm$0.04 & \\ 
5779473405577109376 & 231.3246 & -78.3973 & 20.93 & 1.29$^{a}$ & 1.176 & 3.42 & 23.4 & 1.3 &   &   &   &   &   &   &   & \\ 
5192204419611874688 & 155.9162 & -83.0807 & 16.52 & 2.75 & 1.052 & 3.44 & 20.6 & 3.7 &   &   &   &   &   &   &   & \\ 
4622808716710414848 & 70.0905 & -79.6324 & 12.48 & 1.04 & 38.87 & 3.45 & 21.4 & 6.1 & 35.59$\pm$0.85 & 2 &   &   &   & 1.67 & 0.03$\pm$0.05 & \\ 
5767795660079736320 & 249.5102 & -84.119 & 18.02 & 3.35 & 0.967 & 3.57 & 17.7 & 2.6 &   &   &   &   &   &   &   & \\ 
5199519397456669184 & 174.994 & -80.1998 & 20.2 & 2.75$^{a}$ & 1.093 & 3.59 & 18.9 & 2.9 &   &   &   &   &   &   &   & \\ 
4622284490182043904 & 67.3841 & -80.5925 & 16.89 & 2.98 & 1.08 & 3.62 & 20.8 & 6.0 &   &   &   &   &   &   &   & \\ 
4619829452515725696 & 42.3592 & -80.7851 & 15.59 & 2.74 & 1.176 & 3.71 & 20.2 & 5.9 &   &   &   &   &   &   & 0.16$\pm$0.13 & \\ 
4618418229341698560 & 31.1852 & -81.5554 & 15.21 & 2.91 & 1.17 & 3.76 & 18.7 & 5.6 &   &   &   &   & 1.0 &   & 0.26$\pm$0.09 & \\ 
5195061702439081344 & 138.3218 & -83.1556 & 12.34 & 1.54 & 1.078 & 3.8 & 11.0 & 4.2 & 10.94$\pm$0.63 & 2 &   &   &   & 0.12 & 0.06$\pm$0.04 & \\ 
6349111001031798144 & 310.6164 & -81.5843 & 19.92 & 2.34$^{a}$ & 2.019 & 3.86 & 18.6 & 3.3 &   &   &   &   &   &   &   & \\ 
4621913198849439360 & 65.814 & -81.919 & 15.32 & 3.0 & 1.284 & 3.88 & 18.1 & 5.7 &   &   &   &   & 0.1 & 0.36 & 0.38$\pm$0.07 & \\ 
5801769508521523200 & 259.5494 & -76.5502 & 20.94 & 1.49$^{a}$ & 1.239 & 3.92 & 24.2 & 1.0 &   &   &   &   &   &   &   & \\ 
5790809233025055232 & 199.0133 & -75.6626 & 20.1 & 2.19$^{a}$ & 2.973 & 3.92 & 23.6 & 1.3 &   &   &   &   &   &   &   & \\ 
4625871440709174144 & 62.2652 & -78.8107 & 13.28 & 2.26 & 1.232 & 3.94 & 23.1 & 6.4 &   &   &   &   & 1.0 & 0.26 & 0.28$\pm$0.03 & \\ 
5793192012161711872 & 220.7376 & -74.9443 & 16.22 & 2.92$^{a}$ & 1.176 & 3.98 & 24.6 & 0.6 &   &   &   &   &   &   & 0.41$\pm$0.12 & \\ 
5777277504761508480 & 251.7352 & -78.8342 & 19.81 & 1.47$^{a}$ & 5.438 & 4.0 & 20.4 & 1.4 &   &   &   &   &   &   &   & \\ 
6345553084484695296 & 331.6902 & -81.4623 & 16.06 & 2.66 & 1.141 & 4.08 & 19.3 & 4.0 &   &   &   &   &   &   &   & \\ 
5211271703583439616 & 106.8286 & -79.1943 & 14.08 & 2.18 & 1.249 & 4.14 & 23.8 & 5.5 &   &   &   &   &   &   & 0.18$\pm$0.04 & \\ 
6341876214522656256 & 340.9021 & -86.8051 & 15.3 & 2.64 & 1.148 & 4.16 & 21.5 & 4.1 &   &   &   &   & 0.7 &   & 0.23$\pm$0.11 & \\ 
5210764416405839488 & 93.2639 & -80.3095 & 8.91 & 0.69 & 0.962 & 4.17 & 21.9 & 5.7 & 9.02$\pm$0.23 & 2 & 0.1 &  1 & 2.8 & 11.41 & 0.13$\pm$0.02 & RV-comoving\\ 
5199301694152945920 & 160.8714 & -79.2844 & 12.07 & 1.55 & 1.163 & 4.21 & 17.1 & 3.4 & 32.87$\pm$0.72 & 2 &   &   &   &   & 0.07$\pm$0.03 & \\ 
5766530916766542720 & 225.0778 & -85.6077 & 14.49 & 2.13 & 1.09 & 4.22 & 23.9 & 3.0 &   &   &   &   &   &   & 0.12$\pm$0.1 & \\ 
5785369536682589568 & 211.7935 & -79.1814 & 15.63 & 2.72 & 1.932 & 4.24 & 24.0 & 1.7 &   &   &   &   &   &   &   & \\ 
5768165581318202496 & 244.8881 & -84.3932 & 20.89 & 0.74$^{a}$ & 0.952 & 4.24 & 16.1 & 2.6 &   &   &   &   &   &   &   & \\ 
5769814986200113920 & 218.9572 & -84.8273 & 15.23 & 2.62 & 1.207 & 4.26 & 24.8 & 2.8 &   &   &   &   &   &   & 0.29$\pm$0.1 & \\ 
5771697934222755072 & 221.9322 & -82.7799 & 15.02 & 2.44 & 1.135 & 4.27 & 11.2 & 2.3 &   &   &   &   &   &   & 0.37$\pm$0.11 & \\ 
5197061202988014848 & 137.6753 & -78.9839 & 20.34 & 2.91$^{a}$ & 1.01 & 4.28 & 20.5 & 4.3 &   &   &   &   &   &   &   & \\ 
5769744583095243648 & 208.0466 & -85.2092 & 20.89 & 2.01$^{a}$ & 1.366 & 4.29 & 22.8 & 3.0 &   &   &   &   &   &   &   & \\ 
5773505698840226304 & 224.6774 & -79.1919 & 10.04 & 0.84 & 1.052 & 4.33 & 18.7 & 1.5 & -17.64$\pm$0.29 & 2 &   &   & 10.2 & 3.59 & 0.03$\pm$0.03 & \\ 
5196471516862763904 & 126.9468 & -80.863 & 18.32 & 3.9 & 1.092 & 4.37 & 24.3 & 4.6 &   &   &   &   &   &   &   & \\ 
5780655032687849216 & 243.4951 & -77.3723 & 19.68 & 2.24$^{a}$ & 4.463 & 4.4 & 21.1 & 1.0 &   &   &   &   &   &   &   & \\ 
5209729913403958016 & 122.645 & -78.2472 & 11.54 & 1.18 & 0.931 & 4.41 & 20.8 & 5.0 & 40.44$\pm$0.81 & 2 &   &   &   & 0.53 & 0.05$\pm$0.03 & \\ 
5191383771620758528 & 168.7516 & -84.1247 & 14.93 & 2.7 & 1.214 & 4.41 & 10.0 & 3.5 &   &   &   &   &   &   & 0.26$\pm$0.08 & \\ 
5775994305971486848 & 264.008 & -80.1619 & 17.82 & 3.78 & 0.989 & 4.42 & 18.1 & 1.9 &   &   &   &   &   &   &   & \\ 
5789562897935814272 & 209.3748 & -76.9014 & 18.44 & 3.62 & 0.95 & 4.44 & 20.7 & 1.3 &   &   &   &   &   &   &   & \\ 
4632003206602584960 & 33.8108 & -80.4335 & 18.72 &   & 1.211 & 4.44 & 20.8 & 5.8 &   &   &   &   &   &   &   & \\ 
6345695608678745856 & 279.8669 & -86.0898 & 16.28 &   & 1.324 & 4.49 & 19.4 & 3.2 &   &   &   &   &   &   & 0.42$\pm$0.1 & \\ 
4631316600246944640 & 12.615 & -80.2382 & 16.77 & 2.98 & 0.995 & 4.51 & 21.5 & 5.4 &   &   &   &   &   &   &   & \\ 
5775111230632453248 & 240.8584 & -81.3272 & 11.05 & 1.22 & 0.982 & 4.52 & 22.7 & 1.9 & 4.18$\pm$0.3 & 2 & $<$0.02 &  1 & 9.2 & 0.48 & 0.03$\pm$0.03 & RV-comoving\\ 
5770429166524323200 & 201.5056 & -84.0702 & 10.04 & 0.87 & 3.652 & 4.54 & 8.7 & 2.8 & -11.68$\pm$3.19 & 2 &   &   &   & 4.31 & 0.02$\pm$0.02 & \\ 
4622433400992028288 & 75.2789 & -80.3976 & 19.82 & 1.7$^{a}$ & 5.062 & 4.57 & 21.2 & 5.9 &   &   &   &   &   &   &   & \\ 
5192367525290626432 & 155.8163 & -82.0513 & 13.89 & 1.96 & 1.112 & 4.57 & 23.0 & 3.7 &   &   &   &   &   &   & 0.17$\pm$0.05 & \\ 
5195981237756865536 & 138.7662 & -80.4657 & 15.44 & 2.85 & 2.004 & 4.57 & 16.0 & 4.3 &   &   &   &   &   &   & 0.33$\pm$0.08 & \\ 
5212463093151300480 & 104.0249 & -77.346 & 19.08 &   & 1.138 & 4.58 & 25.0 & 5.8 &   &   &   &   &   &   &   & \\ 
5773505668780214784 & 224.6797 & -79.1959 & 13.72 & 2.04 & 1.132 & 4.59 & 18.6 & 1.5 &   &   &   &   & 9.6 &   & 0.24$\pm$0.05 & \\ 
5769497635359778432 & 186.3938 & -85.3155 & 19.86 & 3.11$^{a}$ & 1.02 & 4.59 & 21.8 & 3.3 &   &   &   &   &   &   &   & \\ 
5766090493638407808 & 181.6605 & -87.2711 & 20.92 & 1.8$^{a}$ & 1.122 & 4.6 & 7.7 & 3.6 &   &   &   &   &   &   &   & \\ 
6355247066190182400 & 331.1366 & -78.7267 & 16.89 & 2.98 & 1.007 & 4.6 & 23.3 & 3.9 &   &   &   &   &   &   &   & \\ 
5203947169437207296 & 158.2174 & -77.7616 & 15.9 & 2.96 & 1.125 & 4.63 & 21.7 & 3.4 &   &   &   &   &   &   & 0.55$\pm$0.1 & \\ 
5770286779768022656 & 189.782 & -84.897 & 16.36 & 2.79 & 10.339 & 4.64 & 15.9 & 3.2 &   &   &   &   &   &   &   & \\ 
5196059749757867264 & 126.1991 & -82.1369 & 17.88 & 3.54 & 0.965 & 4.66 & 18.8 & 4.6 &   &   &   &   &   &   &   & \\ 
5770348794799011712 & 187.5769 & -84.6594 & 14.79 & 1.05 & 27.119 & 4.67 & 14.4 & 3.2 &   &   &   &   & 7.3 & 1.3 &   & \\ 
5787746543379543936 & 195.4347 & -79.0187 & 20.69 & 2.24$^{a}$ & 1.011 & 4.67 & 21.9 & 2.1 &   &   &   &   &   &   & -0.06$\pm$0.05 & \\ 
4620292205175761024 & 47.6485 & -80.1109 & 18.06 & 3.44 & 1.022 & 4.71 & 21.0 & 6.0 &   &   &   &   &   &   &   & \\ 
5765380861963776512 & 259.8728 & -86.7922 & 17.19 & 3.0 & 1.029 & 4.72 & 17.2 & 3.2 &   &   &   &   &   &   &   & \\ 
4622302048008312448 & 66.2987 & -80.3853 & 17.1 & 3.04 & 0.983 & 4.75 & 24.8 & 6.0 &   &   &   &   &   &   &   & \\ 
5202997874288632832 & 142.3315 & -78.507 & 17.7 & 3.28 & 0.987 & 4.76 & 19.1 & 4.1 &   &   &   &   &   &   &   & \\ 
4616786824668104064 & 31.8181 & -84.9722 & 17.5 & 3.19 & 1.004 & 4.77 & 13.2 & 4.9 &   &   &   &   &   &   &   & \\ 
6345695608680536320 & 279.8669 & -86.0903 & 15.92 & 2.76 & 1.264 & 4.8 & 17.5 & 3.2 &   &   &   &   &   &   & 0.42$\pm$0.1 & \\ 
5775084292597850368 & 245.6353 & -81.1553 & 15.98 & 2.8 & 1.16 & 4.8 & 15.4 & 1.9 &   &   &   &   & 2.3 &   &   & \\ 
5837057058610125440 & 182.7018 & -76.6413 & 18.86 & 3.67 & 0.98 & 4.84 & 21.2 & 2.2 &   &   &   &   &   &   &   & \\ 
5225083906094232832 & 174.4109 & -75.2124 & 15.09 & 2.42 & 1.043 & 4.87 & 24.4 & 2.4 &   &   &   &   &   &   & 0.24$\pm$0.1 & \\ 
5792080886934603904 & 225.4489 & -76.2543 & 13.66 &   & 1.772 & 4.97 & 24.0 & 0.8 &   &   &   &   &   & 0.22 & 0.12$\pm$0.03 & \\ 
6353594534572980480 & 347.8998 & -78.9441 & 16.44 & 2.99 & 1.082 & 4.98 & 24.1 & 4.7 &   &   &   &   &   &   &   & \\ 
\enddata
\tablenotetext{a}{Sibling candidate has a $B_p-R_p$ flux excess error, and is excluded from our analysis.}
\tablenotetext{b}{Sibling candidate's rotation period measurement is preliminary, and is excluded from our analysis.}
\tablecomments{RV and \lii\ EW References: (1) This Work, (2) Gaia DR2, (3) \citet{TorresCetal2006}, (4) \citet{Elliotetal2014}}
\end{deluxetable*}
\end{longrotatetable}


\begin{thebibliography}{}
\expandafter\ifx\csname natexlab\endcsname\relax\def\natexlab#1{#1}\fi
\providecommand{\url}[1]{\href{#1}{#1}}
\providecommand{\dodoi}[1]{doi:~\href{http://doi.org/#1}{\nolinkurl{#1}}}
\providecommand{\doeprint}[1]{\href{http://ascl.net/#1}{\nolinkurl{http://ascl.net/#1}}}
\providecommand{\doarXiv}[1]{\href{https://arxiv.org/abs/#1}{\nolinkurl{https://arxiv.org/abs/#1}}}

\bibitem[{{Angus} {et~al.}(2015){Angus}, {Aigrain}, {Foreman-Mackey}, \&
  {McQuillan}}]{Angusetal2015}
{Angus}, R., {Aigrain}, S., {Foreman-Mackey}, D., \& {McQuillan}, A. 2015,
  \mnras, 450, 1787, \dodoi{10.1093/mnras/stv423}

\bibitem[{{Angus} {et~al.}(2018){Angus}, {Morton}, {Aigrain}, {Foreman-Mackey},
  \& {Rajpaul}}]{Angusetal2018}
{Angus}, R., {Morton}, T., {Aigrain}, S., {Foreman-Mackey}, D., \& {Rajpaul},
  V. 2018, \mnras, 474, 2094, \dodoi{10.1093/mnras/stx2109}

\bibitem[{{Angus} {et~al.}(2019){Angus}, {Morton}, {Foreman-Mackey}, {van
  Saders}, {Curtis}, {Kane}, {Bedell}, {Kiman}, {Hogg}, \&
  {Brewer}}]{Angusetal2019}
{Angus}, R., {Morton}, T.~D., {Foreman-Mackey}, D., {et~al.} 2019, \aj, 158,
  173, \dodoi{10.3847/1538-3881/ab3c53}

\bibitem[{{Astropy Collaboration} {et~al.}(2013){Astropy Collaboration},
  {Robitaille}, {Tollerud}, {Greenfield}, {Droettboom}, {Bray}, {Aldcroft},
  {Davis}, {Ginsburg}, {Price-Whelan}, {Kerzendorf}, {Conley}, {Crighton},
  {Barbary}, {Muna}, {Ferguson}, {Grollier}, {Parikh}, {Nair}, {Unther},
  {Deil}, {Woillez}, {Conseil}, {Kramer}, {Turner}, {Singer}, {Fox}, {Weaver},
  {Zabalza}, {Edwards}, {Azalee Bostroem}, {Burke}, {Casey}, {Crawford},
  {Dencheva}, {Ely}, {Jenness}, {Labrie}, {Lim}, {Pierfederici}, {Pontzen},
  {Ptak}, {Refsdal}, {Servillat}, \& {Streicher}}]{astropy2}
{Astropy Collaboration}, {Robitaille}, T.~P., {Tollerud}, E.~J., {et~al.} 2013,
  \aap, 558, A33, \dodoi{10.1051/0004-6361/201322068}

\bibitem[{{Bailer-Jones} {et~al.}(2018){Bailer-Jones}, {Rybizki}, {Fouesneau},
  {Mantelet}, \& {Andrae}}]{Bailer-Jonesetal2018}
{Bailer-Jones}, C.~A.~L., {Rybizki}, J., {Fouesneau}, M., {Mantelet}, G., \&
  {Andrae}, R. 2018, \aj, 156, 58, \dodoi{10.3847/1538-3881/aacb21}

\bibitem[{{Baraffe} {et~al.}(2015){Baraffe}, {Homeier}, {Allard}, \&
  {Chabrier}}]{Baraffeetal2015}
{Baraffe}, I., {Homeier}, D., {Allard}, F., \& {Chabrier}, G. 2015, \aap, 577,
  A42, \dodoi{10.1051/0004-6361/201425481}

\bibitem[{{Barnes}(2007)}]{Barns2007}
{Barnes}, S.~A. 2007, \apj, 669, 1167, \dodoi{10.1086/519295}

\bibitem[{{Barnes} {et~al.}(2015){Barnes}, {Weingrill}, {Granzer}, {Spada}, \&
  {Strassmeier}}]{Barnesetal2015}
{Barnes}, S.~A., {Weingrill}, J., {Granzer}, T., {Spada}, F., \& {Strassmeier},
  K.~G. 2015, \aap, 583, A73, \dodoi{10.1051/0004-6361/201526129}

\bibitem[{{Barrado y Navascu{\'e}s} {et~al.}(1999){Barrado y Navascu{\'e}s},
  {Stauffer}, \& {Patten}}]{BarradoyNavascuesetal1999}
{Barrado y Navascu{\'e}s}, D., {Stauffer}, J.~R., \& {Patten}, B.~M. 1999,
  \apjl, 522, L53, \dodoi{10.1086/312212}

\bibitem[{{Belokurov} {et~al.}(2020){Belokurov}, {Penoyre}, {Oh}, {Iorio},
  {Hodgkin}, {Evans}, {Everall}, {Koposov}, {Tout}, {Izzard}, {Clarke}, \&
  {Brown}}]{Belokurovetal2020}
{Belokurov}, V., {Penoyre}, Z., {Oh}, S., {et~al.} 2020, \mnras, 496, 1922,
  \dodoi{10.1093/mnras/staa1522}

\bibitem[{{Benatti} {et~al.}(2019){Benatti}, {Nardiello}, {Malavolta},
  {Desidera}, {Borsato}, {Nascimbeni}, {Damasso}, {D'Orazi}, {Mesa}, {Messina},
  {Esposito}, {Bignamini}, {Claudi}, {Covino}, {Lovis}, \&
  {Sabotta}}]{Benattietal2019}
{Benatti}, S., {Nardiello}, D., {Malavolta}, L., {et~al.} 2019, \aap, 630, A81,
  \dodoi{10.1051/0004-6361/201935598}

\bibitem[{{Berger} {et~al.}(2018){Berger}, {Howard}, \&
  {Boesgaard}}]{Bergeretal2018}
{Berger}, T.~A., {Howard}, A.~W., \& {Boesgaard}, A.~M. 2018, \apj, 855, 115,
  \dodoi{10.3847/1538-4357/aab154}

\bibitem[{{Bianchi} {et~al.}(2017){Bianchi}, {Shiao}, \&
  {Thilker}}]{Bianchietal2017}
{Bianchi}, L., {Shiao}, B., \& {Thilker}, D. 2017, \apjs, 230, 24,
  \dodoi{10.3847/1538-4365/aa7053}

\bibitem[{{Blackwell} \& {Shallis}(1977)}]{Blackwell1977}
{Blackwell}, D.~E., \& {Shallis}, M.~J. 1977, \mnras, 180, 177

\bibitem[{{Blunt} {et~al.}(2017){Blunt}, {Nielsen}, {De Rosa}, {Konopacky},
  {Ryan}, {Wang}, {Pueyo}, {Rameau}, {Marois}, {Marchis}, {Macintosh},
  {Graham}, {Duch{\^e}ne}, \& {Schneider}}]{Bluntetal2017}
{Blunt}, S., {Nielsen}, E.~L., {De Rosa}, R.~J., {et~al.} 2017, \aj, 153, 229,
  \dodoi{10.3847/1538-3881/aa6930}

\bibitem[{{Boller} {et~al.}(2016){Boller}, {Freyberg}, {Tr{\"u}mper}, {Haberl},
  {Voges}, \& {Nandra}}]{Bolleretal2016}
{Boller}, T., {Freyberg}, M.~J., {Tr{\"u}mper}, J., {et~al.} 2016, \aap, 588,
  A103, \dodoi{10.1051/0004-6361/201525648}

\bibitem[{{Bouma} {et~al.}(2019){Bouma}, {Hartman}, {Bhatti}, {Winn}, \&
  {Bakos}}]{Boumaetal2019}
{Bouma}, L.~G., {Hartman}, J.~D., {Bhatti}, W., {Winn}, J.~N., \& {Bakos},
  G.~{\'A}. 2019, \apjs, 245, 13, \dodoi{10.3847/1538-4365/ab4a7e}

\bibitem[{{Bouma} {et~al.}(2020){Bouma}, {Hartman}, {Brahm}, {Evans},
  {Collins}, {Zhou}, {Sarkis}, {Quinn}, {de Leon}, {Livingston}, {Bergmann},
  {Stassun}, {Bhatti}, {Winn}, {Bakos}, {Abe}, {Crouzet}, {Dransfield},
  {Guillot}, {Marie-Sainte}, {M{\'e}karnia}, {Triaud}, {Tinney}, {Henning},
  {Espinoza}, {Jord{\'a}n}, {Barbieri}, {Nandakumar}, {Trifonov}, {Vines},
  {Vuckovic}, {Ziegler}, {Law}, {Mann}, {Ricker}, {Vanderspek}, {Seager},
  {Jenkins}, {Burke}, {Dragomir}, {Levine}, {Quintana}, {Rodriguez}, {Smith},
  \& {Wohler}}]{Boumaetal2020}
{Bouma}, L.~G., {Hartman}, J.~D., {Brahm}, R., {et~al.} 2020, \aj, 160, 239,
  \dodoi{10.3847/1538-3881/abb9ab}

\bibitem[{{Bouvier} {et~al.}(2018){Bouvier}, {Barrado}, {Moraux}, {Stauffer},
  {Rebull}, {Hillenbrand}, {Bayo}, {Boisse}, {Bouy}, {DiFolco}, {Lillo-Box}, \&
  {Morales Calder{\'o}n}}]{Bouvieretal2018}
{Bouvier}, J., {Barrado}, D., {Moraux}, E., {et~al.} 2018, \aap, 613, A63,
  \dodoi{10.1051/0004-6361/201731881}

\bibitem[{{Bovy}(2015)}]{Bovy2015}
{Bovy}, J. 2015, \apjs, 216, 29, \dodoi{10.1088/0067-0049/216/2/29}

\bibitem[{{Bowler} {et~al.}(2019){Bowler}, {Hinkley}, {Ziegler}, {Baranec},
  {Gizis}, {Law}, {Liu}, {Shah}, {Shkolnik}, {Riaz}, \&
  {Riddle}}]{Bowleretal2019}
{Bowler}, B.~P., {Hinkley}, S., {Ziegler}, C., {et~al.} 2019, \apj, 877, 60,
  \dodoi{10.3847/1538-4357/ab1018}

\bibitem[{{Brandeker} \& {Cataldi}(2019)}]{Brandeker&Cataldi2019}
{Brandeker}, A., \& {Cataldi}, G. 2019, \aap, 621, A86,
  \dodoi{10.1051/0004-6361/201834321}

\bibitem[{{Brandt} \& {Huang}(2015)}]{Brandt&Huang2015}
{Brandt}, T.~D., \& {Huang}, C.~X. 2015, \apj, 807, 24,
  \dodoi{10.1088/0004-637X/807/1/24}

\bibitem[{{Brooke} {et~al.}(2013){Brooke}, {Bernath}, {Schmidt}, \&
  {Bacskay}}]{12C2_data}
{Brooke}, J.~S.~A., {Bernath}, P.~F., {Schmidt}, T.~W., \& {Bacskay}, G.~B.
  2013, \jqsrt, 124, 11, \dodoi{10.1016/j.jqsrt.2013.02.025}

\bibitem[{{Brown} {et~al.}(2010){Brown}, {Portegies Zwart}, \&
  {Bean}}]{Brownetal2010}
{Brown}, A. G.~A., {Portegies Zwart}, S.~F., \& {Bean}, J. 2010, \mnras, 407,
  458, \dodoi{10.1111/j.1365-2966.2010.16921.x}

\bibitem[{{Brown} {et~al.}(2013){Brown}, {Baliber}, {Bianco}, {Bowman},
  {Burleson}, {Conway}, {Crellin}, {Depagne}, {De Vera}, {Dilday}, {Dragomir},
  {Dubberley}, {Eastman}, {Elphick}, {Falarski}, {Foale}, {Ford}, {Fulton},
  {Garza}, {Gomez}, {Graham}, {Greene}, {Haldeman}, {Hawkins}, {Haworth},
  {Haynes}, {Hidas}, {Hjelstrom}, {Howell}, {Hygelund}, {Lister}, {Lobdill},
  {Martinez}, {Mullins}, {Norbury}, {Parrent}, {Paulson}, {Petry}, {Pickles},
  {Posner}, {Rosing}, {Ross}, {Sand}, {Saunders}, {Shobbrook}, {Shporer},
  {Street}, {Thomas}, {Tsapras}, {Tufts}, {Valenti}, {Vander Horst}, {Walker},
  {White}, \& {Willis}}]{Brownetal2013}
{Brown}, T.~M., {Baliber}, N., {Bianco}, F.~B., {et~al.} 2013, \pasp, 125,
  1031, \dodoi{10.1086/673168}

\bibitem[{Campo {et~al.}(2011)Campo, Harrington, Hardy, Stevenson, Nymeyer,
  Ragozzine, Lust, Anderson, Collier-Cameron, Blecic, \& et~al.}]{Campo_2011}
Campo, C.~J., Harrington, J., Hardy, R.~A., {et~al.} 2011, The Astrophysical
  Journal, 727, 125, \dodoi{10.1088/0004-637x/727/2/125}

\bibitem[{{Cannon} \& {Pickering}(1920)}]{CannonPickering1920-12-14}
{Cannon}, A.~J., \& {Pickering}, E.~C. 1920, Annals of Harvard College
  Observatory, 95, 1

\bibitem[{{Cardelli} {et~al.}(1989){Cardelli}, {Clayton}, \&
  {Mathis}}]{Cardellietal1989}
{Cardelli}, J.~A., {Clayton}, G.~C., \& {Mathis}, J.~S. 1989, \apj, 345, 245,
  \dodoi{10.1086/167900}

\bibitem[{{Chatterjee} {et~al.}(2008){Chatterjee}, {Ford}, {Matsumura}, \&
  {Rasio}}]{Chatterjeeetal2008}
{Chatterjee}, S., {Ford}, E.~B., {Matsumura}, S., \& {Rasio}, F.~A. 2008, \apj,
  686, 580, \dodoi{10.1086/590227}

\bibitem[{{Chen} \& {Kipping}(2017)}]{Chen&Kipping2017}
{Chen}, J., \& {Kipping}, D. 2017, \apj, 834, 17,
  \dodoi{10.3847/1538-4357/834/1/17}

\bibitem[{{Claret}(2017)}]{Claret2017}
{Claret}, A. 2017, \aap, 600, A30, \dodoi{10.1051/0004-6361/201629705}

\bibitem[{{Claret} \& {Bloemen}(2011)}]{Claret&Bloemen2011}
{Claret}, A., \& {Bloemen}, S. 2011, \aap, 529, A75,
  \dodoi{10.1051/0004-6361/201116451}

\bibitem[{{Clemens} {et~al.}(2004){Clemens}, {Crain}, \& {Anderson}}]{Goodman}
{Clemens}, J.~C., {Crain}, J.~A., \& {Anderson}, R. 2004, in \procspie, Vol.
  5492, Ground-based Instrumentation for Astronomy, ed. A.~F.~M. {Moorwood} \&
  M.~{Iye}, 331--340

\bibitem[{{Collins} {et~al.}(2017){Collins}, {Kielkopf}, {Stassun}, \&
  {Hessman}}]{Collinsetal2017}
{Collins}, K.~A., {Kielkopf}, J.~F., {Stassun}, K.~G., \& {Hessman}, F.~V.
  2017, \aj, 153, 77, \dodoi{10.3847/1538-3881/153/2/77}

\bibitem[{{Cropper} {et~al.}(2018){Cropper}, {Katz}, {Sartoretti}, {Prusti},
  {de Bruijne}, {Chassat}, {Charvet}, {Boyadjian}, {Perryman}, {Sarri}, {Gare},
  {Erdmann}, {Munari}, {Zwitter}, {Wilkinson}, {Arenou}, {Vallenari},
  {G{\'o}mez}, {Panuzzo}, {Seabroke}, {Allende Prieto}, {Benson}, {Marchal},
  {Huckle}, {Smith}, {Dolding}, {Jan{\ss}en}, {Viala}, {Blomme}, {Baker},
  {Boudreault}, {Crifo}, {Soubiran}, {Fr{\'e}mat}, {Jasniewicz}, {Guerrier},
  {Guy}, {Turon}, {Jean-Antoine-Piccolo}, {Th{\'e}venin}, {David}, {Gosset}, \&
  {Damerdji}}]{Cropperetal2018}
{Cropper}, M., {Katz}, D., {Sartoretti}, P., {et~al.} 2018, \aap, 616, A5,
  \dodoi{10.1051/0004-6361/201832763}

\bibitem[{{Cummings} {et~al.}(2017){Cummings}, {Deliyannis}, {Maderak}, \&
  {Steinhauer}}]{Cummingsetal2017}
{Cummings}, J.~D., {Deliyannis}, C.~P., {Maderak}, R.~M., \& {Steinhauer}, A.
  2017, \aj, 153, 128, \dodoi{10.3847/1538-3881/aa5b86}

\bibitem[{{Cutri} \& {et al.}(2014)}]{allwise}
{Cutri}, R.~M., \& {et al.} 2014, VizieR Online Data Catalog, II/328

\bibitem[{{Cutri} {et~al.}(2012){Cutri}, {Wright}, {Conrow}, {Bauer},
  {Benford}, {Brandenburg}, {Dailey}, {Eisenhardt}, {Evans}, {Fajardo-Acosta},
  {Fowler}, {Gelino}, {Grillmair}, {Harbut}, {Hoffman}, {Jarrett},
  {Kirkpatrick}, {Leisawitz}, {Liu}, {Mainzer}, {Marsh}, {Masci}, {McCallon},
  {Padgett}, {Ressler}, {Royer}, {Skrutskie}, {Stanford}, {Wyatt}, {Tholen},
  {Tsai}, {Wachter}, {Wheelock}, {Yan}, {Alles}, {Beck}, {Grav}, {Masiero},
  {McCollum}, {McGehee}, {Papin}, \& {Wittman}}]{Cutrietal2012}
{Cutri}, R.~M., {Wright}, E.~L., {Conrow}, T., {et~al.} 2012, {Explanatory
  Supplement to the WISE All-Sky Data Release Products}, Explanatory Supplement
  to the WISE All-Sky Data Release Products

\bibitem[{{Czekala} {et~al.}(2015){Czekala}, {Andrews}, {Mandel}, {Hogg}, \&
  {Green}}]{starfish_paper}
{Czekala}, I., {Andrews}, S.~M., {Mandel}, K.~S., {Hogg}, D.~W., \& {Green},
  G.~M. 2015, \apj, 812, 128, \dodoi{10.1088/0004-637X/812/2/128}

\bibitem[{Czekala {et~al.}(2018)Czekala, gully, Gullikson, Andrews, Neal,
  Lucas, Hardegree-Ullman, Rawls, \& Betts}]{starfish_code}
Czekala, I., gully, Gullikson, K., {et~al.} 2018, {iancze/Starfish: ca. Czekala
  et al. 2015 release w/ Zenodo}, \dodoi{10.5281/zenodo.2221006}.
\newblock \url{https://doi.org/10.5281/zenodo.2221006}

\bibitem[{{da Silva} {et~al.}(2009){da Silva}, {Torres}, {de La Reza}, {Quast},
  {Melo}, \& {Sterzik}}]{daSilvaetal2009}
{da Silva}, L., {Torres}, C.~A.~O., {de La Reza}, R., {et~al.} 2009, \aap, 508,
  833, \dodoi{10.1051/0004-6361/200911736}

\bibitem[{{Dahm}(2015)}]{Dahm2015}
{Dahm}, S.~E. 2015, \apj, 813, 108, \dodoi{10.1088/0004-637X/813/2/108}

\bibitem[{{David} {et~al.}(2019){David}, {Petigura}, {Luger}, {Foreman-Mackey},
  {Livingston}, {Mamajek}, \& {Hillenbrand}}]{Davidetal2019}
{David}, T.~J., {Petigura}, E.~A., {Luger}, R., {et~al.} 2019, \apjl, 885, L12,
  \dodoi{10.3847/2041-8213/ab4c99}

\bibitem[{{David} {et~al.}(2016){David}, {Hillenbrand}, {Petigura},
  {Carpenter}, {Crossfield}, {Hinkley}, {Ciardi}, {Howard}, {Isaacson}, {Cody},
  {Schlieder}, {Beichman}, \& {Barenfeld}}]{Davidetal2016}
{David}, T.~J., {Hillenbrand}, L.~A., {Petigura}, E.~A., {et~al.} 2016, \nat,
  534, 658, \dodoi{10.1038/nature18293}

\bibitem[{{David} {et~al.}(2018{\natexlab{a}}){David}, {Crossfield}, {Benneke},
  {Petigura}, {Gonzales}, {Schlieder}, {Yu}, {Isaacson}, {Howard}, {Ciardi},
  {Mamajek}, {Hillenbrand}, {Cody}, {Riedel}, {Schwengeler}, {Tanner}, \&
  {Ende}}]{Davidetal2018a}
{David}, T.~J., {Crossfield}, I. J.~M., {Benneke}, B., {et~al.}
  2018{\natexlab{a}}, \aj, 155, 222, \dodoi{10.3847/1538-3881/aabde8}

\bibitem[{{David} {et~al.}(2018{\natexlab{b}}){David}, {Mamajek}, {Vanderburg},
  {Schlieder}, {Bristow}, {Petigura}, {Ciardi}, {Crossfield}, {Isaacson},
  {Cody}, {Stauffer}, {Hillenbrand}, {Bieryla}, {Latham}, {Fulton}, {Rebull},
  {Beichman}, {Gonzales}, {Hirsch}, {Howard}, {Vasisht}, \&
  {Ygouf}}]{Davidetal2018b}
{David}, T.~J., {Mamajek}, E.~E., {Vanderburg}, A., {et~al.}
  2018{\natexlab{b}}, \aj, 156, 302, \dodoi{10.3847/1538-3881/aaeed7}

\bibitem[{{de Bruijne} {et~al.}(2001){de Bruijne}, {Hoogerwerf}, \& {de
  Zeeuw}}]{deBruijneetal2001}
{de Bruijne}, J.~H.~J., {Hoogerwerf}, R., \& {de Zeeuw}, P.~T. 2001, \aap, 367,
  111, \dodoi{10.1051/0004-6361:20000410}

\bibitem[{{Delorme} {et~al.}(2011){Delorme}, {Collier Cameron}, {Hebb},
  {Rostron}, {Lister}, {Norton}, {Pollacco}, \& {West}}]{Delormeetal2011}
{Delorme}, P., {Collier Cameron}, A., {Hebb}, L., {et~al.} 2011, \mnras, 413,
  2218, \dodoi{10.1111/j.1365-2966.2011.18299.x}

\bibitem[{{D{\'e}sert} {et~al.}(2015){D{\'e}sert}, {Charbonneau}, {Torres},
  {Fressin}, {Ballard}, {Bryson}, {Knutson}, {Batalha}, {Borucki}, {Brown},
  {Deming}, {Ford}, {Fortney}, {Gilliland}, {Latham}, \&
  {Seager}}]{Desertetal2015}
{D{\'e}sert}, J.-M., {Charbonneau}, D., {Torres}, G., {et~al.} 2015, \apj, 804,
  59, \dodoi{10.1088/0004-637X/804/1/59}

\bibitem[{{Douglas} {et~al.}(2019){Douglas}, {Curtis}, {Ag{\"u}eros},
  {Cargile}, {Brewer}, {Meibom}, \& {Jansen}}]{Douglasetal2019}
{Douglas}, S.~T., {Curtis}, J.~L., {Ag{\"u}eros}, M.~A., {et~al.} 2019, \apj,
  879, 100, \dodoi{10.3847/1538-4357/ab2468}

\bibitem[{{Dressing} \& {Charbonneau}(2015)}]{Dressing&Charbonneau2015}
{Dressing}, C.~D., \& {Charbonneau}, D. 2015, \apj, 807, 45,
  \dodoi{10.1088/0004-637X/807/1/45}

\bibitem[{{Eastman} {et~al.}(2013){Eastman}, {Gaudi}, \&
  {Agol}}]{Eastmanetal2013}
{Eastman}, J., {Gaudi}, B.~S., \& {Agol}, E. 2013, \pasp, 125, 83,
  \dodoi{10.1086/669497}

\bibitem[{{Elliott} {et~al.}(2014){Elliott}, {Bayo}, {Melo}, {Torres},
  {Sterzik}, \& {Quast}}]{Elliotetal2014}
{Elliott}, P., {Bayo}, A., {Melo}, C.~H.~F., {et~al.} 2014, \aap, 568, A26,
  \dodoi{10.1051/0004-6361/201423856}

\bibitem[{{Elliott} {et~al.}(2016){Elliott}, {Bayo}, {Melo}, {Torres},
  {Sterzik}, {Quast}, {Montes}, \& {Brahm}}]{Elliottetal2016}
---. 2016, \aap, 590, A13, \dodoi{10.1051/0004-6361/201628253}

\bibitem[{{Evans} {et~al.}(2018){Evans}, {Riello}, {De Angeli}, {Carrasco},
  {Montegriffo}, {Fabricius}, {Jordi}, {Palaversa}, {Diener}, {Busso},
  {Cacciari}, {van Leeuwen}, {Burgess}, {Davidson}, {Harrison}, {Hodgkin},
  {Pancino}, {Richards}, {Altavilla}, {Balaguer-N{\'u}{\~n}ez}, {Barstow},
  {Bellazzini}, {Brown}, {Castellani}, {Cocozza}, {De Luise}, {Delgado},
  {Ducourant}, {Galleti}, {Gilmore}, {Giuffrida}, {Holl}, {Kewley}, {Koposov},
  {Marinoni}, {Marrese}, {Osborne}, {Piersimoni}, {Portell}, {Pulone},
  {Ragaini}, {Sanna}, {Terrett}, {Walton}, {Wevers}, \&
  {Wyrzykowski}}]{Evansetal2018}
{Evans}, D.~W., {Riello}, M., {De Angeli}, F., {et~al.} 2018, \aap, 616, A4,
  \dodoi{10.1051/0004-6361/201832756}

\bibitem[{{Fabrycky} \& {Tremaine}(2007)}]{Fabryckyetal2007}
{Fabrycky}, D., \& {Tremaine}, S. 2007, \apj, 669, 1298, \dodoi{10.1086/521702}

\bibitem[{{Fazio} {et~al.}(2004){Fazio}, {Hora}, {Allen}, {Ashby}, {Barmby},
  {Deutsch}, {Huang}, {Kleiner}, {Marengo}, {Megeath}, {Melnick}, {Pahre},
  {Patten}, {Polizotti}, {Smith}, {Taylor}, {Wang}, {Willner}, {Hoffmann},
  {Pipher}, {Forrest}, {McMurty}, {McCreight}, {McKelvey}, {McMurray}, {Koch},
  {Moseley}, {Arendt}, {Mentzell}, {Marx}, {Losch}, {Mayman}, {Eichhorn},
  {Krebs}, {Jhabvala}, {Gezari}, {Fixsen}, {Flores}, {Shakoorzadeh}, {Jungo},
  {Hakun}, {Workman}, {Karpati}, {Kichak}, {Whitley}, {Mann}, {Tollestrup},
  {Eisenhardt}, {Stern}, {Gorjian}, {Bhattacharya}, {Carey}, {Nelson},
  {Glaccum}, {Lacy}, {Lowrance}, {Laine}, {Reach}, {Stauffer}, {Surace},
  {Wilson}, {Wright}, {Hoffman}, {Domingo}, \& {Cohen}}]{Fazioetal2004}
{Fazio}, G.~G., {Hora}, J.~L., {Allen}, L.~E., {et~al.} 2004, \apjs, 154, 10,
  \dodoi{10.1086/422843}

\bibitem[{{Feinstein} {et~al.}(2019){Feinstein}, {Montet}, {Foreman-Mackey},
  {Bedell}, {Saunders}, {Bean}, {Christiansen}, {Hedges}, {Luger}, {Scolnic},
  \& {Cardoso}}]{Feinsteinetal2019}
{Feinstein}, A.~D., {Montet}, B.~T., {Foreman-Mackey}, D., {et~al.} 2019,
  \pasp, 131, 094502, \dodoi{10.1088/1538-3873/ab291c}

\bibitem[{{Findeisen} \& {Hillenbrand}(2010)}]{Findeisen&Hillenbrand2010}
{Findeisen}, K., \& {Hillenbrand}, L. 2010, \aj, 139, 1338,
  \dodoi{10.1088/0004-6256/139/4/1338}

\bibitem[{{Fischer} \& {Marcy}(1992)}]{Fischer&Marcy1992}
{Fischer}, D.~A., \& {Marcy}, G.~W. 1992, \apj, 396, 178,
  \dodoi{10.1086/171708}

\bibitem[{{Foreman-Mackey} {et~al.}(2017){Foreman-Mackey}, {Agol},
  {Ambikasaran}, \& {Angus}}]{Foreman-Mackeyetal2017}
{Foreman-Mackey}, D., {Agol}, E., {Ambikasaran}, S., \& {Angus}, R. 2017, \aj,
  154, 220, \dodoi{10.3847/1538-3881/aa9332}

\bibitem[{{Foreman-Mackey} {et~al.}(2013){Foreman-Mackey}, {Hogg}, {Lang}, \&
  {Goodman}}]{Foreman-Mackeyetal2013}
{Foreman-Mackey}, D., {Hogg}, D.~W., {Lang}, D., \& {Goodman}, J. 2013, \pasp,
  125, 306, \dodoi{10.1086/670067}

\bibitem[{{Fortney} {et~al.}(2011){Fortney}, {Ikoma}, {Nettelmann}, {Guillot},
  \& {Marley}}]{Fortneyetal2011}
{Fortney}, J.~J., {Ikoma}, M., {Nettelmann}, N., {Guillot}, T., \& {Marley},
  M.~S. 2011, \apj, 729, 32, \dodoi{10.1088/0004-637X/729/1/32}

\bibitem[{{Francic}(1989)}]{Francic1989}
{Francic}, S.~P. 1989, \aj, 98, 888, \dodoi{10.1086/115186}

\bibitem[{{Fulton} \& {Petigura}(2018)}]{Fulton&Petigura2018}
{Fulton}, B.~J., \& {Petigura}, E.~A. 2018, \aj, 156, 264,
  \dodoi{10.3847/1538-3881/aae828}

\bibitem[{{Fulton} {et~al.}(2017){Fulton}, {Petigura}, {Howard}, {Isaacson},
  {Marcy}, {Cargile}, {Hebb}, {Weiss}, {Johnson}, {Morton}, {Sinukoff},
  {Crossfield}, \& {Hirsch}}]{Fultonetal2017}
{Fulton}, B.~J., {Petigura}, E.~A., {Howard}, A.~W., {et~al.} 2017, \aj, 154,
  109, \dodoi{10.3847/1538-3881/aa80eb}

\bibitem[{{Gagn{\'e}} {et~al.}(2018){Gagn{\'e}}, {Mamajek}, {Malo}, {Riedel},
  {Rodriguez}, {Lafreni{\`e}re}, {Faherty}, {Roy-Loubier}, {Pueyo}, {Robin}, \&
  {Doyon}}]{Gagneetal2018}
{Gagn{\'e}}, J., {Mamajek}, E.~E., {Malo}, L., {et~al.} 2018, \apj, 856, 23,
  \dodoi{10.3847/1538-4357/aaae09}

\bibitem[{{Gaia Collaboration} {et~al.}(2016){Gaia Collaboration}, {Prusti},
  {de Bruijne}, {Brown}, {Vallenari}, {Babusiaux}, {Bailer-Jones}, {Bastian},
  {Biermann}, {Evans}, \& et~al.}]{GAIA2016}
{Gaia Collaboration}, {Prusti}, T., {de Bruijne}, J.~H.~J., {et~al.} 2016,
  \aap, 595, A1, \dodoi{10.1051/0004-6361/201629272}

\bibitem[{{Gaia Collaboration} {et~al.}(2018){Gaia Collaboration}, {Brown},
  {Vallenari}, {Prusti}, {de Bruijne}, {Babusiaux}, {Bailer-Jones}, {Biermann},
  {Evans}, {Eyer}, \& et~al.}]{GAIAdr2}
{Gaia Collaboration}, {Brown}, A.~G.~A., {Vallenari}, A., {et~al.} 2018, \aap,
  616, A1, \dodoi{10.1051/0004-6361/201833051}

\bibitem[{{Gaidos} {et~al.}(2014){Gaidos}, {Mann}, {L{\'e}pine}, {Buccino},
  {James}, {Ansdell}, {Petrucci}, {Mauas}, \& {Hilton}}]{Gaidos2014}
{Gaidos}, E., {Mann}, A.~W., {L{\'e}pine}, S., {et~al.} 2014, \mnras, 443, 2561

\bibitem[{{Gaudi} \& {Winn}(2007)}]{Gaudi&Winn2007}
{Gaudi}, B.~S., \& {Winn}, J.~N. 2007, \apj, 655, 550, \dodoi{10.1086/509910}

\bibitem[{{Gentile Fusillo} {et~al.}(2019){Gentile Fusillo}, {Tremblay},
  {G{\"a}nsicke}, {Manser}, {Cunningham}, {Cukanovaite}, {Hollands}, {Marsh},
  {Raddi}, {Jordan}, {Toonen}, {Geier}, {Barstow}, \&
  {Cummings}}]{GentileFusilloetal2019}
{Gentile Fusillo}, N.~P., {Tremblay}, P.-E., {G{\"a}nsicke}, B.~T., {et~al.}
  2019, \mnras, 482, 4570, \dodoi{10.1093/mnras/sty3016}

\bibitem[{{Ginzburg} {et~al.}(2018){Ginzburg}, {Schlichting}, \&
  {Sari}}]{Ginzburgetal2018}
{Ginzburg}, S., {Schlichting}, H.~E., \& {Sari}, R. 2018, \mnras, 476, 759,
  \dodoi{10.1093/mnras/sty290}

\bibitem[{{Goldreich} \& {Soter}(1966)}]{Goldreich&Soter1966}
{Goldreich}, P., \& {Soter}, S. 1966, \icarus, 5, 375,
  \dodoi{10.1016/0019-1035(66)90051-0}

\bibitem[{{Gossage} {et~al.}(2018){Gossage}, {Conroy}, {Dotter}, {Choi},
  {Rosenfield}, {Cargile}, \& {Dolphin}}]{Gossageetal2018}
{Gossage}, S., {Conroy}, C., {Dotter}, A., {et~al.} 2018, \apj, 863, 67,
  \dodoi{10.3847/1538-4357/aad0a0}

\bibitem[{{Gray}(2008)}]{gray_book}
{Gray}, D.~F. 2008, {The Observation and Analysis of Stellar Photospheres}

\bibitem[{{Grevesse} {et~al.}(2007){Grevesse}, {Asplund}, \&
  {Sauval}}]{grevesse_solar_abunds}
{Grevesse}, N., {Asplund}, M., \& {Sauval}, A.~J. 2007, \ssr, 130, 105,
  \dodoi{10.1007/s11214-007-9173-7}

\bibitem[{Guerrero(submitted)}]{Guerrero2020}
Guerrero. submitted

\bibitem[{{Gustafsson} {et~al.}(2008){Gustafsson}, {Edvardsson}, {Eriksson},
  {J{\o}rgensen}, {Nordlund}, \& {Plez}}]{MARCs}
{Gustafsson}, B., {Edvardsson}, B., {Eriksson}, K., {et~al.} 2008, \aap, 486,
  951, \dodoi{10.1051/0004-6361:200809724}

\bibitem[{{Hayashi}(1961)}]{Hayashi1961}
{Hayashi}, C. 1961, \pasj, 13, 450

\bibitem[{{Heiter} {et~al.}(2019){Heiter}, {Lind}, {Bergemann}, {Asplund},
  {Mikolaitis}, {Barklem}, {Masseron}, {de Laverny}, {Magrini}, {Edvardsson},
  {J{\"o}nsson}, {Pickering}, {Ryde}, {Bayo}, {Bensby}, {Casey}, {Feltzing},
  {Jofr{\'e}}, {Korn}, {Pancino}, {Damiani}, {Lanzafame}, {Lardo}, {Monaco},
  {Morbidelli}, {Smiljanic}, {Worley}, {Zaggia}, {Randich}, \& F.}]{Heiter2019}
{Heiter}, U., {Lind}, K., {Bergemann}, M., {et~al.} 2019, \aap

\bibitem[{{Henden} {et~al.}(2015){Henden}, {Levine}, {Terrell}, \&
  {Welch}}]{Hendenetal2015}
{Henden}, A.~A., {Levine}, S., {Terrell}, D., \& {Welch}, D.~L. 2015, in
  American Astronomical Society Meeting Abstracts, Vol. 225, American
  Astronomical Society Meeting Abstracts \#225, 336.16

\bibitem[{{Herczeg} \& {Hillenbrand}(2015)}]{Herczeg&Hillenbrand2015}
{Herczeg}, G.~J., \& {Hillenbrand}, L.~A. 2015, \apj, 808, 23,
  \dodoi{10.1088/0004-637X/808/1/23}

\bibitem[{{Hills}(1980)}]{Hills1980}
{Hills}, J.~G. 1980, \apj, 235, 986, \dodoi{10.1086/157703}

\bibitem[{{H{\o}g} {et~al.}(2000){H{\o}g}, {Fabricius}, {Makarov}, {Urban},
  {Corbin}, {Wycoff}, {Bastian}, {Schwekendiek}, \& {Wicenec}}]{Hog2000}
{H{\o}g}, E., {Fabricius}, C., {Makarov}, V.~V., {et~al.} 2000, \aap, 355, L27

\bibitem[{{Howard} {et~al.}(2012){Howard}, {Marcy}, {Bryson}, {Jenkins},
  {Rowe}, {Batalha}, {Borucki}, {Koch}, {Dunham}, {Gautier}, {Van Cleve},
  {Cochran}, {Latham}, {Lissauer}, {Torres}, {Brown}, {Gilliland}, {Buchhave},
  {Caldwell}, {Christensen-Dalsgaard}, {Ciardi}, {Fressin}, {Haas}, {Howell},
  {Kjeldsen}, {Seager}, {Rogers}, {Sasselov}, {Steffen}, {Basri},
  {Charbonneau}, {Christiansen}, {Clarke}, {Dupree}, {Fabrycky}, {Fischer},
  {Ford}, {Fortney}, {Tarter}, {Girouard}, {Holman}, {Johnson}, {Klaus},
  {Machalek}, {Moorhead}, {Morehead}, {Ragozzine}, {Tenenbaum}, {Twicken},
  {Quinn}, {Isaacson}, {Shporer}, {Lucas}, {Walkowicz}, {Welsh}, {Boss},
  {Devore}, {Gould}, {Smith}, {Morris}, {Prsa}, {Morton}, {Still}, {Thompson},
  {Mullally}, {Endl}, \& {MacQueen}}]{Howardetal2012}
{Howard}, A.~W., {Marcy}, G.~W., {Bryson}, S.~T., {et~al.} 2012, \apjs, 201,
  15, \dodoi{10.1088/0067-0049/201/2/15}

\bibitem[{{Howell} {et~al.}(2011){Howell}, {Everett}, {Sherry}, {Horch}, \&
  {Ciardi}}]{Howelletal2011}
{Howell}, S.~B., {Everett}, M.~E., {Sherry}, W., {Horch}, E., \& {Ciardi},
  D.~R. 2011, \aj, 142, 19, \dodoi{10.1088/0004-6256/142/1/19}

\bibitem[{{Huang} {et~al.}(1993){Huang}, {Lu}, \& {Halpern}}]{CN_data_93}
{Huang}, Y., {Lu}, R., \& {Halpern}, J.~B. 1993, \ao, 32, 981,
  \dodoi{10.1364/AO.32.000981}

\bibitem[{{Husser} {et~al.}(2013){Husser}, {Wende-von Berg}, {Dreizler},
  {Homeier}, {Reiners}, {Barman}, \& {Hauschildt}}]{Husseretal2013}
{Husser}, T.-O., {Wende-von Berg}, S., {Dreizler}, S., {et~al.} 2013, \aap,
  553, A6, \dodoi{10.1051/0004-6361/201219058}

\bibitem[{{Ingalls} {et~al.}(2012){Ingalls}, {Krick}, {Carey}, {Laine},
  {Surace}, {Glaccum}, {Grillmair}, \& {Lowrance}}]{Ingallsetal2012}
{Ingalls}, J.~G., {Krick}, J.~E., {Carey}, S.~J., {et~al.} 2012, Society of
  Photo-Optical Instrumentation Engineers (SPIE) Conference Series, Vol. 8442,
  {Intra-pixel gain variations and high-precision photometry with the Infrared
  Array Camera (IRAC)}, 84421Y

\bibitem[{Ingalls {et~al.}(2012)Ingalls, Krick, Carey, Laine, Surace, Glaccum,
  Grillmair, \& Lowrance}]{IngallsIntrapixel2012}
Ingalls, J.~G., Krick, J.~E., Carey, S.~J., {et~al.} 2012, Space Telescopes and
  Instrumentation 2012: Optical, Infrared, and Millimeter Wave, 8442, 84421Y,
  \dodoi{10.1117/12.926947}

\bibitem[{{Ingalls} {et~al.}(2016){Ingalls}, {Krick}, {Carey}, {Stauffer},
  {Lowrance}, {Grillmair}, {Buzasi}, {Deming}, {Diamond-Lowe}, {Evans},
  {Morello}, {Stevenson}, {Wong}, {Capak}, {Glaccum}, {Laine}, {Surace}, \&
  {Storrie-Lombardi}}]{Ingallsetal2016}
{Ingalls}, J.~G., {Krick}, J.~E., {Carey}, S.~J., {et~al.} 2016, \aj, 152, 44,
  \dodoi{10.3847/0004-6256/152/2/44}

\bibitem[{{Jenkins}(2002)}]{Jenkins2002}
{Jenkins}, J.~M. 2002, \apj, 575, 493, \dodoi{10.1086/341136}

\bibitem[{{Jenkins}(2015)}]{Jenkins2015}
{Jenkins}, J.~M. 2015, in AAS/Division for Extreme Solar Systems Abstracts,
  Vol.~47, AAS/Division for Extreme Solar Systems Abstracts, 106.05

\bibitem[{{Jenkins} {et~al.}(2010){Jenkins}, {Chandrasekaran}, {McCauliff},
  {Caldwell}, {Tenenbaum}, {Li}, {Klaus}, {Cote}, \&
  {Middour}}]{Jenkinsetal2010}
{Jenkins}, J.~M., {Chandrasekaran}, H., {McCauliff}, S.~D., {et~al.} 2010, in
  Society of Photo-Optical Instrumentation Engineers (SPIE) Conference Series,
  Vol. 7740, Software and Cyberinfrastructure for Astronomy, ed. N.~M.
  {Radziwill} \& A.~{Bridger}, 77400D

\bibitem[{{Jenkins} {et~al.}(2016){Jenkins}, {Twicken}, {McCauliff},
  {Campbell}, {Sanderfer}, {Lung}, {Mansouri-Samani}, {Girouard}, {Tenenbaum},
  {Klaus}, {Smith}, {Caldwell}, {Chacon}, {Henze}, {Heiges}, {Latham},
  {Morgan}, {Swade}, {Rinehart}, \& {Vanderspek}}]{Jenkinsetal2016}
{Jenkins}, J.~M., {Twicken}, J.~D., {McCauliff}, S., {et~al.} 2016, Society of
  Photo-Optical Instrumentation Engineers (SPIE) Conference Series, Vol. 9913,
  {The TESS science processing operations center}, 99133E

\bibitem[{{Johnson} \& {Soderblom}(1987)}]{Johnson&Soderblom1987}
{Johnson}, D. R.~H., \& {Soderblom}, D.~R. 1987, \aj, 93, 864,
  \dodoi{10.1086/114370}

\bibitem[{{Johnson} {et~al.}(2018){Johnson}, {Dai}, {Justesen}, {Gandolfi},
  {Hatzes}, {Nowak}, {Endl}, {Cochran}, {Hidalgo}, {Watanabe}, {Parviainen},
  {Hirano}, {Villanueva}, {Prieto-Arranz}, {Narita}, {Palle}, {Guenther},
  {Barrag{\'a}n}, {Trifonov}, {Niraula}, {MacQueen}, {Cabrera}, {Csizmadia},
  {Eigm{\"u}ller}, {Grziwa}, {Korth}, {P{\"a}tzold}, {Smith}, {Albrecht},
  {Alonso}, {Deeg}, {Erikson}, {Esposito}, {Fridlund}, {Fukui}, {Kusakabe},
  {Kuzuhara}, {Livingston}, {Monta{\~n}es Rodriguez}, {Nespral}, {Persson},
  {Purismo}, {Raimundo}, {Rauer}, {Ribas}, {Tamura}, {Van Eylen}, \&
  {Winn}}]{Johnsonetal2018}
{Johnson}, M.~C., {Dai}, F., {Justesen}, A.~B., {et~al.} 2018, \mnras, 481,
  596, \dodoi{10.1093/mnras/sty2238}

\bibitem[{{Kastner} {et~al.}(1997){Kastner}, {Zuckerman}, {Weintraub}, \&
  {Forveille}}]{Kastneretal1997}
{Kastner}, J.~H., {Zuckerman}, B., {Weintraub}, D.~A., \& {Forveille}, T. 1997,
  Science, 277, 67, \dodoi{10.1126/science.277.5322.67}

\bibitem[{{Kennedy} \& {Kenyon}(2008)}]{Kennedy&Kenyon2008}
{Kennedy}, G.~M., \& {Kenyon}, S.~J. 2008, \apj, 682, 1264,
  \dodoi{10.1086/589436}

\bibitem[{{Kipping}(2013)}]{Kipping2013}
{Kipping}, D.~M. 2013, \mnras, 435, 2152, \dodoi{10.1093/mnras/stt1435}

\bibitem[{{Kley} \& {Nelson}(2012)}]{Kley&Nelson2012}
{Kley}, W., \& {Nelson}, R.~P. 2012, \araa, 50, 211,
  \dodoi{10.1146/annurev-astro-081811-125523}

\bibitem[{{Kounkel} {et~al.}(2020){Kounkel}, {Covey}, \&
  {Stassun}}]{Kounkeletal2020}
{Kounkel}, M., {Covey}, K., \& {Stassun}, K.~G. 2020, \aj, 160, 279,
  \dodoi{10.3847/1538-3881/abc0e6}

\bibitem[{{Kov{\'a}cs} {et~al.}(2002){Kov{\'a}cs}, {Zucker}, \&
  {Mazeh}}]{Kovacs2002}
{Kov{\'a}cs}, G., {Zucker}, S., \& {Mazeh}, T. 2002, \aap, 391, 369,
  \dodoi{10.1051/0004-6361:20020802}

\bibitem[{{Kraus} \& {Hillenbrand}(2007)}]{Kraus&Hillenbrand2007}
{Kraus}, A.~L., \& {Hillenbrand}, L.~A. 2007, \aj, 134, 2340,
  \dodoi{10.1086/522831}

\bibitem[{{Kraus} {et~al.}(2014){Kraus}, {Shkolnik}, {Allers}, \&
  {Liu}}]{Krausetal2014}
{Kraus}, A.~L., {Shkolnik}, E.~L., {Allers}, K.~N., \& {Liu}, M.~C. 2014, \aj,
  147, 146, \dodoi{10.1088/0004-6256/147/6/146}

\bibitem[{{Kreidberg}(2015)}]{Kreidberg2015}
{Kreidberg}, L. 2015, \pasp, 127, 1161, \dodoi{10.1086/683602}

\bibitem[{{Kumar} {et~al.}(1998){Kumar}, {Hsiao}, {Hung}, \&
  {Lee}}]{CH_data_98}
{Kumar}, A., {Hsiao}, C.-C., {Hung}, W.-C., \& {Lee}, Y.-P. 1998, \jcp, 109,
  3824, \dodoi{10.1063/1.476982}

\bibitem[{{Lada} \& {Lada}(2003)}]{Lada&Lada2003}
{Lada}, C.~J., \& {Lada}, E.~A. 2003, \araa, 41, 57,
  \dodoi{10.1146/annurev.astro.41.011802.094844}

\bibitem[{{Lammer} {et~al.}(2003){Lammer}, {Selsis}, {Ribas}, {Guinan},
  {Bauer}, \& {Weiss}}]{Lammeretal2003}
{Lammer}, H., {Selsis}, F., {Ribas}, I., {et~al.} 2003, \apjl, 598, L121,
  \dodoi{10.1086/380815}

\bibitem[{{Lanza} {et~al.}(2014){Lanza}, {Das Chagas}, \& {De
  Medeiros}}]{Lanzaetal2014}
{Lanza}, A.~F., {Das Chagas}, M.~L., \& {De Medeiros}, J.~R. 2014, \aap, 564,
  A50, \dodoi{10.1051/0004-6361/201323172}

\bibitem[{{Lasker} {et~al.}(2008){Lasker}, {Lattanzi}, {McLean}, {Bucciarelli},
  {Drimmel}, {Garcia}, {Greene}, {Guglielmetti}, {Hanley}, {Hawkins},
  {Laidler}, {Loomis}, {Meakes}, {Mignani}, {Morbidelli}, {Morrison},
  {Pannunzio}, {Rosenberg}, {Sarasso}, {Smart}, {Spagna}, {Sturch},
  {Volpicelli}, {White}, {Wolfe}, \& {Zacchei}}]{Laskeretal2008}
{Lasker}, B.~M., {Lattanzi}, M.~G., {McLean}, B.~J., {et~al.} 2008, \aj, 136,
  735, \dodoi{10.1088/0004-6256/136/2/735}

\bibitem[{{Lightkurve Collaboration} {et~al.}(2018){Lightkurve Collaboration},
  {Cardoso}, {Hedges}, {Gully-Santiago}, {Saunders}, {Cody}, {Barclay}, {Hall},
  {Sagear}, {Turtelboom}, {Zhang}, {Tzanidakis}, {Mighell}, {Coughlin}, {Bell},
  {Berta-Thompson}, {Williams}, {Dotson}, \& {Barentsen}}]{lightkurve2018}
{Lightkurve Collaboration}, {Cardoso}, J.~V.~d.~M., {Hedges}, C., {et~al.}
  2018, {Lightkurve: Kepler and TESS time series analysis in Python},
  Astrophysics Source Code Library.
\newblock \doeprint{1812.013}

\bibitem[{{Lindegren} {et~al.}(2018){Lindegren}, {Hern{\'a}ndez}, {Bombrun},
  {Klioner}, {Bastian}, {Ramos-Lerate}, {de Torres}, {Steidelm{\"u}ller},
  {Stephenson}, {Hobbs}, {Lammers}, {Biermann}, {Geyer}, {Hilger}, {Michalik},
  {Stampa}, {McMillan}, {Casta{\~n}eda}, {Clotet}, {Comoretto}, {Davidson},
  {Fabricius}, {Gracia}, {Hambly}, {Hutton}, {Mora}, {Portell}, {van Leeuwen},
  {Abbas}, {Abreu}, {Altmann}, {Andrei}, {Anglada}, {Balaguer-N{\'u}{\~n}ez},
  {Barache}, {Becciani}, {Bertone}, {Bianchi}, {Bouquillon}, {Bourda},
  {Br{\"u}semeister}, {Bucciarelli}, {Busonero}, {Buzzi}, {Cancelliere},
  {Carlucci}, {Charlot}, {Cheek}, {Crosta}, {Crowley}, {de Bruijne}, {de
  Felice}, {Drimmel}, {Esquej}, {Fienga}, {Fraile}, {Gai}, {Garralda},
  {Gonz{\'a}lez-Vidal}, {Guerra}, {Hauser}, {Hofmann}, {Holl}, {Jordan},
  {Lattanzi}, {Lenhardt}, {Liao}, {Licata}, {Lister}, {L{\"o}ffler},
  {Marchant}, {Martin-Fleitas}, {Messineo}, {Mignard}, {Morbidelli}, {Poggio},
  {Riva}, {Rowell}, {Salguero}, {Sarasso}, {Sciacca}, {Siddiqui}, {Smart},
  {Spagna}, {Steele}, {Taris}, {Torra}, {van Elteren}, {van Reeven}, \&
  {Vecchiato}}]{Lindegrenetal2018}
{Lindegren}, L., {Hern{\'a}ndez}, J., {Bombrun}, A., {et~al.} 2018, \aap, 616,
  A2, \dodoi{10.1051/0004-6361/201832727}

\bibitem[{{Livingston} {et~al.}(2018){Livingston}, {Dai}, {Hirano}, {Gand
  olfi}, {Nowak}, {Endl}, {Velasco}, {Fukui}, {Narita}, {Prieto-Arranz},
  {Barragan}, {Cusano}, {Albrecht}, {Cabrera}, {Cochran}, {Csizmadia}, {Deeg},
  {Eigm{\"u}ller}, {Erikson}, {Fridlund}, {Grziwa}, {Guenther}, {Hatzes},
  {Kawauchi}, {Korth}, {Nespral}, {Palle}, {P{\"a}tzold}, {Persson}, {Rauer},
  {Smith}, {Tamura}, {Tanaka}, {Van Eylen}, {Watanabe}, \&
  {Winn}}]{Livingstonetal2018}
{Livingston}, J.~H., {Dai}, F., {Hirano}, T., {et~al.} 2018, \aj, 155, 115,
  \dodoi{10.3847/1538-3881/aaa841}

\bibitem[{{Livingston} {et~al.}(2019){Livingston}, {Dai}, {Hirano}, {Gand
  olfi}, {Trani}, {Nowak}, {Cochran}, {Endl}, {Albrecht}, {Barragan},
  {Cabrera}, {Csizmadia}, {de Leon}, {Deeg}, {Eigm{\"u}ller}, {Erikson},
  {Fridlund}, {Fukui}, {Grziwa}, {Guenther}, {Hatzes}, {Korth}, {Kuzuhara},
  {Monta{\~n}es}, {Narita}, {Nespral}, {Palle}, {P{\"a}tzold}, {Persson},
  {Prieto-Arranz}, {Rauer}, {Tamura}, {Van Eylen}, \&
  {Winn}}]{Livingstonetal2019}
---. 2019, \mnras, 484, 8, \dodoi{10.1093/mnras/sty3464}

\bibitem[{{Lopez} \& {Fortney}(2013)}]{Lopez&Fortney2013}
{Lopez}, E.~D., \& {Fortney}, J.~J. 2013, \apj, 776, 2,
  \dodoi{10.1088/0004-637X/776/1/2}

\bibitem[{{Lubow} \& {Ida}(2010)}]{Lubow&Ida2010}
{Lubow}, S.~H., \& {Ida}, S. 2010, {Planet Migration}, ed. S.~{Seager},
  347--371

\bibitem[{{Malo} {et~al.}(2013){Malo}, {Doyon}, {Lafreni{\`e}re}, {Artigau},
  {Gagn{\'e}}, {Baron}, \& {Riedel}}]{Maloetal2013}
{Malo}, L., {Doyon}, R., {Lafreni{\`e}re}, D., {et~al.} 2013, \apj, 762, 88,
  \dodoi{10.1088/0004-637X/762/2/88}

\bibitem[{{Mamajek} \& {Hillenbrand}(2008)}]{Mamajek&Hillenbrand2008}
{Mamajek}, E.~E., \& {Hillenbrand}, L.~A. 2008, \apj, 687, 1264,
  \dodoi{10.1086/591785}

\bibitem[{{Mandel} \& {Agol}(2002)}]{Mandel&Agol2002}
{Mandel}, K., \& {Agol}, E. 2002, \apjl, 580, L171, \dodoi{10.1086/345520}

\bibitem[{{Mann} {et~al.}(2015){Mann}, {Feiden}, {Gaidos}, {Boyajian}, \& {von
  Braun}}]{Mannetal2015a}
{Mann}, A.~W., {Feiden}, G.~A., {Gaidos}, E., {Boyajian}, T., \& {von Braun},
  K. 2015, \apj, 804, 64, \dodoi{10.1088/0004-637X/804/1/64}

\bibitem[{{Mann} {et~al.}(2016{\natexlab{a}}){Mann}, {Gaidos}, {Mace},
  {Johnson}, {Bowler}, {LaCourse}, {Jacobs}, {Vanderburg}, {Kraus}, {Kaplan},
  \& {Jaffe}}]{Mannetal2016a}
{Mann}, A.~W., {Gaidos}, E., {Mace}, G.~N., {et~al.} 2016{\natexlab{a}}, \apj,
  818, 46, \dodoi{10.3847/0004-637X/818/1/46}

\bibitem[{{Mann} {et~al.}(2016{\natexlab{b}}){Mann}, {Newton}, {Rizzuto},
  {Irwin}, {Feiden}, {Gaidos}, {Mace}, {Kraus}, {James}, {Ansdell},
  {Charbonneau}, {Covey}, {Ireland}, {Jaffe}, {Johnson}, {Kidder}, \&
  {Vanderburg}}]{Mannetal2016b}
{Mann}, A.~W., {Newton}, E.~R., {Rizzuto}, A.~C., {et~al.} 2016{\natexlab{b}},
  \aj, 152, 61, \dodoi{10.3847/0004-6256/152/3/61}

\bibitem[{{Mann} {et~al.}(2017){Mann}, {Gaidos}, {Vanderburg}, {Rizzuto},
  {Ansdell}, {Medina}, {Mace}, {Kraus}, \& {Sokal}}]{Mannetal2017}
{Mann}, A.~W., {Gaidos}, E., {Vanderburg}, A., {et~al.} 2017, \aj, 153, 64,
  \dodoi{10.1088/1361-6528/aa5276}

\bibitem[{{Mann} {et~al.}(2018){Mann}, {Vanderburg}, {Rizzuto}, {Kraus},
  {Berlind}, {Bieryla}, {Calkins}, {Esquerdo}, {Latham}, {Mace}, {Morris},
  {Quinn}, {Sokal}, \& {Stefanik}}]{Mannetal2018}
{Mann}, A.~W., {Vanderburg}, A., {Rizzuto}, A.~C., {et~al.} 2018, \aj, 155, 4,
  \dodoi{10.3847/1538-3881/aa9791}

\bibitem[{{Mann} {et~al.}(2020){Mann}, {Johnson}, {Vanderburg}, {Kraus},
  {Rizzuto}, {Wood}, {Bush}, {Rockcliffe}, {Newton}, {Latham}, {Mamajek},
  {Zhou}, {Quinn}, {Thao}, {Benatti}, {Cosentino}, {Desidera}, {Harutyunyan},
  {Lovis}, {Mortier}, {Pepe}, {Poretti}, {Wilson}, {Kristiansen}, {Gagliano},
  {Jacobs}, {LaCourse}, {Omohundro}, {Schwengeler}, {Terentev}, {Kane}, {Hill},
  {Rabus}, {Esquerdo}, {Berlind}, {Collins}, {Murawski}, {Sallam}, {Aitken},
  {Massey}, {Ricker}, {Vanderspek}, {Seager}, {Winn}, {Jenkins}, {Barclay},
  {Caldwell}, {Dragomir}, {Doty}, {Glidden}, {Tenenbaum}, {Torres}, {Twicken},
  \& {Villanueva}}]{Mannetal2020}
{Mann}, A.~W., {Johnson}, M.~C., {Vanderburg}, A., {et~al.} 2020, \aj, 160,
  179, \dodoi{10.3847/1538-3881/abae64}

\bibitem[{{Masseron} {et~al.}(2014){Masseron}, {Plez}, {Van Eck}, {Colin},
  {Daoutidis}, {Godefroid}, {Coheur}, {Bernath}, {Jorissen}, \&
  {Christlieb}}]{CH_data_14}
{Masseron}, T., {Plez}, B., {Van Eck}, S., {et~al.} 2014, \aap, 571, A47,
  \dodoi{10.1051/0004-6361/201423956}

\bibitem[{{Masuda} \& {Winn}(2020)}]{Masuda&Winn2020}
{Masuda}, K., \& {Winn}, J.~N. 2020, \aj, 159, 81,
  \dodoi{10.3847/1538-3881/ab65be}

\bibitem[{{Matson} {et~al.}(2019){Matson}, {Howell}, \&
  {Ciardi}}]{Matsonetal2019}
{Matson}, R.~A., {Howell}, S.~B., \& {Ciardi}, D.~R. 2019, \aj, 157, 211,
  \dodoi{10.3847/1538-3881/ab1755}

\bibitem[{{McCully} {et~al.}(2018){McCully}, {Volgenau}, {Harbeck}, {Lister},
  {Saunders}, {Turner}, {Siiverd}, \& {Bowman}}]{McCullyetal2018}
{McCully}, C., {Volgenau}, N.~H., {Harbeck}, D.-R., {et~al.} 2018, in Society
  of Photo-Optical Instrumentation Engineers (SPIE) Conference Series, Vol.
  10707, \procspie, 107070K

\bibitem[{{McQuillan} {et~al.}(2013){McQuillan}, {Aigrain}, \&
  {Mazeh}}]{McQuillanetal2013}
{McQuillan}, A., {Aigrain}, S., \& {Mazeh}, T. 2013, \mnras, 432, 1203,
  \dodoi{10.1093/mnras/stt536}

\bibitem[{{Meibom} {et~al.}(2011){Meibom}, {Mathieu}, {Stassun}, {Liebesny}, \&
  {Saar}}]{Meibometal2011}
{Meibom}, S., {Mathieu}, R.~D., {Stassun}, K.~G., {Liebesny}, P., \& {Saar},
  S.~H. 2011, \apj, 733, 115, \dodoi{10.1088/0004-637X/733/2/115}

\bibitem[{{Meingast} {et~al.}(2019){Meingast}, {Alves}, \&
  {F{\"u}rnkranz}}]{Meingastetal2019}
{Meingast}, S., {Alves}, J., \& {F{\"u}rnkranz}, V. 2019, \aap, 622, L13,
  \dodoi{10.1051/0004-6361/201834950}

\bibitem[{{Murphy} \& {Lawson}(2015)}]{Murphyetal2015}
{Murphy}, S.~J., \& {Lawson}, W.~A. 2015, \mnras, 447, 1267,
  \dodoi{10.1093/mnras/stu2450}

\bibitem[{{Murphy} {et~al.}(2018){Murphy}, {Mamajek}, \&
  {Bell}}]{Murphyetal2018}
{Murphy}, S.~J., {Mamajek}, E.~E., \& {Bell}, C. P.~M. 2018, \mnras, 476, 3290,
  \dodoi{10.1093/mnras/sty471}

\bibitem[{{Nardiello} {et~al.}(2019){Nardiello}, {Borsato}, {Piotto},
  {Colombo}, {Manthopoulou}, {Bedin}, {Granata}, {Lacedelli}, {Libralato},
  {Malavolta}, {Montalto}, \& {Nascimbeni}}]{Nardielloetal2019}
{Nardiello}, D., {Borsato}, L., {Piotto}, G., {et~al.} 2019, \mnras, 490, 3806,
  \dodoi{10.1093/mnras/stz2878}

\bibitem[{{Nardiello} {et~al.}(2020){Nardiello}, {Piotto}, {Deleuil},
  {Malavolta}, {Montalto}, {Bedin}, {Borsato}, {Granata}, {Libralato}, \&
  {Manthopoulou}}]{Nardielloetal2020}
{Nardiello}, D., {Piotto}, G., {Deleuil}, M., {et~al.} 2020, \mnras, 495, 4924,
  \dodoi{10.1093/mnras/staa1465}

\bibitem[{{Newton} {et~al.}(2019){Newton}, {Mann}, {Tofflemire}, {Pearce},
  {Rizzuto}, {Vanderburg}, {Martinez}, {Wang}, {Ruffio}, {Kraus}, {Johnson},
  {Thao}, {Wood}, {Rampalli}, {Nielsen}, {Collins}, {Dragomir}, {Hellier},
  {Anderson}, {Barclay}, {Brown}, {Feiden}, {Hart}, {Isopi}, {Kielkopf},
  {Mallia}, {Nelson}, {Rodriguez}, {Stockdale}, {Waite}, {Wright}, {Lissauer},
  {Ricker}, {Vanderspek}, {Latham}, {Seager}, {Winn}, {Jenkins}, {Bouma},
  {Burke}, {Davies}, {Fausnaugh}, {Li}, {Morris}, {Mukai}, {Villase{\~n}or},
  {Villeneuva}, {De Rosa}, {Macintosh}, {Mengel}, {Okumura}, \&
  {Wittenmyer}}]{Newtonetal2019}
{Newton}, E.~R., {Mann}, A.~W., {Tofflemire}, B.~M., {et~al.} 2019, \apjl, 880,
  L17, \dodoi{10.3847/2041-8213/ab2988}

\bibitem[{{Newton} {et~al.}(2021){Newton}, {Mann}, {Kraus}, {Livingston},
  {Vanderburg}, {Curtis}, {Thao}, {Hawkins}, {Wood}, {Rizzuto}, {Soubkiou},
  {Tofflemire}, {Zhou}, {Crossfield}, {Pearce}, {Collins}, {Conti}, {Tan},
  {Villeneuva}, {Spencer}, {Dragomir}, {Quinn}, {Jensen}, {Collins},
  {Stockdale}, {Cloutier}, {Hellier}, {Benkhaldoun}, {Ziegler}, {Brice{\~n}o},
  {Law}, {Benneke}, {Christiansen}, {Gorjian}, {Kane}, {Kreidberg}, {Morales},
  {Werner}, {Twicken}, {Levine}, {Ciardi}, {Guerrero}, {Hesse}, {Quintana},
  {Shiao}, {Smith}, {Torres}, {Ricker}, {Vanderspek}, {Seager}, {Winn},
  {Jenkins}, \& {Latham}}]{Newtonetal2021}
{Newton}, E.~R., {Mann}, A.~W., {Kraus}, A.~L., {et~al.} 2021, \aj, 161, 65,
  \dodoi{10.3847/1538-3881/abccc6}

\bibitem[{{Obermeier} {et~al.}(2016){Obermeier}, {Henning}, {Schlieder},
  {Crossfield}, {Petigura}, {Howard}, {Sinukoff}, {Isaacson}, {Ciardi},
  {David}, {Hillenbrand}, {Beichman}, {Howell}, {Horch}, {Everett}, {Hirsch},
  {Teske}, {Christiansen}, {L{\'e}pine}, {Aller}, {Liu}, {Saglia},
  {Livingston}, \& {Kluge}}]{Obermeieretal2016}
{Obermeier}, C., {Henning}, T., {Schlieder}, J.~E., {et~al.} 2016, \aj, 152,
  223, \dodoi{10.3847/1538-3881/152/6/223}

\bibitem[{{Owen} \& {Jackson}(2012)}]{Owen&Jackson2012}
{Owen}, J.~E., \& {Jackson}, A.~P. 2012, \mnras, 425, 2931,
  \dodoi{10.1111/j.1365-2966.2012.21481.x}

\bibitem[{{Owen} \& {Wu}(2017)}]{Owen&Wu2017}
{Owen}, J.~E., \& {Wu}, Y. 2017, \apj, 847, 29,
  \dodoi{10.3847/1538-4357/aa890a}

\bibitem[{{Pearce} {et~al.}(2019){Pearce}, {Kraus}, {Dupuy}, {Ireland },
  {Rizzuto}, {Bowler}, {Birchall}, \& {Wallace}}]{Pearceetal2019}
{Pearce}, L.~A., {Kraus}, A.~L., {Dupuy}, T.~J., {et~al.} 2019, \aj, 157, 71,
  \dodoi{10.3847/1538-3881/aafacb}

\bibitem[{{Pearce} {et~al.}(2020){Pearce}, {Kraus}, {Dupuy}, {Mann}, {Newton},
  {Tofflemire}, \& {Vanderburg}}]{Pearceetal2020}
---. 2020, \apj, 894, 115, \dodoi{10.3847/1538-4357/ab8389}

\bibitem[{{Pecaut} \& {Mamajek}(2013)}]{Pecaut&Mamajek2013}
{Pecaut}, M.~J., \& {Mamajek}, E.~E. 2013, \apjs, 208, 9,
  \dodoi{10.1088/0067-0049/208/1/9}

\bibitem[{{Pepper} {et~al.}(2017){Pepper}, {Gillen}, {Parviainen}, {Hillenbrand
  }, {Cody}, {Aigrain}, {Stauffer}, {Vrba}, {David}, {Lillo-Box}, {Stassun},
  {Conroy}, {Pope}, \& {Barrado}}]{Pepperetal2017}
{Pepper}, J., {Gillen}, E., {Parviainen}, H., {et~al.} 2017, \aj, 153, 177,
  \dodoi{10.3847/1538-3881/aa62ab}

\bibitem[{{Perryman} {et~al.}(1998){Perryman}, {Brown}, {Lebreton}, {Gomez},
  {Turon}, {Cayrel de Strobel}, {Mermilliod}, {Robichon}, {Kovalevsky}, \&
  {Crifo}}]{Perrymanetal1998}
{Perryman}, M.~A.~C., {Brown}, A.~G.~A., {Lebreton}, Y., {et~al.} 1998, \aap,
  331, 81.
\newblock \doarXiv{astro-ph/9707253}

\bibitem[{{Plavchan} {et~al.}(2020){Plavchan}, {Barclay}, {Gagn{\'e}}, {Gao},
  {Cale}, {Matzko}, {Dragomir}, {Quinn}, {Feliz}, {Stassun}, {Crossfield},
  {Berardo}, {Latham}, {Tieu}, {Anglada-Escud{\'e}}, {Ricker}, {Vanderspek},
  {Seager}, {Winn}, {Jenkins}, {Rinehart}, {Krishnamurthy}, {Dynes}, {Doty},
  {Adams}, {Afanasev}, {Beichman}, {Bottom}, {Bowler}, {Brinkworth}, {Brown},
  {Cancino}, {Ciardi}, {Clampin}, {Clark}, {Collins}, {Davison},
  {Foreman-Mackey}, {Furlan}, {Gaidos}, {Geneser}, {Giddens}, {Gilbert},
  {Hall}, {Hellier}, {Henry}, {Horner}, {Howard}, {Huang}, {Huber}, {Kane},
  {Kenworthy}, {Kielkopf}, {Kipping}, {Klenke}, {Kruse}, {Latouf}, {Lowrance},
  {Mennesson}, {Mengel}, {Mills}, {Morton}, {Narita}, {Newton}, {Nishimoto},
  {Okumura}, {Palle}, {Pepper}, {Quintana}, {Roberge}, {Roccatagliata},
  {Schlieder}, {Tanner}, {Teske}, {Tinney}, {Vanderburg}, {von Braun}, {Walp},
  {Wang}, {Wang}, {Weigand }, {White}, {Wittenmyer}, {Wright}, {Youngblood},
  {Zhang}, \& {Zilberman}}]{Plavchanetal2020}
{Plavchan}, P., {Barclay}, T., {Gagn{\'e}}, J., {et~al.} 2020, \nat, 582, 497,
  \dodoi{10.1038/s41586-020-2400-z}

\bibitem[{{Plez}(2012)}]{turbospectrum}
{Plez}, B. 2012, {Turbospectrum: Code for spectral synthesis}.
\newblock \doeprint{1205.004}

\bibitem[{{Price-Whelan} {et~al.}(2018){Price-Whelan}, {Sip{\H{o}}cz},
  {G{\"u}nther}, {Lim}, {Crawford}, {Conseil}, {Shupe}, {Craig}, {Dencheva},
  {Ginsburg}, {VanderPlas}, {Bradley}, {P{\'e}rez-Su{\'a}rez}, {de Val-Borro},
  {Paper Contributors}, {Aldcroft}, {Cruz}, {Robitaille}, {Tollerud},
  {Coordination Committee}, {Ardelean}, {Babej}, {Bach}, {Bachetti}, {Bakanov},
  {Bamford}, {Barentsen}, {Barmby}, {Baumbach}, {Berry}, {Biscani}, {Boquien},
  {Bostroem}, {Bouma}, {Brammer}, {Bray}, {Breytenbach}, {Buddelmeijer},
  {Burke}, {Calderone}, {Cano Rodr{\'\i}guez}, {Cara}, {Cardoso}, {Cheedella},
  {Copin}, {Corrales}, {Crichton}, {D{\textquoteright}Avella}, {Deil},
  {Depagne}, {Dietrich}, {Donath}, {Droettboom}, {Earl}, {Erben}, {Fabbro},
  {Ferreira}, {Finethy}, {Fox}, {Garrison}, {Gibbons}, {Goldstein}, {Gommers},
  {Greco}, {Greenfield}, {Groener}, {Grollier}, {Hagen}, {Hirst}, {Homeier},
  {Horton}, {Hosseinzadeh}, {Hu}, {Hunkeler}, {Ivezi{\'c}}, {Jain}, {Jenness},
  {Kanarek}, {Kendrew}, {Kern}, {Kerzendorf}, {Khvalko}, {King}, {Kirkby},
  {Kulkarni}, {Kumar}, {Lee}, {Lenz}, {Littlefair}, {Ma}, {Macleod},
  {Mastropietro}, {McCully}, {Montagnac}, {Morris}, {Mueller}, {Mumford},
  {Muna}, {Murphy}, {Nelson}, {Nguyen}, {Ninan}, {N{\"o}the}, {Ogaz}, {Oh},
  {Parejko}, {Parley}, {Pascual}, {Patil}, {Patil}, {Plunkett}, {Prochaska},
  {Rastogi}, {Reddy Janga}, {Sabater}, {Sakurikar}, {Seifert}, {Sherbert},
  {Sherwood-Taylor}, {Shih}, {Sick}, {Silbiger}, {Singanamalla}, {Singer},
  {Sladen}, {Sooley}, {Sornarajah}, {Streicher}, {Teuben}, {Thomas},
  {Tremblay}, {Turner}, {Terr{\'o}n}, {van Kerkwijk}, {de la Vega}, {Watkins},
  {Weaver}, {Whitmore}, {Woillez}, {Zabalza}, \& {Contributors}}]{astropy1}
{Price-Whelan}, A.~M., {Sip{\H{o}}cz}, B.~M., {G{\"u}nther}, H.~M., {et~al.}
  2018, \aj, 156, 123, \dodoi{10.3847/1538-3881/aabc4f}

\bibitem[{{Ram} {et~al.}(2014){Ram}, {Brooke}, {Bernath}, {Sneden}, \&
  {Lucatello}}]{12C13C_data}
{Ram}, R.~S., {Brooke}, J.~S.~A., {Bernath}, P.~F., {Sneden}, C., \&
  {Lucatello}, S. 2014, \apjs, 211, 5, \dodoi{10.1088/0067-0049/211/1/5}

\bibitem[{{Ram{\'\i}rez} {et~al.}(2012){Ram{\'\i}rez}, {Fish}, {Lambert}, \&
  {Allende Prieto}}]{Ramirezetal2012}
{Ram{\'\i}rez}, I., {Fish}, J.~R., {Lambert}, D.~L., \& {Allende Prieto}, C.
  2012, \apj, 756, 46, \dodoi{10.1088/0004-637X/756/1/46}

\bibitem[{{Rebull} {et~al.}(2016){Rebull}, {Stauffer}, {Bouvier}, {Cody},
  {Hillenbrand}, {Soderblom}, {Valenti}, {Barrado}, {Bouy}, {Ciardi},
  {Pinsonneault}, {Stassun}, {Micela}, {Aigrain}, {Vrba}, {Somers},
  {Christiansen}, {Gillen}, \& {Collier Cameron}}]{Rebulletal2016}
{Rebull}, L.~M., {Stauffer}, J.~R., {Bouvier}, J., {et~al.} 2016, \aj, 152,
  113, \dodoi{10.3847/0004-6256/152/5/113}

\bibitem[{{Reinhold} {et~al.}(2013){Reinhold}, {Reiners}, \&
  {Basri}}]{Reinholdetal2013}
{Reinhold}, T., {Reiners}, A., \& {Basri}, G. 2013, \aap, 560, A4,
  \dodoi{10.1051/0004-6361/201321970}

\bibitem[{{Ricker} {et~al.}(2015){Ricker}, {Winn}, {Vanderspek}, {Latham},
  {Bakos}, {Bean}, {Berta-Thompson}, {Brown}, {Buchhave}, {Butler}, {Butler},
  {Chaplin}, {Charbonneau}, {Christensen-Dalsgaard}, {Clampin}, {Deming},
  {Doty}, {De Lee}, {Dressing}, {Dunham}, {Endl}, {Fressin}, {Ge}, {Henning},
  {Holman}, {Howard}, {Ida}, {Jenkins}, {Jernigan}, {Johnson}, {Kaltenegger},
  {Kawai}, {Kjeldsen}, {Laughlin}, {Levine}, {Lin}, {Lissauer}, {MacQueen},
  {Marcy}, {McCullough}, {Morton}, {Narita}, {Paegert}, {Palle}, {Pepe},
  {Pepper}, {Quirrenbach}, {Rinehart}, {Sasselov}, {Sato}, {Seager},
  {Sozzetti}, {Stassun}, {Sullivan}, {Szentgyorgyi}, {Torres}, {Udry}, \&
  {Villasenor}}]{Rickeretal2015}
{Ricker}, G.~R., {Winn}, J.~N., {Vanderspek}, R., {et~al.} 2015, Journal of
  Astronomical Telescopes, Instruments, and Systems, 1, 014003,
  \dodoi{10.1117/1.JATIS.1.1.014003}

\bibitem[{{Rizzuto} {et~al.}(2017){Rizzuto}, {Mann}, {Vanderburg}, {Kraus}, \&
  {Covey}}]{Rizzutoetal2017}
{Rizzuto}, A.~C., {Mann}, A.~W., {Vanderburg}, A., {Kraus}, A.~L., \& {Covey},
  K.~R. 2017, \aj, 154, 224, \dodoi{10.3847/1538-3881/aa9070}

\bibitem[{{Rizzuto} {et~al.}(2018){Rizzuto}, {Vanderburg}, {Mann}, {Kraus},
  {Dressing}, {Ag{\"u}eros}, {Douglas}, \& {Krolikowski}}]{Rizzutoetal2018}
{Rizzuto}, A.~C., {Vanderburg}, A., {Mann}, A.~W., {et~al.} 2018, \aj, 156,
  195, \dodoi{10.3847/1538-3881/aadf37}

\bibitem[{{Rizzuto} {et~al.}(2020){Rizzuto}, {Newton}, {Mann}, {Tofflemire},
  {Vanderburg}, {Kraus}, {Wood}, {Quinn}, {Zhou}, {Thao}, {Law}, {Ziegler}, \&
  {Brice{\~n}o}}]{Rizzutoetal2020}
{Rizzuto}, A.~C., {Newton}, E.~R., {Mann}, A.~W., {et~al.} 2020, \aj, 160, 33,
  \dodoi{10.3847/1538-3881/ab94b7}

\bibitem[{{Rodriguez} {et~al.}(2011){Rodriguez}, {Bessell}, {Zuckerman}, \&
  {Kastner}}]{Rodriguezetal2011}
{Rodriguez}, D.~R., {Bessell}, M.~S., {Zuckerman}, B., \& {Kastner}, J.~H.
  2011, \apj, 727, 62, \dodoi{10.1088/0004-637X/727/2/62}

\bibitem[{{Rucinski}(1992)}]{Rucinski1992}
{Rucinski}, S.~M. 1992, \aj, 104, 1968, \dodoi{10.1086/116372}

\bibitem[{{Sanders}(1971)}]{Sanders1971}
{Sanders}, W.~L. 1971, \aap, 14, 226

\bibitem[{{Scargle}(1982)}]{Scargle1982}
{Scargle}, J.~D. 1982, \apj, 263, 835, \dodoi{10.1086/160554}

\bibitem[{{Shkolnik} {et~al.}(2009){Shkolnik}, {Liu}, \&
  {Reid}}]{Shkolniketal2009}
{Shkolnik}, E., {Liu}, M.~C., \& {Reid}, I.~N. 2009, \apj, 699, 649,
  \dodoi{10.1088/0004-637X/699/1/649}

\bibitem[{{Shkolnik} {et~al.}(2017){Shkolnik}, {Allers}, {Kraus}, {Liu}, \&
  {Flagg}}]{Shkolniketal2017}
{Shkolnik}, E.~L., {Allers}, K.~N., {Kraus}, A.~L., {Liu}, M.~C., \& {Flagg},
  L. 2017, \aj, 154, 69, \dodoi{10.3847/1538-3881/aa77fa}

\bibitem[{{Shkolnik} {et~al.}(2011){Shkolnik}, {Liu}, {Reid}, {Dupuy}, \&
  {Weinberger}}]{Shkolniketal2011}
{Shkolnik}, E.~L., {Liu}, M.~C., {Reid}, I.~N., {Dupuy}, T., \& {Weinberger},
  A.~J. 2011, \apj, 727, 6, \dodoi{10.1088/0004-637X/727/1/6}

\bibitem[{{Siverd} {et~al.}(2018){Siverd}, {Brown}, {Barnes}, {Bowman}, {De
  Vera}, {Foale}, {Harbeck}, {Henderson}, {Hygelund}, {Kirby}, {McCully},
  {Nation}, {Smith}, {Taylor}, \& {Tufts}}]{Siverdetal2018}
{Siverd}, R.~J., {Brown}, T.~M., {Barnes}, S., {et~al.} 2018, in Society of
  Photo-Optical Instrumentation Engineers (SPIE) Conference Series, Vol. 10702,
  \procspie, 107026C

\bibitem[{{Skrutskie} {et~al.}(2006){Skrutskie}, {Cutri}, {Stiening},
  {Weinberg}, {Schneider}, {Carpenter}, {Beichman}, {Capps}, {Chester},
  {Elias}, {Huchra}, {Liebert}, {Lonsdale}, {Monet}, {Price}, {Seitzer},
  {Jarrett}, {Kirkpatrick}, {Gizis}, {Howard}, {Evans}, {Fowler}, {Fullmer},
  {Hurt}, {Light}, {Kopan}, {Marsh}, {McCallon}, {Tam}, {Van Dyk}, \&
  {Wheelock}}]{Skrutskie2006}
{Skrutskie}, M.~F., {Cutri}, R.~M., {Stiening}, R., {et~al.} 2006, \aj, 131,
  1163

\bibitem[{{Skumanich}(1972)}]{Skumanich1972}
{Skumanich}, A. 1972, \apj, 171, 565, \dodoi{10.1086/151310}

\bibitem[{Smith {et~al.}(2012)Smith, Stumpe, Van~Cleve, Jenkins, Barclay,
  Fanelli, Girouard, Kolodziejczak, McCauliff, Morris, \&
  Twicken}]{SmithKepler2012}
Smith, J.~C., Stumpe, M.~C., Van~Cleve, J.~E., {et~al.} 2012, Publications of
  the Astronomical Society of the Pacific, 124, 1000, \dodoi{10.1086/667697}

\bibitem[{{Sneden} {et~al.}(2014){Sneden}, {Lucatello}, {Ram}, {Brooke}, \&
  {Bernath}}]{CN_data_14}
{Sneden}, C., {Lucatello}, S., {Ram}, R.~S., {Brooke}, J.~S.~A., \& {Bernath},
  P. 2014, \apjs, 214, 26, \dodoi{10.1088/0067-0049/214/2/26}

\bibitem[{{Spada} \& {Lanzafame}(2020)}]{Spada&Lanzafame2020}
{Spada}, F., \& {Lanzafame}, A.~C. 2020, \aap, 636, A76,
  \dodoi{10.1051/0004-6361/201936384}

\bibitem[{{Stanford-Moore} {et~al.}(2020){Stanford-Moore}, {Nielsen}, {De
  Rosa}, {Macintosh}, \& {Czekala}}]{Stanford-Mooreetal2020}
{Stanford-Moore}, S.~A., {Nielsen}, E.~L., {De Rosa}, R.~J., {Macintosh}, B.,
  \& {Czekala}, I. 2020, \apj, 898, 27, \dodoi{10.3847/1538-4357/ab9a35}

\bibitem[{{Stassun} {et~al.}(2018){Stassun}, {Oelkers}, {Pepper}, {Paegert},
  {De Lee}, {Torres}, {Latham}, {Charpinet}, {Dressing}, {Huber}, {Kane},
  {L{\'e}pine}, {Mann}, {Muirhead}, {Rojas-Ayala}, {Silvotti}, {Fleming},
  {Levine}, \& {Plavchan}}]{Stassunetal2018}
{Stassun}, K.~G., {Oelkers}, R.~J., {Pepper}, J., {et~al.} 2018, \aj, 156, 102,
  \dodoi{10.3847/1538-3881/aad050}

\bibitem[{{Stassun} {et~al.}(2019){Stassun}, {Oelkers}, {Paegert}, {Torres},
  {Pepper}, {De Lee}, {Collins}, {Latham}, {Muirhead}, {Chittidi},
  {Rojas-Ayala}, {Fleming}, {Rose}, {Tenenbaum}, {Ting}, {Kane}, {Barclay},
  {Bean}, {Brassuer}, {Charbonneau}, {Ge}, {Lissauer}, {Mann}, {McLean},
  {Mullally}, {Narita}, {Plavchan}, {Ricker}, {Sasselov}, {Seager}, {Sharma},
  {Shiao}, {Sozzetti}, {Stello}, {Vanderspek}, {Wallace}, \&
  {Winn}}]{Stassunetal2019}
{Stassun}, K.~G., {Oelkers}, R.~J., {Paegert}, M., {et~al.} 2019, \aj, 158,
  138, \dodoi{10.3847/1538-3881/ab3467}

\bibitem[{{Stauffer} {et~al.}(1998){Stauffer}, {Schultz}, \&
  {Kirkpatrick}}]{Staufferetal1998}
{Stauffer}, J.~R., {Schultz}, G., \& {Kirkpatrick}, J.~D. 1998, \apjl, 499,
  L199, \dodoi{10.1086/311379}

\bibitem[{{Stevenson} {et~al.}(2010){Stevenson}, {Harrington}, {Nymeyer},
  {Madhusudhan}, {Seager}, {Bowman}, {Hardy}, {Deming}, {Rauscher}, \&
  {Lust}}]{stevenson2010}
{Stevenson}, K.~B., {Harrington}, J., {Nymeyer}, S., {et~al.} 2010, \nat, 464,
  1161, \dodoi{10.1038/nature09013}

\bibitem[{Stevenson {et~al.}(2012)Stevenson, Harrington, Fortney, Loredo,
  Hardy, Nymeyer, Bowman, Cubillos, Bowman, \& Hardin}]{StevensonTransit2012}
Stevenson, K.~B., Harrington, J., Fortney, J.~J., {et~al.} 2012, Astrophysical
  Journal, 754, 136, \dodoi{10.1088/0004-637X/754/2/136}

\bibitem[{{Stumpe} {et~al.}(2014){Stumpe}, {Smith}, {Catanzarite}, {Van Cleve},
  {Jenkins}, {Twicken}, \& {Girouard}}]{StumpeMultiscale2014}
{Stumpe}, M.~C., {Smith}, J.~C., {Catanzarite}, J.~H., {et~al.} 2014, \pasp,
  126, 100, \dodoi{10.1086/674989}

\bibitem[{Stumpe {et~al.}(2012)Stumpe, Smith, Van~Cleve, Twicken, Barclay,
  Fanelli, Girouard, Jenkins, Kolodziejczak, McCauliff, \&
  Morris}]{StumpeKepler2012}
Stumpe, M.~C., Smith, J.~C., Van~Cleve, J.~E., {et~al.} 2012, Publications of
  the Astronomical Society of the Pacific, 124, 985, \dodoi{10.1086/667698}

\bibitem[{{Takeda} {et~al.}(2013){Takeda}, {Honda}, {Ohnishi}, {Ohkubo},
  {Hirata}, \& {Sadakane}}]{Takedaetal2013}
{Takeda}, Y., {Honda}, S., {Ohnishi}, T., {et~al.} 2013, \pasj, 65, 53,
  \dodoi{10.1093/pasj/65.3.53}

\bibitem[{{Tang} {et~al.}(2019){Tang}, {Pang}, {Yuan}, {Chen}, {Hong},
  {Goldman}, {Just}, {Shukirgaliyev}, \& {Lin}}]{Tangetal2019}
{Tang}, S.-Y., {Pang}, X., {Yuan}, Z., {et~al.} 2019, \apj, 877, 12,
  \dodoi{10.3847/1538-4357/ab13b0}

\bibitem[{{Taylor}(2008)}]{Taylor2008}
{Taylor}, B.~J. 2008, \aj, 136, 1388, \dodoi{10.1088/0004-6256/136/3/1388}

\bibitem[{{Tian} {et~al.}(2020){Tian}, {El-Badry}, {Rix}, \&
  {Gould}}]{Tianetal2020}
{Tian}, H.-J., {El-Badry}, K., {Rix}, H.-W., \& {Gould}, A. 2020, \apjs, 246,
  4, \dodoi{10.3847/1538-4365/ab54c4}

\bibitem[{{Tofflemire} {et~al.}(2019){Tofflemire}, {Mathieu}, \&
  {Johns-Krull}}]{Tofflemireetal2019}
{Tofflemire}, B.~M., {Mathieu}, R.~D., \& {Johns-Krull}, C.~M. 2019, \aj, 158,
  245, \dodoi{10.3847/1538-3881/ab4f7d}

\bibitem[{{Tokovinin} {et~al.}(2013){Tokovinin}, {Fischer}, {Bonati},
  {Giguere}, {Moore}, {Schwab}, {Spronck}, \& {Szymkowiak}}]{Tokovininetal2013}
{Tokovinin}, A., {Fischer}, D.~A., {Bonati}, M., {et~al.} 2013, \pasp, 125,
  1336, \dodoi{10.1086/674012}

\bibitem[{{Torres} {et~al.}(2006){Torres}, {Quast}, {da Silva}, {de La Reza},
  {Melo}, \& {Sterzik}}]{TorresCetal2006}
{Torres}, C.~A.~O., {Quast}, G.~R., {da Silva}, L., {et~al.} 2006, \aap, 460,
  695, \dodoi{10.1051/0004-6361:20065602}

\bibitem[{{Torres} {et~al.}(2003){Torres}, {Quast}, {de La Reza}, {da Silva},
  {Melo}, \& {Sterzik}}]{TorresCetal2003}
{Torres}, C. A.~O., {Quast}, G.~R., {de La Reza}, R., {et~al.} 2003,
  Astrophysics and Space Science Library, Vol. 299, {SACY - a Search for
  Associations Containing Young stars}, ed. J.~{L{\'e}pine} \&
  J.~{Gregorio-Hetem}, 83

\bibitem[{{Torres} {et~al.}(2010){Torres}, {Andersen}, \&
  {Gim{\'e}nez}}]{Torresetal2010}
{Torres}, G., {Andersen}, J., \& {Gim{\'e}nez}, A. 2010, \aapr, 18, 67,
  \dodoi{10.1007/s00159-009-0025-1}

\bibitem[{{Tutukov}(1978)}]{Tutukov1978}
{Tutukov}, A.~V. 1978, \aap, 70, 57

\bibitem[{{Vanderburg} {et~al.}(2019){Vanderburg}, {Huang}, {Rodriguez},
  {Becker}, {Ricker}, {Vanderspek}, {Latham}, {Seager}, {Winn}, {Jenkins},
  {Addison}, {Bieryla}, {Brice{\~n}o}, {Bowler}, {Brown}, {Burke}, {Burt},
  {Caldwell}, {Clark}, {Crossfield}, {Dittmann}, {Dynes}, {Fulton}, {Guerrero},
  {Harbeck}, {Horner}, {Kane}, {Kielkopf}, {Kraus}, {Kreidberg}, {Law}, {Mann},
  {Mengel}, {Morton}, {Okumura}, {Pearce}, {Plavchan}, {Quinn}, {Rabus},
  {Rose}, {Rowden}, {Shporer}, {Siverd}, {Smith}, {Stassun}, {Tinney},
  {Wittenmyer}, {Wright}, {Zhang}, {Zhou}, \& {Ziegler}}]{Vanderburgetal2019}
{Vanderburg}, A., {Huang}, C.~X., {Rodriguez}, J.~E., {et~al.} 2019, \apjl,
  881, L19, \dodoi{10.3847/2041-8213/ab322d}

\bibitem[{{Virtanen} {et~al.}(2020){Virtanen}, {Gommers}, {Oliphant},
  {Haberland}, {Reddy}, {Cournapeau}, {Burovski}, {Peterson}, {Weckesser},
  {Bright}, {van der Walt}, {Brett}, {Wilson}, {Jarrod Millman}, {Mayorov},
  {Nelson}, {Jones}, {Kern}, {Larson}, {Carey}, {Polat}, {Feng}, {Moore}, {Vand
  erPlas}, {Laxalde}, {Perktold}, {Cimrman}, {Henriksen}, {Quintero}, {Harris},
  {Archibald}, {Ribeiro}, {Pedregosa}, {van Mulbregt}, \&
  {Contributors}}]{scipy2}
{Virtanen}, P., {Gommers}, R., {Oliphant}, T.~E., {et~al.} 2020, Nature
  Methods, 17, 261, \dodoi{https://doi.org/10.1038/s41592-019-0686-2}

\bibitem[{{Weber} \& {Davis}(1967)}]{Weber&Davis1967}
{Weber}, E.~J., \& {Davis}, Leverett, J. 1967, \apj, 148, 217,
  \dodoi{10.1086/149138}

\bibitem[{{Williams} \& {Cieza}(2011)}]{Williams&Cieza2011}
{Williams}, J.~P., \& {Cieza}, L.~A. 2011, \araa, 49, 67,
  \dodoi{10.1146/annurev-astro-081710-102548}

\bibitem[{{Wolf} {et~al.}(2018){Wolf}, {Onken}, {Luvaul}, {Schmidt}, {Bessell},
  {Chang}, {Da Costa}, {Mackey}, {Martin-Jones}, {Murphy}, {Preston}, {Scalzo},
  {Shao}, {Smillie}, {Tisserand}, {White}, \& {Yuan}}]{Wolfetal2018}
{Wolf}, C., {Onken}, C.~A., {Luvaul}, L.~C., {et~al.} 2018, \pasa, 35, e010,
  \dodoi{10.1017/pasa.2018.5}

\bibitem[{{Ziegler} {et~al.}(2018){Ziegler}, {Law}, {Baranec}, {Morton},
  {Riddle}, {De Lee}, {Huber}, {Mahadevan}, \& {Pepper}}]{Ziegleretal2018a}
{Ziegler}, C., {Law}, N.~M., {Baranec}, C., {et~al.} 2018, \aj, 156, 259,
  \dodoi{10.3847/1538-3881/aad80a}

\end{thebibliography}
\end{document}